\documentclass[useAMS,usenatbib]{mn2e}
\usepackage{graphicx}
\usepackage{epsfig}

\title[Spectra Study of Type 1 AGN - III. Broadband SED Properties]{A combined Optical and X-ray Spectra Study for Type 1 AGN. III. Broadband SED Properties}

\author[C. Jin, M. Ward, C. Done]
{Chichuan Jin\thanks{E-mail: chichuan.jin@durham.ac.uk},
Martin Ward, Chris Done\\
Department of Physics, University of Durham, South Road, Durham, DH1 3LE, UK}

\begin{document}

\date{Accepted by MNRAS}

\maketitle

\label{firstpage}

\begin{abstract}
In this third paper in a series of three,
we present a detailed study of the broadband spectral energy
distribution (SED) of active galactic nuclei (AGN) based on a nearby
unobscured Type 1 AGN sample.
We perform a systematic cross-correlation study of 
several key parameters, i.e. $\Gamma_{2-10keV}$, $L_{2-10keV}$,
$L_{bol}/L_{Edd}=\lambda_{Edd}$, $L_{bol}/L_{2-10keV}=
\kappa_{2-10keV}$, $L_{bol}/L_{5100A}= \kappa_{5100A}$, 
FWHM$_{H\beta}$, M$_{BH}$, $\alpha_{ox}$, $\alpha_{X}$ and $\alpha_{UV}$.
The well defined
spectral properties of the sample enable us to improve existing
relations and to identify new correlations among these parameters. We
confirm a break region around FWHM$_{H\beta}$ $\simeq$ 4000 km s$^{-1}$ in
the $\Gamma_{2-10keV}$ vs. FWHM$_{H\beta}$ correlation and
Log(M$_{BH}$) $\simeq$ 8.0 in the $\Gamma_{2-10keV}$ vs. M$_{BH}$
correlation, where these correlations appear to change form. Beyond
the break point the intrinsic $\Gamma_{2-10keV}$ index is dispersed
around 1.8.  Several new correlations are also reported in this paper
e.g. strong correlations in $\kappa_{5100}$ vs. $\lambda_{Edd}$,
$\kappa_{5100}$ vs. $\kappa_{2-10keV}$ and $\kappa_{2-10keV}$
vs. M$_{BH}$. The principal component analysis (PCA) is performed on
the correlation matrix of the above parameters. This shows that
the three physical parameters, i.e. black hole mass, mass accretion rate
and Eddington ratio, drive the majority of the correlations.
This is consistent with PCA results found from previous optical spectral studies.

For each key parameter, we split the AGN into three sub-samples, binned
based on increasing value of that parameter. We co-add the model SEDs for
each object in the sub-sample to see how the SED changes with that
parameter. Most parameters, except L$_{bol}$, show similar systematic changes in the
SED such that the temperature at which the disc peaks is correlated
with the ratio of power in the disc versus the Comptonised components
and the hard X-ray spectral index. This underlying change in SED shape
shows that AGN do exhibit intrinsically different spectral states. This
is superficially similar to the SED differences in BHB (black hole binary) seen as
$\lambda_{Edd}$ increases, but the analogy does not hold in detail. 
Only objects with the highest $\lambda_{Edd}$  appear to correspond to a BHB
spectral state (the disc dominated high/soft state). The AGN with typical mass
accretion rates have spectra which do not match well 
with any state observed in BHB. We speculate that this could be due to the 
presence of a powerful UV line driven disc wind, which complicates simple mass 
scaling between stellar and supermassive black holes. 

\end{abstract}


\section{Introduction}
\label{section:introduction}

The broadband spectral energy distribution (SED) of active galactic
nuclei (AGN) has been studied for many years. The unobservable gap
between the extreme UV and soft X-rays, imposed by Galactic
photoelectric absorption sets a major barrier to reconstructing the
overall SED. This is particularly problematic because it is in this
energy range that the peak of the SED often occurs. Due to this
difficulty, most previous work focused on spectral properties of
either the optical/UV side (e.g. Sanders et al. 1989; Francis et al. 1991;
Zheng et al. 1997; Vanden Berk et al. 2001), or the X-ray side
(e.g. Wilkes \& Elvis 1987; Green et al. 1995; George et al. 2000).
However, the multi-wavelength combination is
required to see the overall SED behaviour
(e.g. Elvis et al. 1994; Laor et al. 1997; Richards et al. 2006;
Shang et al. 2005; Lusso et al. 2010, hereafter: Lusso10).
In order to recover the spectral shape in the
UV/soft X-ray gap, some studies simply connected the optical/UV and
X-ray spectra to estimate the SED (e.g. Elvis et al. 1994;
Richards et al. 2006; Shang et al. 2005; Lusso10).
As the origin of the optical and hard X-ray emission became
clearer, more broadband SED models with multiple physical components
were used to fit the multi-waveband spectra. These were used to
extrapolate across the UV gap based on the model assumptions, but with
better justification than before (e.g. Middleton, Done \& Gierli\'{n}ski 2007;
Vasudevan \& Fabian 2007, hereafter: VF07; Vasudevan \& Fabian 2009,
hereafter: VF09; Jin et al. 2009; Grupe et al. 2010, hereafter:
Grupe10). For example, VF07 used a power law plus accretion disc model
to construct the broadband SED for their sample. This technique was
repeated in VF09 based on 29 AGNs from the reverberation mapping
sample (Peterson et al. 2004), and using simultaneous
optical/UV and X-ray data from XMM-Newton. Grupe10 used an absorbed
power law for the X-ray spectrum, and an exponentially cut-off power
law (or a broken power law) for the optical/UV spectrum, to construct
the broadband SED for their sample, for which simultaneous optical,
UV, X-ray data are available from SWIFT. Further discussion of some
previous SED studies is given in Jin et al. (2011) (hereafter: Paper-I).

Within these works, some specific AGN parameters were studied,
including black hole mass (M$_{BH}$), Eddington ratio
($L/L_{Edd}=\lambda_{Edd}$), bolometric luminosity (L$_{bol}$), 2-10 keV
luminosity (L$_{2-10keV}$), 2-10 keV photon index
($\Gamma_{2-10keV}$), H$\beta$ FWHM. Also, some parameters directly
related to the broadband SED shape were proposed, such as the optical
to X-ray spectral index
($\alpha_{ox}$\footnote{$\alpha_{ox}=-\frac{Log(F(2keV)/F(2500\AA))}{2.605}$,
assuming $F(\nu) \propto \nu^{-\alpha}$ to ensure a non-negative $\alpha_{ox}$.}, Tananbaum et al. 1979),
2-10 keV bolometric correction ($\kappa_{2-10keV}$, defined as L$_{bol}$/L$_{2-10keV}$,
e.g. VF07), 5100{\AA} luminosity scaling factor ($\kappa_{5100}$,
defined as L$_{bol}$/L$_{5100}$, where L$_{5100}$ is the monochromatic
luminosity at 5100{\AA}, Kaspi et al. 2000).

Since the number of AGN with both high quality optical/UV and X-ray spectra
is relatively small,  much effort  is devoted to search for correlations among
the key SED parameters, especially those parameters capable of representing
the broadband SED shape. Then, for those many AGNs lacking sufficient spectral
information, these correlations can be used to predict the SEDs that cannot be
defined from direct observation. Indeed, many such parameter correlations have
been proposed. For example, VF07 reported a strong correlation between
$\kappa_{2-10keV}$ and $\lambda_{Edd}$ (VF07; VF09; Lusso10). Correlations
were also found between  H$\beta$ FWHM and $\Gamma_{2-10keV}$
(e.g. Leighly 1999; Reeves \& Turner 2000; Shemmer et al. 2006,
hereafter: S06; Shemmer et al. 2008, hereafter: S08;
Zhou \& Zhang 2010, hereafter: Zhou10a), between $\lambda_{Edd}$
and $\Gamma_{2-10keV}$ (e.g. Lu \& Yu 1999; Porquet et al. 2004;
Wang, Watarai \& Mineshige 2004; Bian 2005; S06,08),
and between H$\beta$ FWHM and $\lambda_{Edd}$ (e.g. Grupe10; Paper-I).
The correlation between L$_{2500}$ and L$_{2keV}$ is another important result,
which led to further correlation studies related to $\alpha_{ox}$, L$_{2500}$,
L$_{2keV}$ and redshift (e.g. Green et al. 2009, see references given in
Table~\ref{slope:compare:table}). In table~\ref{former:study:summary:table},
we list some of the principal parameters, and list the relevant papers
discussing the correlations.  Later in the papers we will discuss some
additional correlations e.g. M$_{BH}$ vs. $\Gamma_{2-10keV}$ and M$_{BH}$ vs.
$\kappa_{2-10keV}$, in the context that changes in these parameters are  caused by
changes in the fundamental physical processes.

However, due to the difficulty in obtaining and analyzing both optical/UV
and X-ray spectra for a large sample, these parameter correlations are
reported separately and based on different samples, rather than being studied
systematically for a single well-defined sample. Furthermore, most of the
previous samples were not selected based on their spectral properties,
so effects such as reddening will introduce biases into the cross-correlations.
The lack of a self-consistent physically motivated broadband model,
has also been a constraint to performing a more detailed SED study.

In Paper-I, we defined a sample of 51 Type 1 AGNs with both optical/UV
and X-ray spectra which are of high quality, and without evidence of
complex spectral absorption features e.g. a warm absorber. Based on
this bright and unobscured Type 1 AGN sample, we applied our latest
optical and broadband SED model to perform the spectral fitting, and
so matched the optical spectrum and produced a broadband SED for each
AGN in the sample. This is so far the most detailed spectral analysis
for a medium sized sample of AGNs, with such well defined high quality
spectra.  In our preceding paper (Jin, Ward \& Done 2012, hereafter:
Paper-II), we studied the Balmer emission line properties and the
Relation between the optical and X-ray emission, based on
the same sample and the spectral fitting results given in Paper-I.
This paper is based on the sample of Paper-I,
but focuses on the shape of the broadband SED, and how this
relates to a range of parameters from the model. We will approach
this problem by first investigating the numerous correlations
previously reported.  Then a set of mean SEDs based on key parameters
are constructed and studied in detail.

\begin{table*}
 \centering
  \begin{minipage}{175mm}
   \caption{The mean parameter values with one standard deviation for our sample, together with some recently published samples. N: sample size. $\kappa_{2-10}$: the 2-10 keV bolometric correction. Lusso10: Lusso et al. (2010); Grupe10: Grupe et al. (2010); Zhou10a: Zhou \& Zhang (2010); VF07,09: Vasudevan \& Fabian (2007, 2009). {\it m\/}: the Eddington ratios were calculated using the luminosity dependent 2-10 keV bolometric correction in Marconi et al. (2004); {\it r\/}: the black hole masses are from the reverberation mapping study in Peterson et al. (2004). $*$: there are 114 sources in Zhou10a, but we only consider the 69 NLS1s and BLS1s with both H$\beta$ FWHM and $\lambda_{Edd}$ measurements.}
   \label{sample-compare}
     \begin{tabular}{@{}ccccccccccc@{}}
\hline
&N&Redshift&$\Gamma_{2-10keV}$&$\kappa_{2-10}$&$\kappa_{5100}$&$\lambda_{Edd}$&FWHM$_{H\beta}$&M$_{BH}$&$\alpha_{ox}$&L$_{bol}$\\
Sample&&&&&&&$km~s^{-1}$&{\it log\/}($M_{\sun}$)&&{\it log\/}({\it erg~s$^{-1}$\/})\\
\hline
This Work&51&0.137$^{+0.158}_{-0.073}$&1.91$\pm$0.26&38$^{+58}_{-23}$&15$^{+14}_{-7}$&0.27$^{+0.61}_{-0.19}$&3560$^{+3880}_{-1860}$&7.93$\pm$0.52&1.35$\pm$0.14&45.47$\pm$0.57\\
Lusso10&545&1.440$^{+1.020}_{-0.597}$&---&27$^{+28}_{-14}$&---&---&---&---&1.40$\pm$0.16&45.54$\pm$0.57\\
Grupe10&92&0.112$\pm$0.077&---&---&21$\pm$3&1.87$\pm$3.26&2670$\pm$1670&7.37$\pm$0.73&1.42$\pm$0.17&45.00$\pm$0.96\\
Zhou10a&$^{*}$69&0.050$^{+0.103}_{-0.034}$&1.97$\pm$0.29&---&---&$^{m}$0.24$^{+0.76}_{-0.18}$&2600$^{+2500}_{-1280}$&---&---&---\\
VF09&29&0.033$^{+0.074}_{-0.023}$&1.85$\pm$0.32&28$^{+74}_{-20}$&---&0.18$\pm$0.16&---&$^{r}$7.93$\pm$0.66&1.39$\pm$0.24&44.89$\pm$1.00\\
VF07&54&0.064$^{+0.147}_{-0.044}$&---&26$^{+39}_{-16}$&---&0.15$^{+0.76}_{-0.13}$&---&7.89$\pm$0.82&---&45.20$\pm$1.01\\
\hline
   \end{tabular}
 \end{minipage}
\end{table*}

These mean SEDs show a clear change in shape
as a function of the AGN parameters, 
probably most fundamentally driven by a change in
$\lambda_{Edd}$. The SED becomes more disc dominated and the X-ray
tail softens as $\lambda_{Edd}$ increases (e.g. Grupe et al. 2010).
This is superficially similar to the changes seen in the Galactic black
hole binary (BHB) systems as the mass accretion rate increases.
In BHB at low Eddington ratio ($\lambda_{Edd}\le 0.02$) the X-ray
spectra are hard, $\Gamma_{2-10keV}<2$, and the disc emission is weak
(low/hard state). As the mass accretion rate increases the disc
increases in importance relative to the tail, and the tail steepens
until it reaches the disc dominated `thermal state' (also known as the
high/soft state) where the power law tail is somewhat steeper
$\Gamma_{2-10keV}\sim 2-2.2$. At high Eddington fractions the source
can also show a `very high' or steep power law state, 
where both disc and tail contribute to the spectra, but the
spectra are typically steep ($\Gamma_{2-10keV}\ge 2.2$)
(Remillard \& McClintock 2006; Meyer-Hofmeister, Liu \& Meyer 2009).
It has been proposed that AGNs are simply
scaled up counterparts to the BHBs, so they should also show similar
spectral states (Done \& Gierli\'{n}ski 2005; McHardy et al. 2006;
Gierli\'{n}ski et al. 2008; Middleton et al. 2009; Jin et al. 2009).
Therefore the Eddington ratio may also be a good
indicator an AGN's accretion state, and so determine the shape of its
SED, explaining the multiple strong correlations between various SED
parameters.

This paper is organized as follows. Section 2 gives a brief
description of the main characteristics of the sample, and the method
used for fitting the broadband SED. Further details of this are
presented in Paper-I. In Section 3 we examine the parameter
correlations across five parameter groups. Some significant new
correlations are proposed. In Section 4 we perform a systematic study
of all correlations among several selected key parameters, and build
a cross-correlation matrix. The principal component analysis technique
is used on this correlation matrix in order to derive the
eigenvectors. In Section 5 various mean SEDs are constructed, based on
the mean values of the key parameters. In Section 6 we investigate
further the reliability of these correlations, including the effect of
a correction to the black hole mass from 
radiation pressure.  In Section 7 we
summarize our results, and propose topics for further study. A flat
universe model with a Hubble constant of H$_{0} = 72$ km s$^{-1}$
Mpc$^{-1}$, $\Omega_{M} = 0.27$ and $\Omega_{\Lambda} = 0.73$, is
adopted throughout the paper.

\section{Properties of the Sample and Broadband SED Modeling}
\label{sample:property}

\subsection{Sample Selection and Properties}
The AGN sample used in this paper is reported in Paper-I, based on the
cross-correlation of SDSS DR7 and XMM-Newton catalogs.
The complete source list, detailed sample selection criteria and
sample properties can be found in Paper-I.
The main selection steps are summarized below:\\
(1) We searched the 2XMMi and SDSS DR7 catalogs and identified
3342 extragalactic sources having both X-ray and optical spectra.\\
(2) Within these sources, we selected those with $H\beta$ in emission and
redshift $z < 0.4$, so that both the H$\alpha$ and H$\beta$ emission
lines are covered by the SDSS spectra. This assists with modelling
of the Balmer lines (see Paper-I).  This selection resulted in 802
unique X-ray sources.\\
(3) Within this subsample, we identified 96
Type 1 AGNs all with a minimum of 2000 counts in at least one of the three
XMM-Newton EPIC cameras, to ensure high X-ray spectral quality.\\
(4) We then excluded 23 sources whose $H\beta$ line was modified due to
high reddening, low S/N or a data gap in the SDSS spectra. The
resulting sample contains 73 AGNs.\\
(5) For each of the 73 sources, a
power law model was fitted to the rest-frame 2-10 keV X-ray
spectra. The 16 objects with photon index uncertainties greater than 0.5
were excluded, leaving 57 Type 1 AGNs with relatively well
constrained 2-10 keV spectra.\\
(6) A further 6 objects were excluded
because of an obvious signature of a warm absorber at $\sim$0.7 keV.
This criterion means that the observed spectra are very likely directly
related to emission from the bare central core.\\

The final sample contains 51 AGNs, with 12 AGNs classified as
NLS1 using the conventional definition (Goodrich 1989), while the
others are all BLS1s. The BAL quasar PG 1004+130 is also included
in our sample whose weak and featureless X-ray emission is still
under debate (Miller et al. 2006)
The vast majority of the sample are radio-quiet with only 
three radio-loud sources i.e. PG 1004+130, RBS 0875 and PG 1512+370.\\
The H$\beta$ FWHM of the sample ranges from
600 km s$^{-1}$ to 13000 km s$^{-1}$.

\subsection{The Spectral Modelling}
Based on the high quality spectra, Paper-I conducted detailed spectral
analysis for each source in the sample. In the optical spectral fitting,
the H$\alpha$ and H$\beta$ lines were fitted using three components.
The narrow component has the same profile as the entire [OIII]$\lambda$5007 line,
i.e. including both the central and blue component in [OIII]$\lambda$5007.
The intermediate and broad components are assumed have the shape of a Gaussian.
All other strong nearby emission lines are included by adding more
Gaussian profiles into the whole model. Then a complete model with
multiple components were used to fit the whole SDSS spectra, including
the underlying continuum approximated by a power law, the Balmer
continuum, the FeII `false' continuum and all strong emission lines. 
The broadband SED model used in Paper-I ({\it optxagn\/}, hereafter: Model-A)
consists of the following three continuum components:
1. emission from a modified standard accretion disc, whose energy within
the corona radius is completely reprocessed into the other two high energy
Comptonisation components.
2. emission from the low temperature, optically thick
Comptonisation, which mainly accounts for the soft X-ray excess.
3. emission from the high temperature, optically thin Comptonisation
which gives the power law shape of the hard X-ray spectrum above 2 keV.
Both Galactic extinction and the small amount of intrinsic
reddening/extinction are included in the model.

\begin{figure}
\centering
\includegraphics[scale=0.5,clip=1]{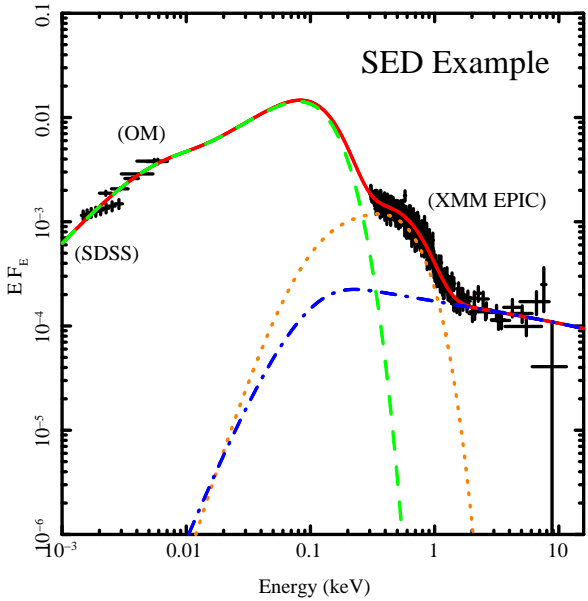}
\caption{An example of the broadband SED fitting using {\it optxagnf \/} model in {\tt Xspec\/} v12 which includes the effect of a colour correction. The data is taken from SDSS and XMM-Newton observations of RBS 769. The solid red line shows the total model; the dashed green line shows the colour corrected and truncated accretion disc emission; the dotted orange line shows the low temperature optically thick Compotonisation; the dot-dash blue line shows the high temperature optically thin Compotonisation. The reduced $\chi^{2}$ is 1.16 for this spectral fitting.}
\label{sed:example:plot}
\end{figure}

Because the model was in the process of development,
Paper-I did not consider the effect of a colour temperature
correction in the accretion disc model
(e.g. Ross, Fabian \& Mineshige 1992; Davis \& Hubeny 2006). 
This effect is due to the fact that
the absorption opacity decreases significantly as the black hole mass increases 
($\kappa_{abs}~\propto~M^{-1/8}$, Done et al. 2011), so for the same Eddington ratio,
AGN's disc has a lower absorption opacity than BHB. Then for some AGNs the accretion
disc may no longer be locally thermalised. The higher temperature photons
can emerge from regions deeper in the disc, and so the disc spectrum extends to higher
energy than for standard accretion disc spectrum, producing the effect of
a colour temperature correction.
However, this can only happen for disc regions of $T~>~3{\times}10^4$ K
where sufficient hydrogen atoms are ionized.
In addition, the electron scattering opacity also becomes important in these regions.
The lower energy electrons in the surface regions Compton down-scatter
the higher temperature photons from deeper disc regions, thus reducing the
colour temperature correction. These two competing effects lead to an effective
colour temperature correction (Davis et al. 2006).

The maximum effective temperature of the accretion disc is
$kT~\sim~10(\dot{m}/M_{8})^{1/4}$ eV
(where $\dot{m}=L_{bol}/L_{Edd}$, $M_{8}=M/10^{8}$ M$_{\sun}$),
so only for AGNs with low mass black holes and high mass accretion
rates such as the NLS1s (e.g. Boller, Brandt \& Fink 1996; Paper-I),
electron scattering opacity will dominate,
and the effect of colour temperature correction becomes important.
A typical colour temperature correction of 2.6 is predicted for
an AGN with M$_{BH}$=10$^{6}$M$_{\sun}$, $\lambda_{Edd}$=1.0
(Davis, Done \& Blaes 2006; Done et al. 2011).
This effect combines with an already hot disc due to the low black hole mass,
resulting in a disc spectrum that
extends significantly into the soft X-ray range.

A more detailed explanation of this colour temperature correction
in AGN can be found in Done et al. (2011), where a model
({\it optxagnf} in {\tt Xspec} v12) is proposed which applies a colour 
temperature correction for the Comptonised accretion disc 
model (hereafter: Model-B). In this paper we
use this more advanced model to fit the optical, UV and X-ray 
data so as to construct the broadband SED for each source. 
Figure~\ref{sed:example:plot} shows an example of the broadband 
SED fitting to the multi-waveband spectra of RBS 769. 
A detailed description of Model-B in comparison with the earlier
model {\it optxagn\/} (Model-A), can be found in Done et al. (2011).
The set of fitting parameters for each source, based on Model-B, is given in 
Table~\ref{app:SED-fitting-parameters}.
Section~\ref{section:color:correction} discusses the statistical
differences resulting from using Model-B and Model-A for our sample.

All the principal SED parameters such as $\lambda_{Edd}$,
$\kappa_{2-10keV}$ and $\alpha_{ox}$ are calculated from 
the new model fitting (see Table~\ref{app:SED-key-parameters}). 
A cross-correlation study of the various parameters is then  
conducted. In Table~\ref{sample-compare} we compare the mean 
values of some SED parameters of our sample with those samples 
used in previous work. The result of correlations established in these previous
studies will be compared with ours in the following sections.

There are two sources that we treated as being anomalous in our study.
The first is PG 1004+130, a broad absorption line (BAL) quasar, whose
X-ray was reported as being extraordinarily weak.  Although its X-ray
spectrum does not show clear absorption edges, it is nevertheless
likely to be heavily absorbed, so that the remaining X-rays may have a different
origin such as a sub-parsec-scale jet. It has been suggested that the
X-ray emission from PG 1004+130, after correcting for its intrinsic
absorption, is 0.73 dex weaker than normal PG
radio loud quasars (PG RLQs) normalized to similar optical/UV luminosities
(Miller et al. 2006).  Due to its distinct X-ray spectrum
(and correspondingly different X-ray parameters), we did not include
this source in our regression analysis. The other anomalous source is
Mrk 110. This source shows
strong optical variability (Kollatschny et al. 2001; Kollatschny 2003)
and its SDSS spectrum has a different slope from the (non-simultaneous)
XMM-Newton OM data (see Paper-I).
However, the optical spectrum obtained using FAST shown in
Landt et al. (2011), is consistent with the OM data, and is also an
order of magnitude more brighter than the SDSS spectrum. Therefore the
SDSS spectrum of Mrk 110 is not consistent with its broadband SED
parameters, and so we exclude Mrk 110 from any correlation which
depends on optical parameters, but include it for any UV/X-ray 
correlations.

\begin{table*}
  \begin{minipage}{175mm}
  \centering
    \label{slope:compare:table}
  \caption{Comparison of regression line coefficients for L$_{2keV}$, L$_{2500}$, $\alpha_{ox}$ and redshift Correlations. Lusso10: Lusso et al. (2010); Grupe10: Grupe et al. (2010); Green09: Green et al. (2009); Just07: Just et al. (2007); Steffen06: Steffen et al. (2006); Strateva05: Strateva et al. (2005); Hasinger05: Hasinger (2005). {\it opt\/}: optically selected sample. {\it xray\/}: X-ray selected sample. $^{1}$linear regression results using L$_{2keV}$ and L$_{2500}$ from the reconstructed broadband SED corrected for both intrinsic and Galactic reddening/extinction.
   $^{2}$linear regression when L$_{2keV}$ and L$_{2500}$ were not corrected for the best-fit intrinsic reddening/extinction, but corrected for the Galactic value.
   $^{a}$for the SDSS main sample $+$ high-z sample $+$ Sy 1 Sample (see Strateva05).
   $^{*}$measured directly from the regression line in Fig.5(b) of Hasinger05.}
     \begin{tabular}{lcccccc}
\hline
 & \multicolumn{2}{c}{L$_{2keV}$ vs. L$_{2500}$}&\multicolumn{2}{c}{$\alpha_{ox}$ vs. L$_{2500}$}&\multicolumn{2}{c}{$\alpha_{ox}$ vs. L$_{2keV}$}\\
 Sample& $\beta_{1}^{bi}$ & $\xi_{1}^{bi}$& $\beta_{2}^{em}$ & $\xi_{2}^{em}$& $\beta_{3}^{em}$ & $\xi_{3}^{em}$\\
\hline
$^{1}$This Work&0.95$\pm$0.06&-2.04$\pm$1.77&0.07$\pm$0.02&-0.61$\pm$0.60&-0.03$\pm$0.03&2.17$\pm$0.76\\
$^{2}$This Work-r$_{int}$&0.91$\pm$0.05&-0.69$\pm$1.68&0.08$\pm$0.02&-0.92$\pm$0.57&-0.01$\pm$0.03&1.55$\pm$0.76\\
\hline
Lusso10$^{xray}$&0.76$\pm$0.02&3.51$\pm$0.64&0.15$\pm$0.01&-3.18$\pm$0.22&0.02$\pm$0.01&0.86$\pm$0.34\\
Grupe10$^{opt}$&---&---&0.11$\pm$0.01&-1.18$\pm$0.31&---&---\\
Green09$^{opt}$&1.12$\pm$0.02&-7.59$\pm$0.64&0.06$\pm$0.01&-0.32$\pm$0.26&0.10$\pm$0.01&1.38$\pm$0.21\\
Just07$^{opt}$&0.71$\pm$0.01&4.88$\pm$0.63&0.14$\pm$0.01&-2.71$\pm$0.21&0.09$\pm$0.01&-0.90$\pm$0.36\\
Steffen06$^{opt}$&0.72$\pm$0.01&4.53$\pm$0.69&0.14$\pm$0.01&-2.64$\pm$0.24&0.08$\pm$0.02&-4.1$\pm$0.39\\
Strateva05$^{opt}$&0.65$\pm$0.02$^{a}$&6.73$\pm$0.64$^{a}$&0.14$\pm$0.01$^{a}$&-2.62$\pm$0.25$^{a}$&---&---\\
Hasinger05$^{xray}$&1.0$^{*}$&-3.7$^{*}$&---&---&---&---\\
\hline
   \end{tabular}
 \end{minipage}
\end{table*}

\section{Investigating the Correlations for Various SED Parameter Groups}
In this section we divide the SED parameters into several sub-groups,
based on correlations reported in the literature (see
Section~\ref{section:introduction}), and then perform a
cross-correlation analysis within each group.

\subsection{Group 1: L$_{2500}$, L$_{2keV}$ and $\alpha_{ox}$}
\label{group1}

The $\alpha_{ox}$ index has been used as the indicator of the SED
shape for more than thirty years. It is often used as an indication of
the AGN's broadband SED, and to convert between the AGNs' optical
luminosity function (OLF) and X-ray luminosity function (XLF)
(e.g. Hopkins, Richards \& Hernquist 2007, hereafter: Hopkins07 ; Tang, Zhang \& Hopkins 2007, hereafter: Tang07).

Many studies have been carried out on the evolution of $\alpha_{ox}$
with luminosity (e.g. Avni \& Tananbaum 1982;
Wilkes et al. 1994; Strateva et al. 2005; Steffen et al. 2006;
Just et al. 2007; Green et al. 2009, hereafter: Green09; Lusso10; Grupe10),
which may provide clues on the emission mechanism. The value
of $\alpha_{ox}$ has been found in the range 1.2$\sim$1.8, with a mean
value of $\sim$1.5. Correlations have also been found between
L$_{2keV}$, L$_{2500}$ and $\alpha_{ox}$, with the primary
correlation being L$_{2keV}\propto~$L$_{2500}^{\beta}$. The slope
index $\beta$ was often found to deviate from unity for both optically
selected (e.g. Strateva et al. 2005; Steffen et al. 2006; Just et al. 2007)
and X-ray selected AGN samples (e.g. Lusso10).
However, La Franca, Franceschini \& Cristiani (1995)
re-analyzed Wilkes et al. (1994)'s sample by
considering both variables and intrinsic scattering, and found that
$\beta$ was consistent with unity. Green09 collected a large,
well-defined sample of 2308 SDSS/ChaMP QSOs in the redshift range
0.2$\sim$5.4, and concluded that $\beta$ is not less than unity.

The reality of a non-linear correlation in L$_{2keV}$ vs. L$_{2500}$
remains an open question, but one possible explanation could be a
selection effect in a flux limited sample for which dispersions in
the optical and X-ray luminosity are not equal, or which change with
cosmic time (Yuan, Siebert \& Brinkmann 1998, hereafter: Yuan98; Tang07). However,
the possibility of a truely intrinsic non-linear correlation cannot
be ruled out. A non-linear L$_{2keV}$ vs. L$_{2500}$ correlation
implies that there is a dependence of $\alpha_{ox}$ on L$_{2keV}$ and
L$_{2500}$ (e.g. Vignali, Brandt \& Schneider 2003; Just et al. 2007; Lusso10),
but this is still a matter of debate (Yuan98; Tang07).

To further test the basis of these correlations, 
we also calculated the values of  L$_{2keV}$, L$_{2500}$ and $\alpha_{ox}$ from our 
best-fit model of the SEDs, and then preformed the same cross-correlation analysis. 
The limitation of our results arises from the lack of actual  
spectral coverage at 2500{\AA} for the 16 sources without 
OM UVW1 and UVM2 data. The luminosity and redshift range of 
our sample is also relatively small. But on the merit side  
we have included two inputs of reddening/absorption to model 
both the Galactic and the AGN's intrinsic extinction during the 
broadband SED fitting, so our values of  L$_{2keV}$, L$_{2500}$ and 
$\alpha_{ox}$ should be closer to those of the intrinsic source. The unobscured 
nature of our sample and the exclusion of warm absorber 
sources also helps reduce uncertainties in the corrections 
caused by reddening/absorption.

\subsubsection{The L$_{2keV}$ vs. L$_{2500}$ Correlation}
\label{l210:l2500:section}
\begin{figure}
\centering
\includegraphics[bb=33 120 570 650,scale=0.46,clip=1]{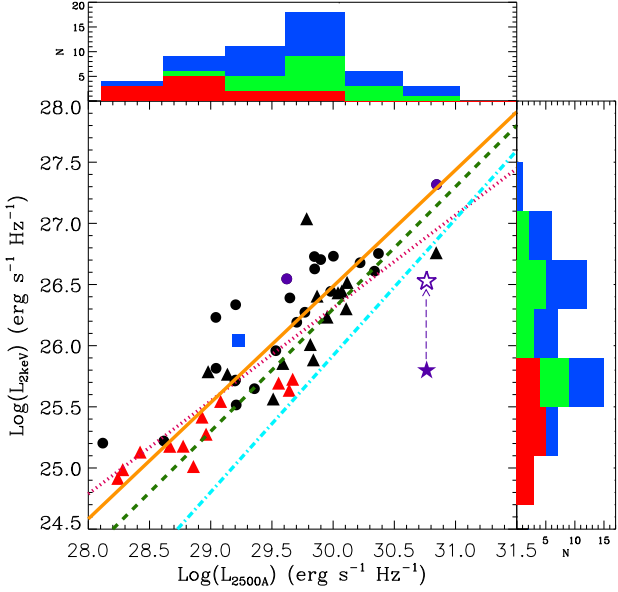}
\caption{The cross-correlatin between L$_{2500}$ and L$_{2keV}$. 
The solid orange line is the bisector regression line for our 
sample. The red triangle symbols represent NLS1s; 
purple symbols show the radio loud AGN; the blue square symbol is Mrk 110. 
The filled purple star is the BAL-quasar PG 1004+130, and the open 
purple star is the position if its intrinsic X-ray flux was 
0.73 dex higher (Miller et al. 2006). All triangle symbols represent 
Population A sources whose H$\beta$ FWHM is less than 
4000 km s$^{-1}$. In the two histograms the green and red regions 
are for the Population A sources, and the red region indicates  
the 12 NLS1s. The dashed green line is based on Hasinger05; the dash-dotted
cyan line is based on Green09; the dotted pink line is based on Lusso10.}
\label{l2keV:l2500:plot}
\end{figure}

Figure~\ref{l2keV:l2500:plot} shows our L$_{2keV}$ vs. L$_{2500}$
correlation. The statistical methods used are the same as in
Lusso10, i.e. we use the 
full parametric estimate and maximized regression (EM)
algorithm. We use this to derive two
regression lines assuming, first L$_{2keV}$, then L$_{2500}$ to be the
independent variable. Then the bisector of the two regression lines is
calculated using the equations in Isobe et al. (1990). This method is more
appropriate in cases where the cross-correlations are dominated by
intrinsic scatter. The correlations found are as follows:
\\ (i) the EM regression line, when L$_{2500}$ is assumed to be the
independent variable:
\begin{equation}
\label{l2kev:l2500:eqn:1}
Log(L_{2keV})=(0.83{\pm}0.05)Log(L_{2500}) + (1.59{\pm}1.55)
\end{equation}
(ii) the EM regression line, when L$_{2keV}$ is assumed to be the independent variable:
\begin{equation}
\label{l2kev:l2500:eqn:2}
Log(L_{2keV})=(1.09{\pm}0.09)Log(L_{2500}) - (6.17{\pm}2.66)
\end{equation}
(iii) the bisector of the above two regression lines (the solid orange line in Figure~\ref{l2keV:l2500:plot})
\begin{equation}
\label{l2kev:l2500:eqn:3}
Log(L_{2keV})=(0.95{\pm}0.06)Log(L_{2500}) - (2.04{\pm}1.77)
\end{equation}
The Spearman's rank test gives a rank coefficient of $\rho_{s}=0.87$, 
and the probability of deviation from a random distribution is 
$d_{s}=1.2{\times}10^{-16}$, confirming a very high level of
significance. 
We superimpose 
PG 1004+130 (filled purple star) on 
Figure~\ref{l2keV:l2500:plot}, showing that it lies far from the
correlation due to 
its unusual X-ray weakness. It matches much better to
the regression line if corrected  in L$_{2keV}$ by 0.73 dex 
(Miller et al. 2006, the empty purple star in 
Figure~\ref{l2keV:l2500:plot}).

Our correlation between L$_{2keV}$ vs. L$_{2500}$  is close to linear, but 
previous studies have found a wide range of values as listed in 
Table~\ref{slope:compare:table}. Figure~\ref{l2keV:l2500:plot} plots
these results for comparison. The correlation found by
Hasinger05 (green dashed line) which is based on an X-ray selected
sample, is the most consistent with our sample, whereas the slope
found by Lusso10 (pink dashed line) is significantly flatter. Our
slope is also consistent with Green09 (cyan
dashed line) in which a large sample of optically selected quasars is
analyzed. We note that our sample only covers the low luminosity
region of the sample in Green09. The NLS1s are the least luminous
sources. The different value of the Y-axis intercept in Green09 may be
due to their larger sample and larger luminosity dispersion.

There can be several reasons for the difference between our results
and Lusso10. Firstly there may be a selection effect of a flux limited
sample if there are different amounts of dispersion in optical and
X-ray luminosities (Yuan98; Tang07). More importantly, our sample is
corrected for both Galactic and intrinsic reddening/absorption in the
host Galaxy through the spectral fitting whereas that of Lusso10 is
only corrected for Galactic absorption. We remove the intrinsic
reddening correction and re-compute the EM regression, with results
given in Table~\ref{slope:compare:table} under the row `This
Work-r$_{int}$'
The dust reddening and gas absorption column are related by,
E(B-V)=1.7$\times$(N$_{H}$/10$^{22}$)cm$^{-2}$ (Bessell 1991), which
means that L$_{2500}$ is suppressed much more severely than
L$_{2keV}$.  Hence the removal of intrinsic reddening correction decreases
our correlation slope from 0.95 $(\pm$0.06) to 0.91 ($\pm$0.05).  
Thus the intrinsic reddening can flatten the correlation, but it
does not seem to be enough on its own to explain the difference with
Lusso10, unless their sample is strongly reddened in the optical.

\subsubsection{The $\alpha_{ox}$ vs. L$_{2keV}$ and L$_{2500}$ Correlations}

\begin{figure*}
\centering
\includegraphics[scale=0.52,clip=1,angle=90]{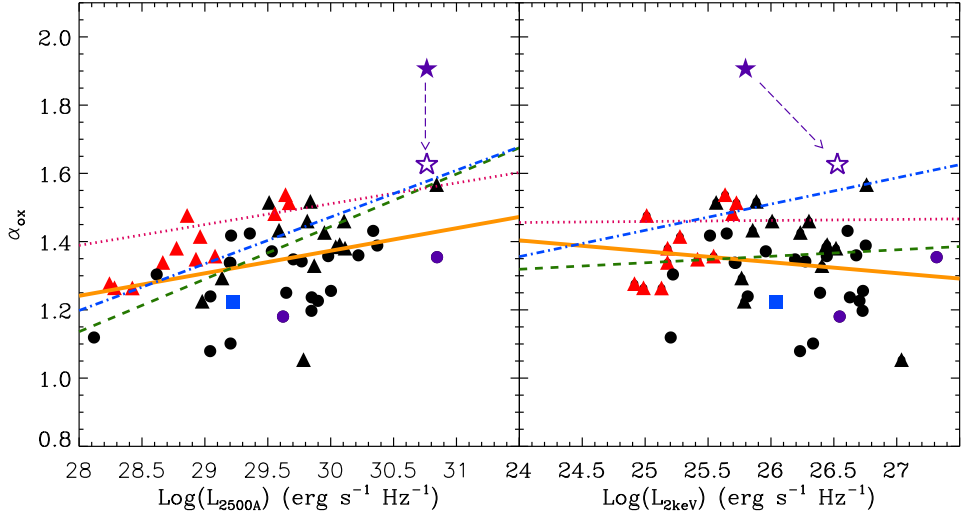}
\caption{$\alpha_{ox}$ vs. L$_{2500}$ and L$_{2keV}$. 
Each symbol represents the same type of source as in 
Figure~\ref{l2keV:l2500:plot}. In each panel the solid orange 
line is the OLS regression line for our sample, assuming the 
X-axis variable to be the independent variable. The dash-dotted blue line is 
based on Steffen06; the dotted pink line on Green09; and the dashed green line on Lusso10.}
\label{aox:selfcor:plot}
\end{figure*}

To further investigate the correlation between the optical/UV 
and X-ray continua, we adopt the same approach as in previous work to 
produce $\alpha_{ox}$ vs. L$_{2500}$ and $\alpha_{ox}$ vs. L$_{2keV}$ correlations.
If we assume L$_{2keV}\propto$ L$_{2500}^{\beta}$, then $\alpha_{ox}\propto$ 
L$_{2500}^{1-\beta}$ and $\alpha_{ox}\propto$ L$_{2keV}^{\beta(1-\beta)}$ 
are expected by definition.
However, the Spearman's rank test does not imply very strong correlations:
$\rho_{s}=0.31,d_{s}=0.03$ for $\alpha_{ox}$ vs. L$_{2500}$; 
$\rho_{s}=-0.13,d_{s}=0.35$, for $\alpha_{ox}$ vs. L$_{2keV}$.
The regression lines were derived but with large uncertainty. 
The results are presented in Figure~\ref{aox:selfcor:plot} and listed in
Table~\ref{slope:compare:table}. The solid orange line in each panel of 
Figure~\ref{aox:selfcor:plot} is our EM regression line, compared with 
some previous work shown by dashed lines. 
Note that our results are limited to redshift $z<0.4$, 
L$_{2500}<10^{+31}$ erg s$^{-1}$ Hz$^{-1}$ and
L$_{2keV}<10^{+27.5}$ erg s$^{-1}$ Hz$^{-1}$.
Our results also suggest that the cross-correlations in
$\alpha_{ox}$ vs. L$_{2500}$ and L$_{2keV}$ are 
dominated by the AGN's intrinsic dispersion.

\subsection{Group 2: $\alpha_{ox}$, $\kappa_{2-10keV}$ and $\lambda_{Edd}$}
\label{group2}
As mentioned in the previous section $\alpha_{ox}$ is often used as a proxy for the broadband SED shape. Since L$_{bol}$ is often dominated by the big blue bump (BBB) peaking in the unobservable EUV region (Walter \& Fink 1993). $\kappa_{2-10keV}$, defined as L$_{bol}$/L$_{2-10keV}$, is also an indicator of the SED shape. $\lambda_{Edd}$ is an important parameter which relates directly to  the accretion processes close to the central SMBH. Therefore, correlations are to be expected between $\lambda_{Edd}$, $\kappa_{2-10keV}$ and $\alpha_{ox}$.

\begin{figure}
\centering
\includegraphics[bb=54 144 594 650,scale=0.48,clip=]{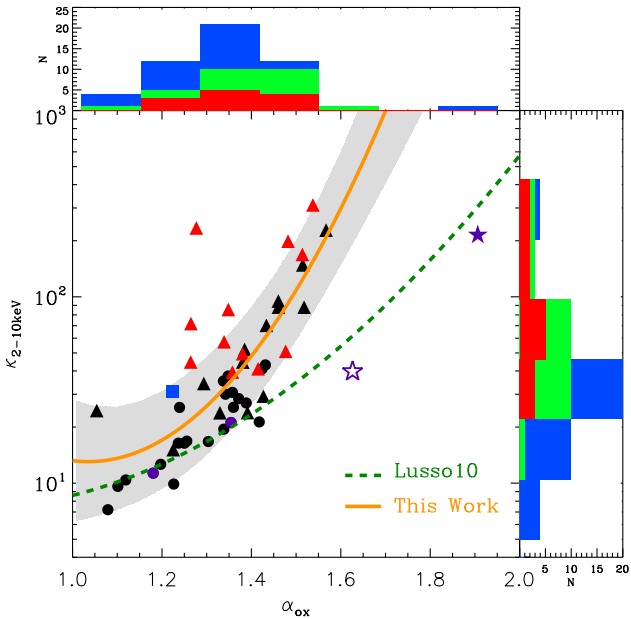}
\caption{$\kappa_{2-10keV}$ vs. $\alpha_{ox}$. Each symbol represents the same type of source as in Figure~\ref{l2keV:l2500:plot}. The solid orange line is the best fit line found using a second order polynomial, and the shaded area is the $\pm$1$\sigma$ zone.}
\label{aox:k210:plot}
\end{figure}

\subsubsection{The $\alpha_{ox}$ vs. $\kappa_{2-10keV}$ Correlation}
Lusso10 reported a tight second-order polynomial correlation
for $\kappa_{2-10keV}$ vs. $\alpha_{ox}$. We find a similar correlation 
for our sample.  Spearman's rank test shows $\rho_{s}$ = 0.73 
and $d_{s}$ = $2{\times}10^{-9}$. We also fitted a second-order 
polynomial to the correlation and obtained the following equation:
\begin{equation}
\label{k210-aox-eqn}
Log(\kappa_{2-10})=(5.7{\pm}3.4)-(8.8{\pm}5.2)\alpha_{ox}+(4.3{\pm}2.0)\alpha_{ox}^{2}
\end{equation}
Figure~\ref{aox:k210:plot} shows our best-fit polynomial (solid orange line) 
with $\pm$1$\sigma$ dispersion region (the shaded region). 
Note that our fit excludes BAL quasar PG 1004+130
(purple star in Figure~\ref{aox:k210:plot}).
The best-fit polynomial from Lusso10 is plotted as the 
dashed green line, which is not as steep as ours. 
The reason is that our value of $\kappa_{2-10keV}$ (and L$_{bol}$) 
is higher than found by Lusso10, especially for the narrow line 
objects (the average $\kappa_{2-10keV}$ for our 12 NLS1s 
is 86$^{+96}_{-45}$). Lusso10 constructed their 
broadband SEDs by first assuming a power law extending from the optical 
to 500{\AA}, then connecting the continuum at 500{\AA} 
linearly to that at 1 keV, and finally by extrapolating from 1 keV towards 
higher energies, using an exponentially cut-off power law. 
This model substantially underestimates L$_{bol}$ for narrow line objects 
because such objects often have strong soft-X-ray excesses which 
contain a large fraction of the L$_{bol}$ (Middleton et al. 2009;
Jin et al. 2009; Paper-I). Our detailed broadband SED fitting 
has modeled this soft-excess feature by including a low temperature optically 
thick Comptonization component. We claim that this results in a more accurate 
estimate of L$_{bol}$ (Paper-I). So certainly for the  
nearby Type 1 AGNs (redshift $< 0.4$), the $\kappa_{2-10keV}$ 
vs. $\alpha_{ox}$ correlation we find should be more reliable. 
How the correlation behaves at high redshift requires further study,
but Lusso10 has shown that such a second-order polynomial correlation
still holds for Type 1 AGNs up to redshifts $z~=~$4.

\subsubsection{The $\alpha_{ox}$ vs. $\lambda_{Edd}$ Correlation}
\begin{figure}
\centering
\includegraphics[bb=54 144 594 650,scale=0.48,clip=]{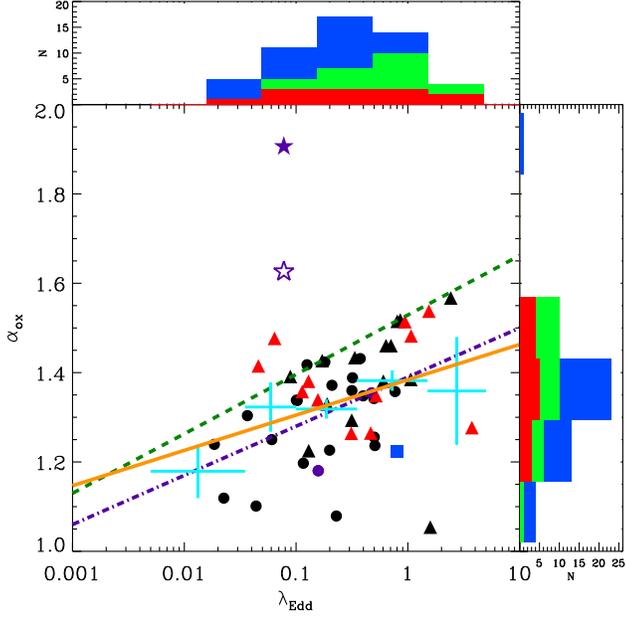}
\caption{$\lambda_{Edd}$ vs. $\alpha_{ox}$. Each symbol represents the same type of source as in Figure~\ref{l2keV:l2500:plot}. The solid orange line is the bisector regression line determined by our sample. The cyan crosses are the binned data points of our sample. The dashed green line is from Lusso10; the dash-dotted purple line is from Grupe10.}
\label{aox:eddr}
\end{figure}

The existence of a correlation of $\alpha_{ox}$ vs. $\lambda_{Edd}$ remains unclear. 
VF07 found no correlation between these quantities, and so they proposed that 
$\alpha_{ox}$ did not provide useful information on the 
broadband SED shape. S08 confirmed VF07's result 
for their sample of 35 moderate to high luminosity radio-quiet AGN. 
On the contrary, Lusso10 did find a correlation between 
$\alpha_{ox}$ and $\lambda_{Edd}$, although with a large dispersion. 
This was confirmed by Grupe10 for their sample containing 
92 soft X-ray selected AGNs, but their correlation was both flatter and 
stronger than that of Lusso10. We use our sample to investigate this
situation, and our results are shown in 
Figure~\ref{aox:eddr}. The Spearman's rank test gives 
$\rho_{s}=0.35$ and $d_{s}=1{\times}10^{-2}$, suggesting that  
a correlation does exist. We then applied the ordinary least squares
(OLS) regression, assuming $\lambda_{Edd}$ to be the independent variable, 
and found the following relation:
\begin{equation}
\label{aox:eddr:eqn}
\alpha_{ox}=(0.079{\pm}0.038)Log(\lambda_{Edd})+(1.384{\pm}0.029)
\end{equation}

In Figure~\ref{aox:eddr} we show our results. 
Our correlation has less dispersion than found by VF07 and Lusso10, 
but has larger dispersion than that from Grupe10. The exclusion of a correlation 
is at the $\sim$2$\sigma$ significance level, which is less significant 
than in Grupe10. Our regression line slope is consistent
with, but slightly flatter, than that in Lusso10 ($\beta$ = 0.133$\pm$0.023) 
and Grupe10 ($\beta$ = 0.11$\pm$0.02). This is partly because our 
estimation of L$_{bol}$ is higher than in previous studies, due to the inclusion 
of a soft X-ray excess in our model. Therefore our value of $\lambda_{Edd}$ is also 
higher for the NLS1s and other relatively narrow line objects. Another 
reason could be a selection effect. We have shown in the previous section that
$\alpha_{ox}$ has a luminosity dependence, thus at higher redshift we tend to
detect more luminous sources with steeper $\alpha_{ox}$.
While both our sample and Grupe10's have low redshifts (us: $z<0.4$; Grupe10: $z<0.3$),
Lusso10's sample covers a much larger range in redshift ($0.04<z<4.25$),
and indeed Lusso10's sample contains many objects with 
$\alpha_{ox}>1.5$ and $z>0.4$, which populate the empty region above 
$\alpha_{ox}=1.5$ in Figure~\ref{aox:eddr}, and create a larger dispersion.
The large dispersion observed in our study and previous work suggest that
one should be cautious about using the $\alpha_{ox}$ vs. $\lambda_{Edd}$ relation,
because the $\lambda_{Edd}$ inferred by $\alpha_{ox}$ may contain considerable uncertainties.

\subsubsection{The $\kappa_{2-10keV}$ vs. $\lambda_{Edd}$ Correlation}
\label{subsection:k210:eddr}
\begin{figure}
\centering
\includegraphics[bb=54 144 594 650,scale=0.48,clip=]{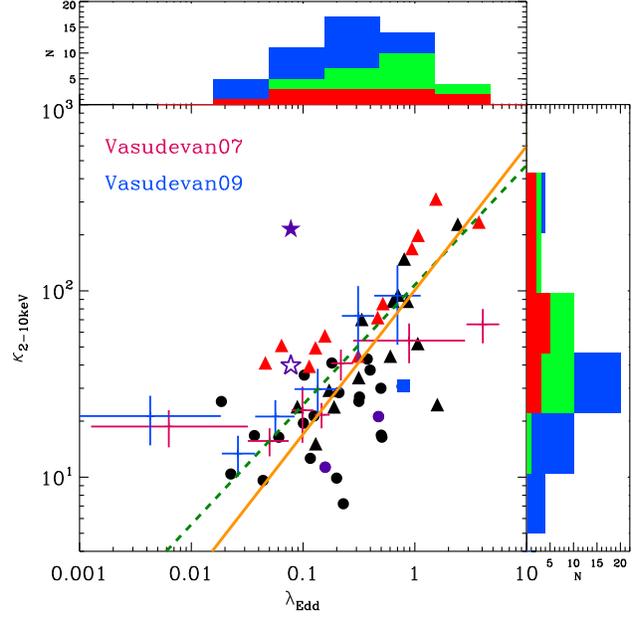}
\caption{$\lambda_{Edd}$ vs. $\kappa_{2-10keV}$. Each symbol represents the same type of source as in Figure~\ref{l2keV:l2500:plot}. The solid orange line is the bisector regression line determined for our sample. The binned data points are from VF07 (pink) and VF09 (blue). The dashed green line is from Lusso10.}
\label{k210:eddr}
\end{figure}

Wang, Watarai \& Mineshige (2004) reported the correlation between $\kappa_{2-10keV}$ and
$\lambda_{Edd}$, which was later confirmed by VF07,09. Most recently,
Lusso10 also found this correlation for the 545 X-ray selected type 1
AGNs from the XMM-COSMOS survey. They suggested that the observed
step change in this correlation does not result from the L$_{bol}$
dependence on both $\kappa_{2-10keV}$ and $\lambda_{Edd}$. In our
study we also find that $\lambda_{Edd}$ is correlated with
$\kappa_{2-10keV}$. A Spearman's rank test resulted in
$\rho_{s}=0.60$, $d_{s}=5{\times}10^{-6}$ for the whole sample, and
$\rho_{s}=0.60$, $d_{s}=5{\times}10^{-6}$ for the 12
NLS1s. Figure~\ref{k210:eddr} shows our results, together with the
results from VF07,09 and Lusso10. We performed an EM regression
analysis and derived the
following equations:\\ 
(i) An EM regression with $\lambda_{Edd}$ being
the independent variable 
\begin{equation}
\label{k210-eddr-eqn-1}
Log(\kappa_{2-10})=(0.482{\pm}0.088)Log(\lambda_{Edd})+(1.840{\pm}0.071)
\end{equation}
(ii) An EM regression with $\kappa_{2-10keV}$ being the independent
\begin{equation}
\label{k210-eddr-eqn-2}
Log(\kappa_{2-10})=(1.179{\pm}0.166)Log(\lambda_{Edd})+(2.232{\pm}0.090)
\end{equation}
(iii) bisector of the above two lines (solid orange line in Figure~\ref{k210:eddr}):
\begin{equation}
\label{k210-eddr-eqn-3}
Log(\kappa_{2-10})=(0.773{\pm}0.069)Log(\lambda_{Edd})+(2.004{\pm}0.049)
\end{equation}
Our regression lines are highly consistent with 
the binned points from VF07,09 and also the regression line 
reported by Lusso10. The two lowest data bins from VF07,09 
seem to have a relatively high deviation from the correlation lines, 
which may imply a change in slope of the correlation for sources with 
$\lambda_{Edd}~<$ 0.01. But we cannot test this possibility from our data 
due to the exclusion of sources with low $\lambda_{Edd}$ resulting from 
our sample selection (Paper-I).

The results show that the $\kappa_{2-10keV}$ vs. 
$\lambda_{Edd}$ correlation extends up to high 
$\kappa_{2-10keV}$ (${\sim}100$) and super Eddington 
accretion rates (${\sim}10$); such objects are mainly NLS1s 
(red square symbols in Figure~\ref{k210:eddr}) and some other 
relatively narrow line sources (black square symbols). 
We also note that the dispersion in our regression line is smaller 
than that in VF07,09 and Lusso10, in spite of the different methods 
used in deriving L$_{bol}$ and the different redshift ranges. 
This suggests that the dispersion observed in the correlation 
is intrinsic.
In Figure~\ref{k210:eddr}, we see that PG 1004+130 
(filled purple star) deviates far from the regression line 
(also more than 3$\sigma$ from VF07,09's binned data points), 
confirming its anomalously weak L$_{2-10keV}$.
Increasing its L$_{2-10keV}$ by 0.73 dex (open purple star) moves
it much closer to the correlation line.

\begin{figure*}
 \centering
  \begin{minipage}{185mm}
    \begin{tabular}{cc}
    \includegraphics[bb=54 144 594 650,scale=0.47,clip=1]{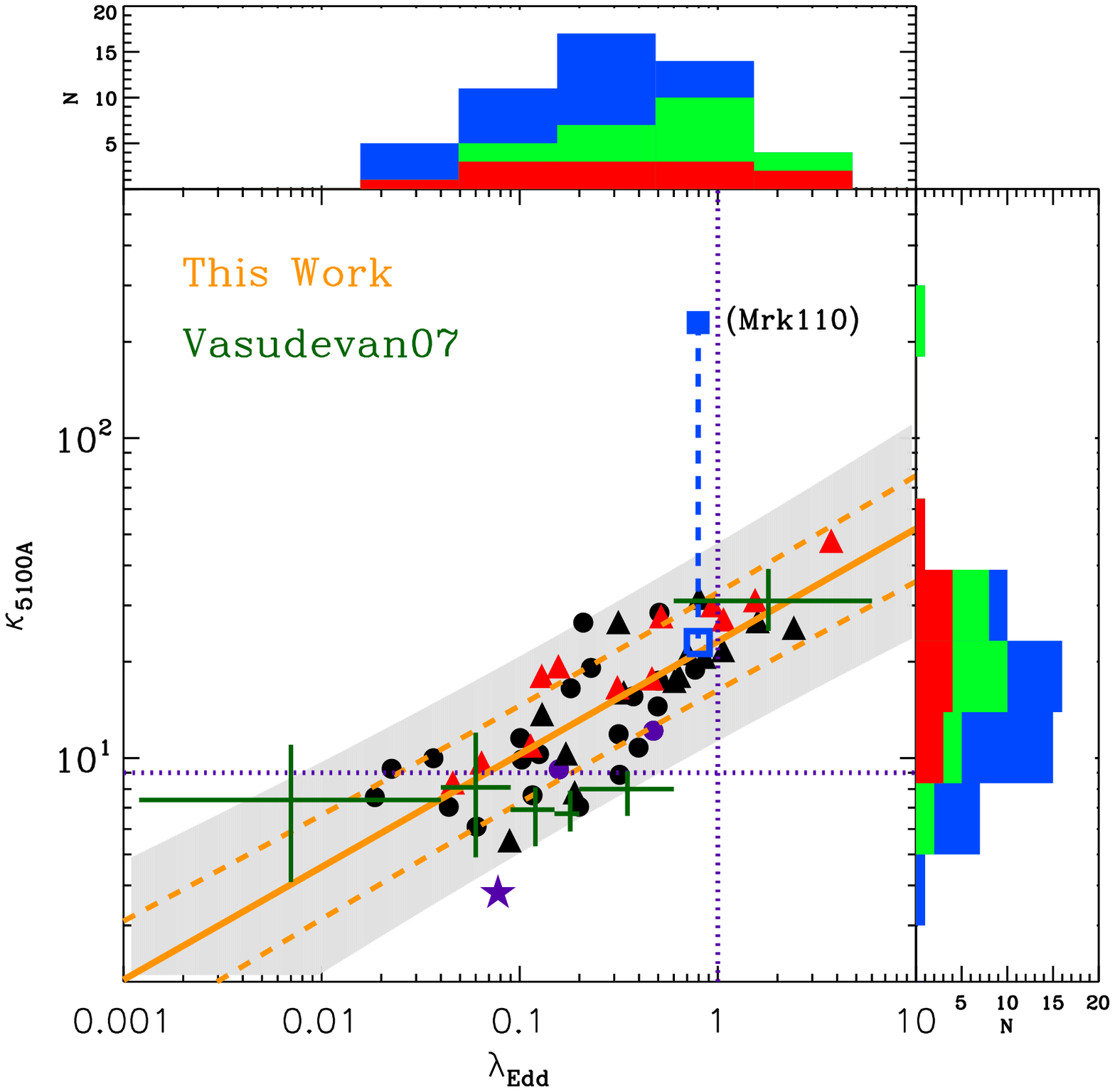}&\includegraphics[bb=60 144 594 650,scale=0.47,clip=1]{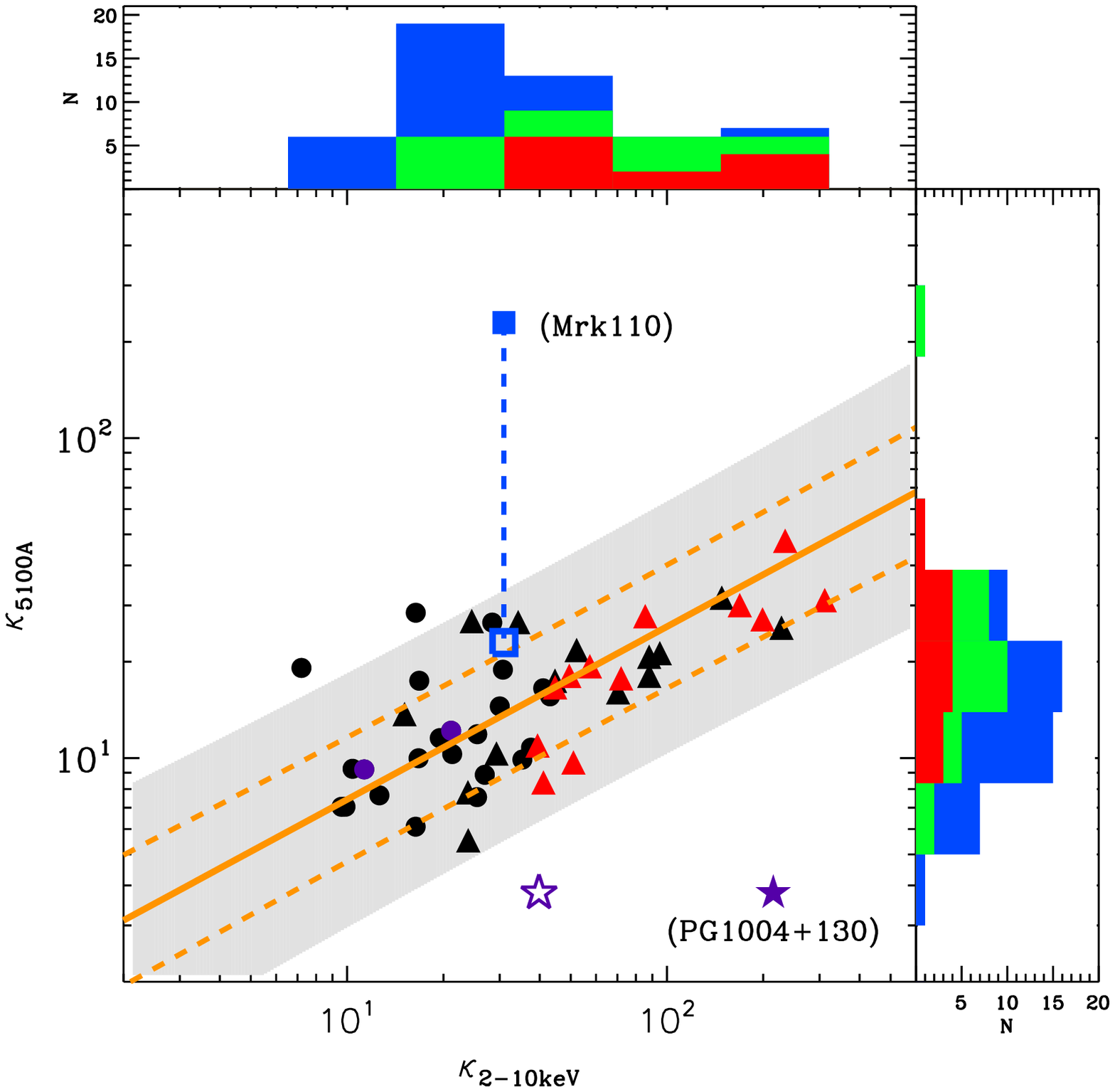}\\
    \end{tabular}
 \end{minipage}
 \caption{$\kappa_{5100}$ vs. $\lambda_{Edd}$ and
 $\kappa_{2-10keV}$. Each symbol represent the same type of source as
 in Figure~\ref{l2keV:l2500:plot}. In the left panel, the solid orange
 line is the OLS line assuming $\lambda_{Edd}$ to be the independent
 variable. The two dashed orange lines show the $\pm$1$\sigma$ region,
 and the shaded region is the $\pm$2$\sigma$ region. The blue open
 square symbol is Mrk 110 reported by Landt et al. (2011),
 which is connected by the blue dashed line to the
 filled blue square point measured from SDSS spectrum.
 The vertical and horizontal purple lines are for $\kappa_{5100}$=9 and
 $\lambda_{Edd}$=1. The symbols and lines in the right panel have the
 same meaning as those in the left panel.}
 \label{k5100:eddr:k210:plot}
\end{figure*}

\subsection{Group 3: $\kappa_{5100}$, $\lambda_{Eddr}$ and $\kappa_{2-10keV}$}
\label{group3}
The 5100{\AA} monochromatic continuum luminosity (L$_{5100}$) is often
used to estimate L$_{bol}$, particularly for very large  samples of AGN,
when broadband SED modeling for every source is not practical.
The conventional method is to use a constant scaling factor
$\kappa_{5100}$ = 9 (Kaspi et al. 2000; Richards et al. 2006: 10.3$\pm$2.1),
or $\kappa_{5100}$ value that is anti-corrected with L$_{bol}$
(Marconi et al. 2004, hereafter: Marconi04). VF07 showed that for high $\lambda_{Edd}$
sources such as many of the NLS1s, there is a clear  deviation from
constant $\kappa_{5100}$ = 9. In addition, potential contamination
from the host galaxy will introduce dispersion into the
$\kappa_{5100}$ vs. $\lambda_{Edd}$ correlation for low luminosity
sources. However, this should not be a severe problem for our
sample since in our sample host galaxy is not dominating (Paper-I).
In our study a much stronger correlation was found in 
$\kappa_{5100}$ vs. $\lambda_{Edd}$ as the Spearman's rank test
gives $\rho_{s}$=0.81 ($d_{s}$=4$\times$10$^{-13}$).
Motivated by the strong correlations between $\lambda_{Eddr}$
and $\kappa_{2-10keV}$, we also found a strong correlation
between $\kappa_{5100}$ and $\kappa_{2-10keV}$, with a Spearman's
rank test of  $\rho_{s}$=0.64 ($d_{s}$=9$\times$10$^{-7}$).

The left panel of Figure~\ref{k5100:eddr:k210:plot} shows the
correlation between $\kappa_{5100}$ and $\lambda_{Edd}$. The solid
orange line is the OLS regression line, the two dashed orange lines
show the $\pm$1$\sigma$ region, and the shaded region is the
$\pm$2$\sigma$ region. For a specific $\lambda_{Edd}$ value, the
1$\sigma$ dispersion of $\kappa_{5100}$ is $\sim$ 0.17
dex. The binned data points from VF07 are also shown in the plot for
comparison. VF07's results are consistent with ours within
$\pm$2$\sigma$, but our correlation is much stronger. This may be
attributed to the high spectral quality of our sample and the
carefully derived $\kappa_{5100}$ and $\lambda_{Edd}$, based on our
detailed broadband SED fitting. It also shows that the distribution of
$\kappa_{5100}$ peaks at 10$\sim$20, with a 1$\sigma$ dispersion of
0.29 dex. For the NLS1s, the mean $\kappa_{5100}$ increases to 20
(1$\sigma$ = 0.23 dex). This means that using a $\kappa_{5100}$ = 9
(the horizontal purple line in the left panel
Figure~\ref{k5100:eddr:k210:plot}) would
underestimate the intrinsic L$_{bol}$ and $\lambda_{Edd}$,
especially for samples containing sources with high $\lambda_{Edd}$ e.g. the
NLS1s. The OLS regression line that assumes $\lambda_{Edd}$ to be the
independent variable can be expressed by the following equation:
\begin{equation}
\label{k5100:eddr:eqn:OLS}
Log(\kappa_{5100}) = (0.467\pm0.045)Log(\lambda_{Edd})+(1.430\pm0.027)
\end{equation}
We superimpose the SDSS and FAST (Landt et al. 2011) data from Mrk 110
(filled and open green square, respectively) on
Figure~\ref{k5100:eddr:k210:plot}. This shows the large optical
variability in the spectrum. The FAST data is much more
consistent with the XMM-Newton OM and also matches very well with the
regression line. This supports the reliability of the correlation.

The strong correlation between $\kappa_{5100}$ and $\kappa_{2-10keV}$
(shown in the right panel of Figure~\ref{k5100:eddr:k210:plot}) is an
expected result, given that both $\kappa_{5100}$ and
$\kappa_{2-10keV}$ strongly correlate with $\lambda_{Edd}$. Such a
correlation is also reasonable considering the strong correlation
between L$_{2-10keV}$ and L$_{5100}$ (Paper-II). The 1$\sigma$
dispersion of $\kappa_{2-10keV}$ is 0.40 dex, which is larger than
that 0.29 dex for $\kappa_{5100}$. The $\pm$1$\sigma$,
$\pm$2$\sigma$ zones are also shown in the Figure. 
The bisector regression line can be expressed by the
following equation:
\begin{equation}
\label{k5100:eddr:eqn:bisector}
Log(\kappa_{5100}) = (0.593\pm0.053)Log(\kappa_{210})+(0.239\pm0.086)
\end{equation}
Mrk 110 and PG 1004+130 are superimposed on the plot, and their
corrected positions are much more consistent with the
regression lines. 

Marconi04 proposed that $\kappa_{5100}$ anti-correlated with
L$_{bol}$, but our study does not support such an anti-correlation,
although our sample only occupies the L$_{bol}$ region above
10$^{10.7}$ L$_{\sun}$ in Fig. 3 left panel of Marconi04. A
Spearman's rank test for our sample gives $\rho_{s}$=0.12
($d_{s}$=0.39), suggesting no correlation.
The sources lying between 10$^{11}\sim$10$^{12.5}$
L$_{\sun}$ have a mean $\kappa_{5100}$= 16, with a 1$\sigma$
dispersion of 0.29 dex. So we find that our $\kappa_{5100}$ values
for these objects are much higher than reported in
Elvis et al. (1994) and Marconi04.

\begin{figure*}
 \centering
  \begin{minipage}{185mm}
    \begin{tabular}{cc}
    \includegraphics[bb=54 144 594 650,scale=0.47,clip=1]{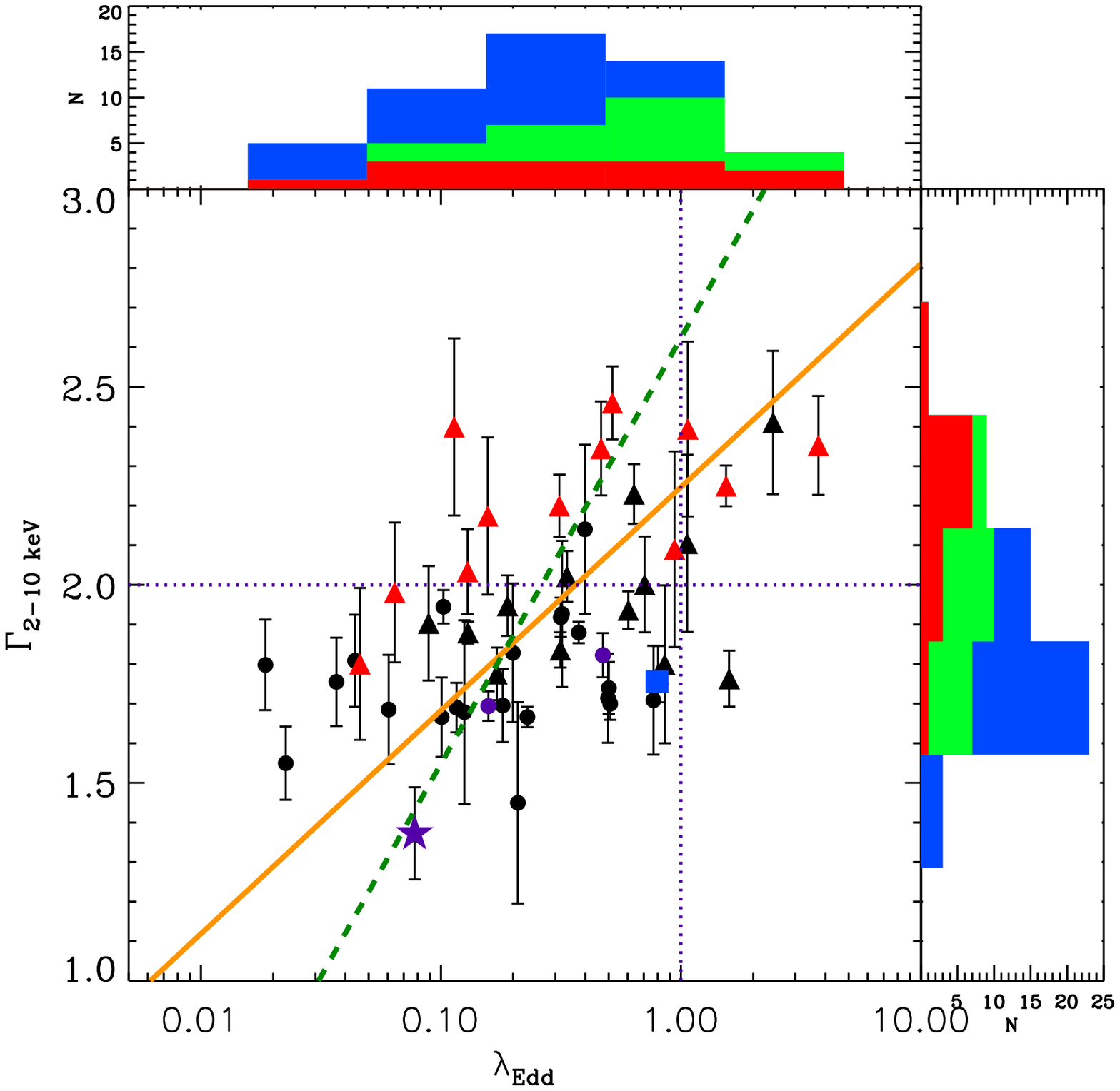}&\includegraphics[bb=54 144 594 650,scale=0.47,clip=1]{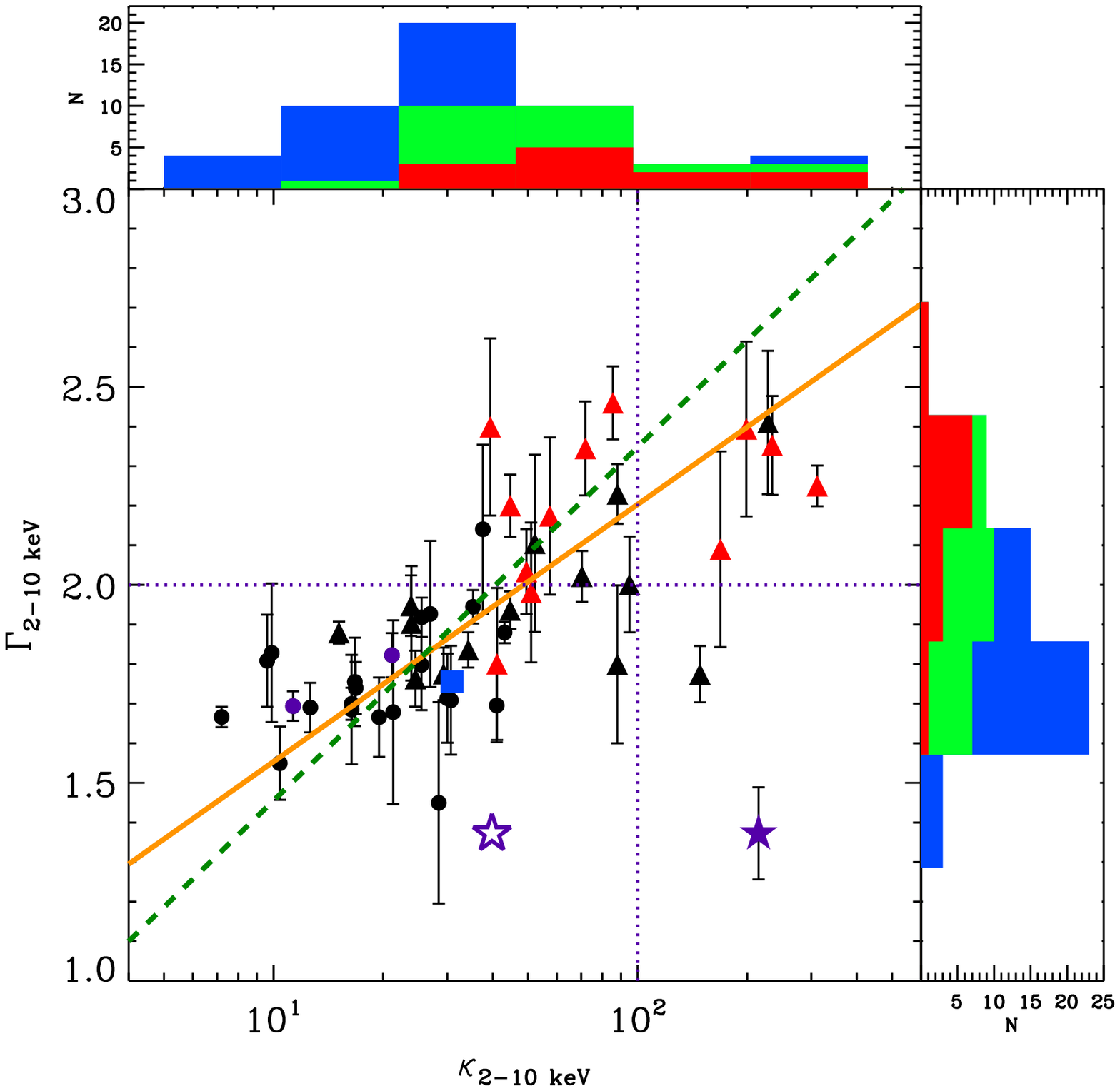}\\
    \end{tabular}
 \end{minipage}
 \caption{$\Gamma_{2-10keV}$ vs. $\lambda_{Edd}$ and $\kappa_{2-10keV}$.
 Each symbol represents the 
 same type of source as in Figure~\ref{l2keV:l2500:plot}.
 In the left panel, the solid orange line is the bisector regression line.
 The dashed green line is that reported by Zhou10b.
 The vertical and horizontal purple lines are for 
 $\Gamma_{2-10keV}$=2 and $\lambda_{Edd}$=1. The symbols 
 and lines in the right panel have the same meaning as those 
 in the left panel. The vertical purple line is for 
 $\kappa_{2-10keV}$=100.}
 \label{gamma:eddr:k210:plot}
\end{figure*}

\subsection{Group 4: $\Gamma_{2-10keV}$, $\lambda_{Edd}$ and $\kappa_{2-10keV}$}
\label{group4}

The strong correlation between 2-10 keV photon index
($\Gamma_{2-10keV}$) and $\lambda_{Edd}$ has been studied in detail
for the past ten years (e.g. Lu \& Yu 1999; Wang, Watarai \& Mineshige 2004;
S06,08; Zhou10a). It is proposed that increasing the mass accretion rate
leads to enhanced emission from the accretion disc, resulting in
more seed photons from the disc, which then increases the Compton
cooling of the corona, and softens the Comptonized hard X-ray
spectrum, i.e. the slope of $\Gamma_{2-10keV}$ increases. It was also
reported that both $\Gamma_{2-10keV}$ and $\lambda_{Edd}$ strongly
correlate with the FWHM of H$\beta$ (e.g. Brandt, Mathur \& Elvis 1997; S06,08;
Grupe10), therefore these three parameters all strongly correlate
with each other. However, S06,08 found that the correlation of $\Gamma_{2-10keV}$
vs. FWHM$_{H\beta}$ is weakened by the inclusion of highly luminous
sources, but that the correlation of $\Gamma_{2-10keV}$
vs. $\lambda_{Edd}$ still exists. This implies that the correlation of
$\Gamma_{2-10keV}$ vs. $\lambda_{Edd}$ is more fundamental. We also
mentioned in Section~\ref{group2} that the strong correlation between
$\lambda_{Edd}$ and $\kappa_{2-10keV}$ is confirmed, thus a
strong correlation between $\Gamma_{2-10keV}$ and $\kappa_{2-10keV}$ is expected.
Indeed, such a correlation has been reported
recently by Zhou \& Zhao (2010), hereafter: Zhou10b.
In this section we carry out a similar cross-correlation study,
to test the robustness of previous claims.

\subsubsection{The Correlations and Regression Lines}

The two panels in Figure~\ref{gamma:eddr:k210:plot} show  
our correlations between  $\Gamma_{2-10keV}$, $\lambda_{Edd}$ 
and $\kappa_{2-10keV}$. Table~\ref{gamma:eddr:k210:table} 
summarizes the numerical results. Significant correlations are 
confirmed based on the  Spearman's rank test: $\rho_{s}$=0.40
($d_{s}$=4$\times$10$^{-3}$) for $\Gamma_{2-10keV}$ vs.
$\lambda_{Edd}$, and $\rho_{s}$=0.73
($d_{s}$=4$\times$10$^{-9}$) for $\Gamma_{2-10keV}$
vs. $\kappa_{2-10keV}$.

Following S08's approach, we applied the $\chi^2$ minimization method for
$\Gamma_{2-10keV}$ vs. $\lambda_{Edd}$ correlation, assuming $\Gamma_{2-10keV}={\beta}Log(\lambda_{Edd})+\xi$.
A typical error of 10\%  was assumed for $\lambda_{Edd}$.
The small error in $\Gamma_{2-10keV}$ for Mrk 110 (the blue square
symbol) caused the slope $\beta$ to be 0.018$\pm$0.019,
which is clearly not the best-fit line for the whole sample.
We therefore excluded Mrk 110 and so found a more reasonable
slope of 0.189$\pm$0.026, but this is still $\sim$5$\sigma$
away from 0.31$\pm$0.01 reported by S08 using the same method.
It implies that the $\chi^2$ minimization technique may not be
an appropriate method for quantifying this correlation,
because it can be strongly biased by sources with small error
in the $\Gamma_{2-10keV}$ measurement (if the 2-10 keV
spectrum has high S/N). The $\chi^2$/$\nu$ = 6.5 in our fitting
means that this correlation contains a big intrinsic dispersion
along with the observational dispersion, thus the method of assuming
$\chi^2$/$\nu~\sim$1 by taking intrinsic dispersion into account
is more appropriate. This method gives slopes of 0.202$\pm$0.061
and 0.226$\pm$0.026 before and after excluding Mrk 110, so the
results are less sensitive to individual sources of much smaller
error bars. The intrinsic dispersion is 0.18, which is 86\% of the
total dispersion, and is also consistent with
$\Delta\Gamma_{2-10keV}\sim$0.1$\times\Gamma_{2-10keV}$
reported in S08. The bisector regression line is derived.
The result is plotted as a solid orange line in the left
panel of Figure~\ref{gamma:eddr:k210:plot}, and it can
be expressed by the following equation:
\begin{equation}
\label{gamma:eddr:eqn}
Log(\lambda_{Edd}) = (1.773\pm0.238)\Gamma_{2-10keV}-(3.983\pm0.469)
\end{equation}
The slope is $\sim$2$\sigma$ steeper than the bisector slope
of 0.9$\pm$0.3 reported by S08 (dashed green line
in the left panel of Figure~\ref{gamma:eddr:k210:plot}).
This discrepancy may be due to the different methods used to
estimate the bolometric luminosity.
We will discuss this point in Section~\ref{gamma:eddr:k210:subsubsection2}.

Similar analytical methods were applied to the relation of
$\Gamma_{2-10keV}$ vs. $\kappa_{2-10keV}$. Zhou10b reported
a slope of 2.52$\pm$0.08, using standard $\chi^2$ minimization
and assuming $\kappa_{2-10keV}={\beta}Log(\Gamma_{2-10keV})+\xi$.
This is consistent with our value of $\beta=$ 2.620$\pm$0.184
with a $\chi^2$/$\nu$ = 2.78. Considering the intrinsic scatter,
Zhou10b reported a slope of 1.12$\pm$0.30 by adding 0.32 dex of
intrinsic dispersion to reduce $\chi^2$/$\nu$ to unity (the dashed
green line in the right panel of Figure~\ref{gamma:eddr:k210:plot}).
Applying the same method to our sample resulted in a slope
of 1.529$\pm$0.183, which is steeper than found by Zhou10b.
The intrinsic dispersion found by us is 80\% of the total dispersion.
The bisector regression method gives a slope of 1.533$\pm$0.153
(the solid orange line in the right panel of
Figure~\ref{gamma:eddr:k210:plot}) and can be expressed by the following equation:
\begin{equation}
\label{gamma:k210:eqn}
Log(\kappa_{2-10keV}) = (1.533\pm0.153)\Gamma-(1.376\pm0.288)
\end{equation}
This is also consistent with the results found by assuming $\chi^2$/$\nu~\sim$1.

\begin{table}
 \centering
 \caption{The line coefficients found using different regression
methods for the correlations of $\Gamma_{2-10keV}$ vs. $\lambda_{Edd}$
and $\kappa_{2-10keV}$.}
     \begin{tabular}{lccc}
     \hline
     &&$Log(\lambda_{Edd})$=$\beta\Gamma$+$\xi$&$Log(\kappa_{210})$=$\beta\Gamma$+$\xi$\\
     \hline
     OLS(Y$\mid$X)&$\beta$&0.918$\pm$0.269&1.115$\pm$0.172\\
     &$\xi$&-2.325$\pm$0.527&-0.573$\pm$0.329\\
     OLS(X$\mid$Y)&$\beta$&4.730$\pm$1.217&2.209$\pm$0.283\\
     &$\xi$&-9.650$\pm$2.304&-2.675$\pm$0.525\\
     Bisector&$\beta$&1.753$\pm$0.239&1.533$\pm$0.153\\
     &$\xi$&-3.931$\pm$0.471&-1.376$\pm$0.288\\
     \hline
     $\chi^{2}$ min&$\beta$&4.416$\pm$0.510&2.620$\pm$0.184\\
     &$\xi$&-8.761$\pm$0.946&-3.447$\pm$0.343\\
     \hline
     $\chi^2$/$\nu\sim$1&$\beta$&1.274$\pm$0.283&1.529$\pm$0.183\\
     &$\xi$&-3.003$\pm$0.544&-1.372$\pm$0.350\\
     \hline
     \end{tabular}
  \label{gamma:eddr:k210:table}
\end{table}

\begin{figure*}
\centering
\includegraphics[scale=0.65,clip=]{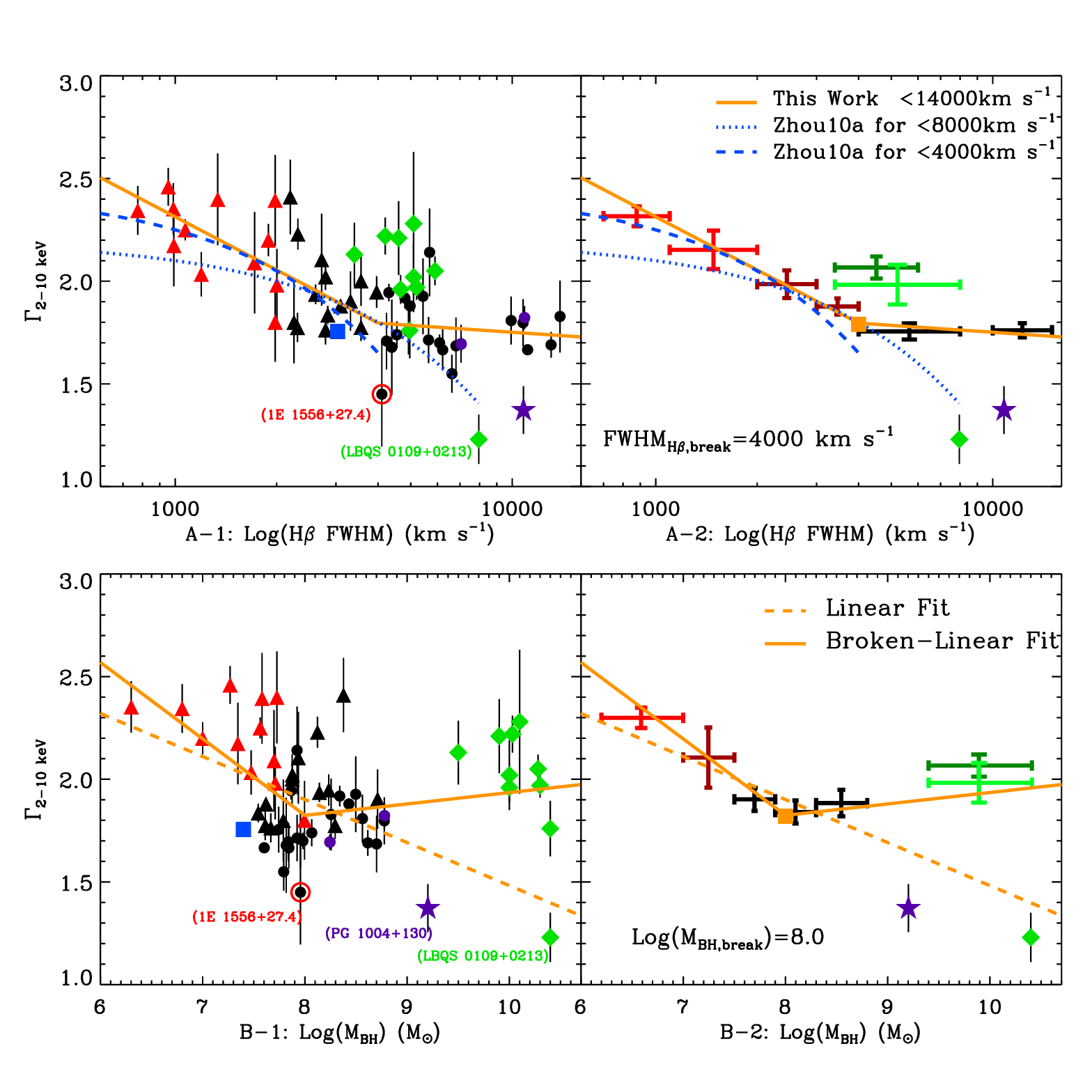}
\caption{$\Gamma_{2-10keV}$ vs. FWHM of H$\beta$ and M$_{BH}$. In the upper left panel, a broken line is fitted to the sample using the minimum $\chi^{2}$ method, with the break point chosen to be FWHM $=$ 4000 km s$^{-1}$ (the orange square point). S06,08 proposed the linear correlation between $\Gamma_{2-10keV}$ and Log(H$\beta$ FWHM) was not followed  by their 10 extremely high luminosity sources, so we plot their sample as blue diamond symbols for comparison. In the right panel binned points are plotted with 1 standard error of $\Gamma_{2-10keV}$ in order to show the break points more clearly. The two red points only include the NLS1s, the two dark points are the broadest BLS1s. The cyan point is the binned point for the whole sample of S06,08. The blue point is the binned point for S06,08's sample but excluding LBQS 0109+0213 whose $\Gamma_{2-10keV}$ is anomalously low. 1E 1556+27.4 shown by the red circle is another source with $\Gamma_{2-10keV}<$1.5. In the second row, the symbols all have the sample meaning as in the first row, the break point is chosen to be Log(M$_{BH}$) $=$ 8.0 (the orange square point). We plot the linear regression line as the dashed orange line.}
\label{pi210-bhfwhm-bhmass}
\end{figure*}

\subsubsection{Advances from Our Correlation Analysis}
\label{gamma:eddr:k210:subsubsection2}

Compared with the results found by S08 and Zhou10b, our study
of $\Gamma_{2-10keV}$ vs. $\lambda_{Edd}$ and $\kappa_{2-10keV}$
provides the following advances. \\ \\ (1) We have confirmed these
correlations based on sample of twice the size of those
in S08 and Zhou10b, including more sources
with high values of $\lambda_{Edd}$, $\Gamma_{2-10keV}$ and
$\kappa_{2-10keV}$, which significantly extend the previous
correlations (see Figure~\ref{gamma:eddr:k210:plot}). The regression
line fits are better constrained and cover wider parameter space. The
difference between our regression lines and those of previous studies
may be partially due to the fact that we have more sources
of most extreme $\lambda_{Edd}$. \\ \\ (2) Our sample has
been carefully screened to exclude sources with a strong warm
absorber. These sources may have $\Gamma_{2-10keV}$
and higher $\kappa_{2-10keV}$ artificially lower than the intrinsic values.
Our sample quality is essential to reduce the dispersion and so reveal
intrinsic correlations. \\ \\ (3) Our estimates of L$_{bol}$ were derived from the
broadband SED fitting, which was based on high quality spectra and a
new multi-component model.  We claim this to be more reliable than the
procedure used in previous studies.  A conventional method is to apply
a multiplication factor to L$_{5100}$ to estimate L$_{bol}$. However,
we showed in Section~\ref{group3} that $\kappa_{5100}$ is well
correlated with $\lambda_{Edd}$, rather than being constant or
dependent on L$_{bol}$, consequently the conventional scaling from
L$_{5100}$ to L$_{bol}$ is likely to result in poor accuracy in some
cases.  The L$_{bol}$ used in Zhou10b does come from VF09's broadband
SED model for the reverberation mapped sample, but it does not take
account of the `soft X-ray excess' component or where the disc peaks
in the EUV. Therefore the L$_{bol}$ we
calculate will be larger than previous works, especially for those
sources with a strong `soft excess', Our
$\lambda_{Edd}$ and $\kappa_{2-10keV}$ will also be higher, which could
account for the differences in slope between our regression lines and
those reported in previous work.

\subsection{Group 5: $\Gamma_{2-10keV}$, H$\beta$ FWHM and M$_{BH}$}
\label{group5}
\subsubsection{The $\Gamma_{2-10keV}$ vs. H$\beta$ FWHM Correlation}
The correlations between the soft/hard X-ray slopes and the Balmer
line velocity width have been the subject of many previous studies.
Puchnarewicz et al. (1992) studied 53 AGNs with ultra-soft X-ray excess and 
noticed that these ultra-soft AGNs tend to have narrower optical permitted lines.
Laor et al. (1994) studied 23 ROSAT selected bright quasars, and found an
anti-correlation between the 0.2-2 keV slope ($\alpha_{X}$) and the
H$\beta$ line width. Later Boller, Brandt \& Fink (1996) showed that NLS1s tend to
have softer X-ray spectra. Brandt, Mathur \& Elvis (1997) extended this
anti-correlation to include the 2-10 keV slope, by showing that NLS1s
also have steeper hard X-ray continua than BLS1s, a result which was
confirmed and extended by other studies (e.g. Grupe et al. 1999;
Leighly 1999; Piconcelli et al. 2005; Brocksopp et al. 2006; S08; Zhou10a).
However, there is large scatter within this
correlation, and the trend seems to invert below $\sim$1000 km
s$^{-1}$ for NLS1s (Zhou et al. 2006). The observed large scatter is
to be expected since we know that it is not a single variable that
determines the spectral slope in either the soft or the hard X-ray
region. In the soft X-ray region, the
extinction, a soft-excess component and a warm absorber will all
influence the spectral shape, which would require very detailed
modeling. The situation for the 2-10 keV region is somewhat less
complicated since often a single power law dominates
(e.g. Middleton, Done \& Gierli\'{n}ski 2007), but a warm absorber and reflection may
still modify the hard X-ray spectral shape. In summary, the true
correlation can only be found when the intrinsic X-ray continuum is used.

\begin{figure}
\centering
\includegraphics[bb=54 144 594 660, scale=0.48,clip=]{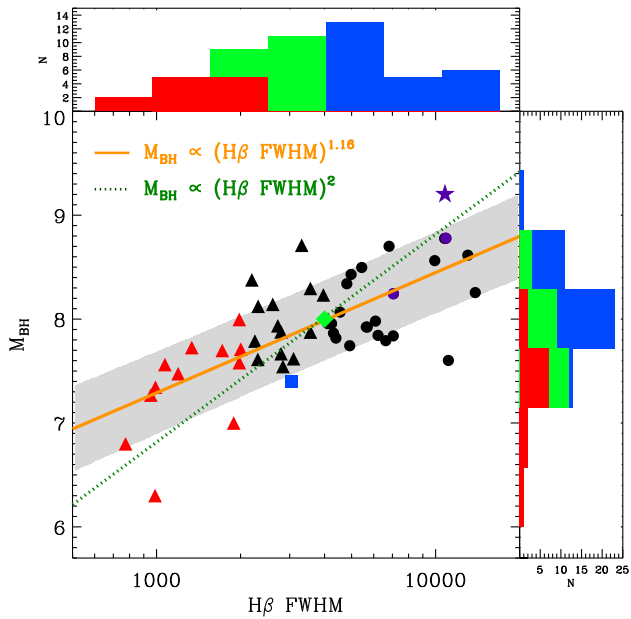}
\caption{M$_{BH}$ vs. H$\beta$ FWHM.
Note that the black hole mass M$_{BH}$ was constrained by H$\beta$
FWHM and obtained from the broadband SED fitting (i.e. the M$_{BH,FIT}$ in Paper-I),
rather than directly calculated from the H$\beta$ FWHM.
The symbols used represent the
different types of source as in Figure~\ref{l2keV:l2500:plot}. The
solid orange line is the OLS regression line, assuming H$\beta$ FWHM
to be the independent variable. The shaded region is the
$\pm$1$\sigma$ region of the regression line. The cyan triangle shows
the position of (FWHM$_{H{\beta},break}$, Log(M$_{BH,break}$)) in
Figure~\ref{pi210-bhfwhm-bhmass}.}
\label{hbfwhm:bhmass:plot}
\end{figure}

Our sample selection has ensured that every object in the sample has
high quality 2-10 keV spectra, without significant cold gas
absorption or a warm absorber (Paper-I).
We confirm that there is an anti-correlation
between $\Gamma_{2-10keV}$ and H$\beta$ FWHM, see
Figure~\ref{pi210-bhfwhm-bhmass}. The Spearman's rank test gives
$\rho_{s}$=-0.72 ($d_{s}$=4.9$\times$10$^{-9}$). The best-fit lines
from Zhou10a are also plotted in Figure~\ref{pi210-bhfwhm-bhmass} as
the cyan lines. Compared with their linear correlation in
using FWHM$_{H\beta}$, we find that the linear
correlation using Log(FWHM$_{H\beta}$) is better.
Previous work also noted that the correlation may change form
at $\sim$4000 km s$^{-1}$ (Sulentic et al. 2008). 
Therefore we fit a broken power law to the 
data points in the left panel by assuming a break
point at FWHM$_{H\beta}$ $=$ 4000 km s$^{-1}$.
Intrinsic dispersion will dominate this correlation
as shown in previous studies (e.g. Grupe et al. 1999; Grupe10; Zhou10a),
so we performed this fitting without considering
the error in $\Gamma_{2-10keV}$ associated with each point,
otherwise the the fitting would be biased by the few points with best
constrained $\Gamma_{2-10keV}$ rather than revealing the distribution
of the whole sample. The best-fit parameters are shown below:\\
(i) when FWHM$_{H\beta}$ $\leqslant$ 4000 km s$^{-1}$,
\begin{equation}
\label{gamma:hbfwhm:eqn:1}
\Gamma = (-0.86\pm0.01)Log(FWHM_{H\beta})+(4.89\pm0.03)
\end{equation}
(ii) when FWHM$_{H\beta}$ $>$ 4000 km s$^{-1}$,
\begin{equation}
\label{gamma:hbfwhm:eqn:2}
\Gamma = (-0.11\pm0.01)Log(FWHM_{H\beta})+(2.20\pm0.04)
\end{equation}

Sources with FWHM$_{H\beta}$ $>$ 4000 km s$^{-1}$ have
an average $\Gamma_{2-10keV}~=~$1.78$\pm$0.12.
The only source included in the correlation
whose $\Gamma_{2-10keV}<$ 1.5, is 1E 1556+27.4,
(the data for PG 1004+130 
is superimposed but not used for the regression). 
A closer examination of the spectrum of this AGN 
shows that it probably has a strong 
reflection component modifying its intrinsic hard X-ray power law
slope (Paper-I). All other objects have values consistent with
$\Gamma_{2-10keV}~>$ 1.5. The differences in the results of Zhou10a
and our work are not only because we performed our correlation fitting
using Log(FWHM$_{H\beta}$), but also because their
sample did not exclude BAL quasars and warm absorbers, whose low
values of $\Gamma_{2-10keV}$ are probably not intrinsic. This
will bias the correlation and increase the scatter. 

Our sample includes 
six objects with FWHM$_{H\beta}>$10000 km s$^{-1}$. These 
are the sources have average 
$<\Gamma_{2-10keV}>=1.76{\pm}0.14$ 
independent of the FWHM$_{H\beta}$. This is slightly lower but still
consistent with the $\Gamma_{2-10keV}= 1.97 \pm 0.31$ index found by 
S06,08, who included more high redshift,
high luminosity sources, with FWHM$_{H\beta}~>$ 10000 km s$^{-1}$.

\subsubsection{The $\Gamma_{2-10keV}$ vs. M$_{BH}$ Correlation}
The H$\beta$ FWHM is frequently used to estimate the M$_{BH}$, using
the relation M$_{BH}\propto$ FWHM$_{H\beta}^{2}$ (Wandel, Peterson
\& Malkan 1999; Woo \& Urry 2002). The correlation of $\Gamma_{2-10keV}$
vs. FWHM$_{H\beta}$ implies a similar correlation in $\Gamma_{2-10keV}$
vs. M$_{BH}$. This is confirmed in our study as shown in the second
row of Figure~\ref{pi210-bhfwhm-bhmass}. The Spearman's rank test gives
$\rho_{s}$=-0.3 ($d_{s}$=3$\times$10$^{-2}$).
We also plot FWHM$_{H\beta}$ vs. M$_{BH}$ in Figure~\ref{hbfwhm:bhmass:plot}.
This is an independent plot
as our  M$_{BH}$ are derived from the SED continuum fits rather than 
directly measured from FWHM$_{H\beta}$.
The OLS regression 
gives M$_{BH}\propto$ FWHM$_{H\beta}^{1.16}$, with a 1${\sigma}=$ 0.4 dex.
The cyan triangle symbol in the figure shows the
position of the break in FWHM$_{H\beta}$ vs. M$_{BH}$.
The correlation in Figure~\ref{hbfwhm:bhmass:plot} suggests
FWHM$_{H\beta}$ $=$ 4000 km s$^{-1}$ corresponds to a black hole
mass of 10$^8$ M$_{\sun}$.
Indeed, the correlation in the second row of Figure~\ref{pi210-bhfwhm-bhmass}
shows a change in slope at Log(M$_{BH}$) $\sim$ 8.0.
A broken power law fitting, assuming the break point
at log(M$_{BH}$) $=$ 8.0, can be expressed as follows:\\
(i) when Log(M$_{BH}$) $\leqslant$ 8.0,
\begin{equation}
\label{gamma:bhmass:eqn:1}
\Gamma_{2-10keV} = (-0.372\pm0.005)Log(M_{BH})+(4.802\pm0.037)
\end{equation}
(ii) when Log(M$_{BH}$) $>$ 8.0,
\begin{equation}
\label{gamma:bhmass:eqn:2}
\Gamma_{2-10keV} = (0.056\pm0.007)Log(M_{BH})+(1.380\pm0.052)
\end{equation}
Sources with Log(M$_{BH}$) $>$ 8.0 have
an average $\Gamma_{2-10keV}~=~$1.86$\pm$0.21.

\subsection{Other Strong Correlations}
\label{section:other:correlations}
The diverse correlations found in small parameter groups imply 
more correlations among all these parameters. We show some
significant correlations in Figure~\ref{the-rest-keypar_cor}.
These include decreasing $\lambda_{Edd}$, $\kappa_{2-10keV}$, 
$\kappa_{5100}$ with increasing H$\beta$ FWHM and M$_{BH}$. 
Binned data points are shown as cyan crosses. 
The dashed orange line in each panel is the bisector regression line.

\subsubsection{The $\kappa_{2-10keV}$ vs. L$_{2-10keV}$ correlation}
\label{section:k210:l210}
\begin{figure}
\centering
\includegraphics[bb=54 144 594 660, scale=0.48,clip=]{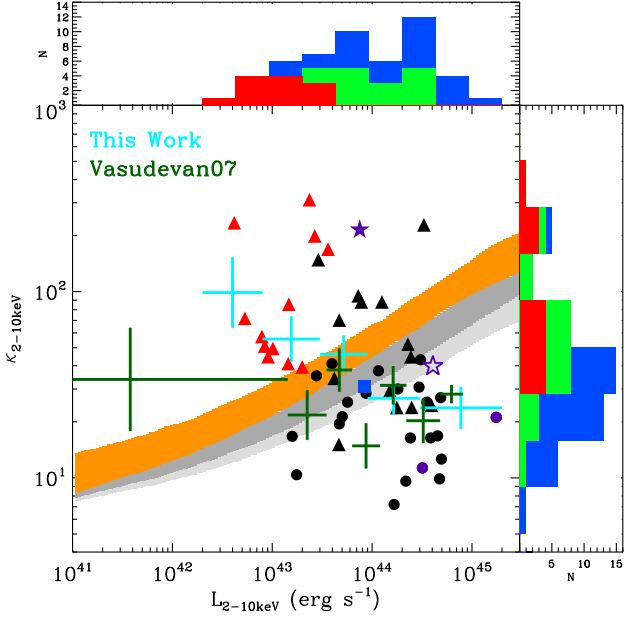}
\caption{$\kappa_{2-10keV}$ vs. L$_{2-10keV}$. Different symbols represent the same type of sources as in Figure~\ref{l2keV:l2500:plot}. The orange and gray shaded regions represent the theoretical $\kappa_{2-10keV}$ with $\pm$1$\sigma$ scattering at each L$_{2-10keV}$ in Hopkins07 and Marconi04. The green data points are reproduced from Fig.3 in VF07.}
\label{l210:k210:plot}
\end{figure}

Marconi04 and Hopkins07 reported a strong
positive correlation in $\kappa_{2-10keV}$ vs. L$_{2-10keV}$, based on
a quasar SED template and the $\alpha_{ox}$ vs. L$_{2500}$ correlation
reported by Steffen et al. (2006). VF07 tested the same correlation in
their low redshift AGN sample but found no correlation
(see Figure~\ref{l210:k210:plot}). We tested this correlation
in our sample and confirmed VF07's result (see
Figure~\ref{l210:k210:plot}). A highly dispersed anti-correlation of
$\kappa_{2-10keV}$ vs. L$_{2-10keV}$ is found in our study.

For the well sampled high X-ray luminosity region
(L$_{2-10keV}>$10$^{43}$ erg s$^{-1}$), both VF07 and our samples show
very similar distribution and dispersion. NLS1s in both samples mainly
populate the regions of $\kappa_{2-10keV}\ga$100 and 10$^{42}$ erg
s$^{-1}<$L$_{2-10keV}<$10$^{43}$ erg s$^{-1}$, and deviate from the
correlation suggested by Marconi04 and Hopkins07. The
main difference from VF07 is that our sample have more sources within
the range of 3${\times}$10$^{42}$ erg
s$^{-1}<$L$_{2-10keV}<$10$^{43}$ erg s$^{-1}$, while VF07 sample
contains three extraordinary weak X-ray sources whose
L$_{2-10keV}<~$2${\times}$10$^{42}$ erg s$^{-1}$. Thus we think those
intrinsically X-ray weak (L$_{2-10keV}\sim$10$^{42}$ ergs$^{-1}$)
sources may populate the low L$_{2-10keV}$, small $\kappa_{2-10keV}$
region, creating a totally dispersed distribution in the $\kappa_{2-10keV}$
vs. L$_{2-10keV}$ plot. The correlations from Marconi04 and
Hopkins07 may have underestimated the uncertainties in using
$\alpha_{ox}$ vs. L$_{2500}$ correlation (see discussion in
Section~\ref{group1}) and the universal quasar SED template
(e.g. Elvis et al. 1994; VF07; Paper-I). However, we cannot rule out
the possibility that the behaviors of nearby Seyfert AGNs are different
from quasars at higher redshift.

\subsubsection{$\alpha_{UV}$ and $\alpha_{X}$ related correlations}
\label{section:auv:ax}
While $\Gamma_{2-10keV}$ is the 2-10 keV photon-index measured directly from the X-ray data,
we can also measure the optical/UV slope and soft X-ray slope from our reconstructed SED.
Assuming F$_{\nu}~\propto~{\nu}^{-\alpha}$, we define $\alpha_{UV}$ as the spectral slope
between 1700-6500 {\AA} (the {\it Swift\/} UVOT wavelength coverage),
and $\alpha_{X}$ as the soft X-ray slope between 0.2-2 keV, so as
to be comparable with the results in Grupe10.
Note that we do not have complete OM data for every source, and we only use X-ray data
from XMM-Newton above 0.3 keV, so our $\alpha_{UV}$ and $\alpha_{X}$ values are model dependent.
The soft X-ray and optical/UV regions cannot be simply fitted by a single power law
(see the SEDs in Paper-I),
and thus $\alpha_{UV}$ and $\alpha_{X}$ are just rough estimates of the spectral shape.
We measure $\alpha_{UV}$ and $\alpha_{X}$ from the best-fit SED in Paper-I corrected for
Galactic and intrinsic reddening/absorption, and list the values in Table~\ref{app:SED-key-parameters}.

It was found previously that AGNs with bluer optical/UV spectra
tend to have softer X-ray spectra (Walter \& Fink 1993; Grupe et al. 1998;
Grupe10). We found $\rho_{s}=-0.65$ ($d_{s}=4{\times}10^{-7}$) for
$\alpha_{UV}$ vs. $\Gamma_{2-10keV}$, and $\rho_{s}=-0.41$ ($d_{s}=3{\times}10^{-3}$) for
$\alpha_{UV}$ vs. $\alpha_{X}$, so confirm the results from other studies.
We also tested the results of $\alpha_{UV}$ and $\alpha_{X}$ without correction for
intrinsic reddening/absorption. The Spearman's rank coefficients remain similar which is
because of the unobscured nature of our sample. Another correlation found in previous work
is that AGNs with steeper X-ray spectra tend to be weak at hard X-ray energies
(Atlee \& Mathur 2009; Grupe10; but see also Young, Elvis \& Risaliti 2009).
This can be directly confirmed in our study by the anti-correlation found
between $\alpha_{X}$ and L$_{2-10keV}$ ($\rho_{s}=-0.60$, $d_{s}=6{\times}10^{-6}$),
and the correlation between $\alpha_{X}$ and $\alpha_{ox}$
($\rho_{s}=0.65$, $d_{s}=5{\times}10^{-7}$, Figure~\ref{the-rest-keypar_cor2}).
However, we only find a marginal anti-correlation between $\alpha_{UV}$ and $\alpha_{ox}$
(see Table~\ref{sum-cor-keypar:table}), which is much less
significant than found by Grupe10. This seems to be due to the fact that our sample
has few sources with $\alpha_{ox}~>~1.6$ and $\alpha_{UV}~<~0$.

\begin{figure*}
\centering
\includegraphics[scale=0.55,clip=1,angle=90]{keypar_cor5.ps}
\caption{Examples of some good correlations not reported previously. In each panel the various symbols represent the same types of source as in Figure~\ref{l2keV:l2500:plot}. The cyan symbols are the binned data points over the X-axis with a 1 standard error on the Y-axis.}
\label{the-rest-keypar_cor}
\end{figure*}

\begin{figure*}
\centering
\includegraphics[scale=0.6,clip=1,angle=90]{keypar_cor6.ps}
\caption{Examples of some good correlations not reported previously. In each panel the various symbols represent the same types of source as in Figure~\ref{l2keV:l2500:plot}. The cyan symbols are the binned data points over the X-axis with a 1 standard error on the Y-axis.}
\label{the-rest-keypar_cor2}
\end{figure*}

Similar to the $\Gamma_{2-10keV}$ vs. $\lambda_{Edd}$ correlation
discussed in Section~\ref{group4}, $\alpha_{X}$ and $\alpha_{UV}$ were also found
to (anti-)correlate with $\lambda_{Edd}$ (Grupe 2004; S08; Grupe10).
This is confirmed in our study as we find $\rho_{s}=0.49$ ($d_{s}=4{\times}10^{-4}$) for
$\alpha_{X}$ vs. $\lambda_{Edd}$ which is stronger than the
correlation between $\Gamma_{2-10keV}$ and $\lambda_{Edd}$,
and $\rho_{s}=-0.55$ ($d_{s}=4{\times}10^{-5}$) for $\alpha_{UV}$ vs. $\lambda_{Edd}$.
In addition, we also find strong (anti-)correlations such as
$\alpha_{X}$ (or $\alpha_{UV}$) vs. $\kappa_{2-10keV}$,
$\alpha_{X}$ (or $\alpha_{UV}$) vs. $\kappa_{5100A}$,
$\alpha_{X}$ (or $\alpha_{UV}$) vs. FWHM$_{H\beta}$
and $\alpha_{X}$ (or $\alpha_{UV}$) vs. M$_{BH}$ 
(see Figure~\ref{the-rest-keypar_cor2}, Table~\ref{sum-cor-keypar:table}).

\begin{table*}
 \centering
   \centering
   \caption{ The cross-correlation matrix of the 11 key parameters. $\rho_{s}$ is the Spearman's rank coefficient. $d_{s}^{1}$ is the logarithm of the significance level of being random distribution.}
   \label{sum-cor-keypar:table}
     \begin{tabular}{lcccccccccccc}
\hline

Parameters& &$\Gamma_{2-10keV}$&$\kappa_{2-10keV}$&$\kappa_{5100A}$& $\lambda_{Edd}$ &H$\beta$ FWHM &M$_{BH}$&$\alpha_{ox}$&L$_{bol}$&L$_{2-10keV}$&$\alpha_{X}$&$\alpha_{UV}$\\
&&&&&&{\it km s$^{-1}$\/}&{\it M$_{\sun}$\/}&&{\it 10\/}$^{+44}$&{\it 10\/}$^{+44}$\\
&&&{\it log\/}&{\it log\/}&{\it log\/}&{\it log\/}&{\it log\/}&&{\it log\/}&{\it log\/}&&\\

\hline
$\Gamma_{2-10keV}$ & $\rho_{s}$ & 1 & 0.73 & 0.32 & 0.40 & -0.72 & -0.33 & 0.39 & 0.05 & -0.38 & 0.63 & -0.41 \\
 & $d_{s}^{1}$ & -$\infty$ & -8. & -2. & -2. & -8. & -2. & -2. & -0. & -2. & -6. & -3. \\
\hline
$\kappa_{2-10keV}$ & $\rho_{s}$ & 0.73 & 1 & 0.64 & 0.60 & -0.81 & -0.45 & 0.74 & 0.12 & -0.49 & 0.89 & -0.59 \\
 & $d_{s}^{1}$ & -8. & -$\infty$ & -6. & -5. & -12. & -3. & -9. & -0. & -3. & -17. & -5. \\
\hline
$\kappa_{5100}$ & $\rho_{s}$ & 0.32 & 0.64 & 1 & 0.80 & -0.56 & -0.60 & 0.32 & 0.20 & -0.24 & 0.64 & -0.70 \\
 & $d_{s}^{1}$ & -2. & -6. & -$\infty$ & -11. & -5. & -5. & -2. & -1. & -1. & -6. & -8. \\
\hline
$\lambda_{Edd}$ & $\rho_{s}$ & 0.40 & 0.60 & 0.80 & 1 & -0.40 & -0.24 & 0.38 & 0.62 & 0.16 & 0.49 & -0.55 \\
 & $d_{s}^{1}$ & -2. & -5. & -11. & -$\infty$ & -2. & -1. & -2. & -6. & -1. & -4. & -4. \\
\hline
H$\beta$ FWHM & $\rho_{s}$ & -0.72 & -0.81 & -0.56 & -0.40 & 1 & 0.64 & -0.48 & 0.17 & 0.63 & -0.78 & 0.64 \\
 & $d_{s}^{1}$ & -8. & -12. & -5. & -2. & -$\infty$ & -6. & -3. & -1. & -6. & -11. & -6. \\
\hline
M$_{BH}$ & $\rho_{s}$ & -0.33 & -0.45 & -0.60 & -0.24 & 0.64 & 1 & -0.03 & 0.55 & 0.78 & -0.56 & 0.66 \\
 & $d_{s}^{1}$ & -2. & -3. & -5. & -1. & -6. & -$\infty$ & -0. & -4. & -10. & -5. & -7. \\
\hline
$\alpha_{ox}$ & $\rho_{s}$ & 0.39 & 0.74 & 0.32 & 0.38 & -0.48 & -0.03 & 1 & 0.33 & -0.17 & 0.65 & -0.27 \\
 & $d_{s}^{1}$ & -2. & -9. & -2. & -2. & -3. & -0. & -$\infty$ & -2. & -1. & -6. & -1. \\
\hline
L$_{bol}$ & $\rho_{s}$ & 0.05 & 0.12 & 0.20 & 0.62 & 0.17 & 0.55 & 0.33 & 1 & 0.78 & -0.06 & 0.03 \\
 & $d_{s}^{1}$ & -0. & -0. & -1. & -6. & -1. & -4. & -2. & -$\infty$ & -10. & -0. & -0. \\
\hline
L$_{2-10keV}$ & $\rho_{s}$ & -0.38 & -0.49 & -0.24 & 0.16 & 0.63 & 0.78 & -0.17 & 0.78 & 1 & -0.60 & 0.38 \\
 & $d_{s}^{1}$ & -2. & -3. & -1. & -1. & -6. & -10. & -1. & -10. & -$\infty$ & -5. & -2. \\
\hline
$\alpha_{X}$ & $\rho_{s}$ & 0.63 & 0.89 & 0.64 & 0.49 & -0.78 & -0.56 & 0.65 & -0.06 & -0.60 & 1 & -0.65 \\
 & $d_{s}^{1}$ & -6. & -17. & -6. & -4. & -11. & -5. & -6. & -0. & -5. & -$\infty$ & -7. \\
\hline
$\alpha_{UV}$ & $\rho_{s}$ & -0.41 & -0.59 & -0.70 & -0.55 & 0.64 & 0.66 & -0.27 & 0.03 & 0.38 & -0.65 & 1 \\
 & $d_{s}^{1}$ & -3. & -5. & -8. & -4. & -6. & -7. & -1. & -0. & -2. & -7. & -$\infty$ \\
\hline
   \end{tabular}
\end{table*}

\section{A Systematic Correlation Study on The Key Parameters}
\label{SED:parameter:matrix}
To summarize the various correlations discussed in the previous section, 
we performed a systematic correlation study of the following key
parameters: $\Gamma_{2-10keV}$, $\kappa_{2-10keV}$, 
$\kappa_{5100A}$, $\lambda_{Edd}$, FWHM$_{H\beta}$, M$_{BH}$, 
$\alpha_{ox}$, L$_{bol}$, L$_{2-10keV}$, $\alpha_{UV}$ and $\alpha_{X}$.

First, a correlation matrix was constructed
as shown in Table~\ref{sum-cor-keypar:table}. The Spearman's 
rank coefficient and probability of a null hypothesis are given. 
The table shows that there are some 
sub-groups of parameters which are strongly coupled with each other. 
For example, $\kappa_{2-10keV}$, $\kappa_{5100A}$ and 
$\lambda_{Edd}$ are coupled; $\Gamma_{2-10keV}$, 
H$\beta$ FWHM, $\alpha_{X}$ and $\kappa_{2-10keV}$ are also coupled with 
each other. The strong correlation between H$\beta$ FWHM and 
$\kappa_{2-10keV}$ can be expressed by the following equation 
derived from a bisector regression analysis:
\begin{equation}
\label{k210:hbfwhm:eqn}
Log(\kappa_{2-10}) = (-1.22{\pm}0.12)Log(FWHM) + (5.88{\pm}0.45)
\end{equation}
There also appears to be a sub-group consisting of M$_{BH}$, 
L$_{bol}$ and L$_{2-10keV}$, and a sub-group consisting of 
M$_{BH}$, H$\beta$ FWHM and L$_{2-10keV}$. However, correlations 
within these sub-groups are probably a result of selection 
effects arising from our sample selection criteria. For example, 
he inclusion of extremely weak L$_{2-10keV}$ sources may weaken or
eliminate the correlations between L$_{2-10keV}$ and other parameters.

The observed properties of AGN should be ultimately 
driven by the black hole mass, 
mass accretion rate, black hole spin and orientation angle. 
We have assumed the simplest Schwarzschild black hole in our model and so  
its spin is not considered.
Uncertainties introduced by orientation angle should also be small 
since our sample only contains unobscured Type 1 AGNs.
Therefore, the remaining intrinsic parameters are just the black 
hole mass and mass accretion rate (or equivalently, Eddington ratio)

We can examine the correlations further by performing a principal component analysis
(PCA) on the correlation matrix formed by Pearson's correlation
coefficient (Pearson 1901; Boroson \& Green 1992; Francis \& Wills 1999).
First, we include all the 11 parameters and so the dimension of the
correlation matrix is 11. Therefore the outcome of the
PCA must contain 11 normalized eigenvectors (principal components: PCs),
each associated with a positive eigenvalue. Each PC is a linear
combination of the 11 parameters, and is orthogonal to all the
other PCs. The sum of the 11 eigenvalues equals 11. A higher eigenvalue
would suggest a larger fraction of correlations contained in the
direction of the corresponding eigenvector.

The {\it EIGENQL\/} program in {\tt IDL}
(Interactive Data Language) v6.2 was used to perform
the PCA.  We found that the first three eigenvectors contain 87\% of
the total correlations in the matrix, i.e. 53\% in eigenvector 1
(PC1), 25\% in eigenvector 2 (PC2) and 9\% in eigenvector 3 (PC3). To
determine the actual contributors of these eigenvectors, we
cross-correlated them with the 11 key parameters. Table~\ref{pca:table}
lists the Spearman's rank coefficients. It is clear that PC1
correlates/anti-correlates very well with most key parameters, except
for L$_{bol}$. The highest correlation strength is for
$\kappa_{2-10keV}$, $\alpha_{X}$ and H$\beta$ FWHM. These in turn are probably proxies for the 
physical variables of M$_{BH}$ and $\lambda_{Edd}$. 
PC2 is dominated by L$_{bol}$ which confirms that
L$_{bol}$ is a relatively independent variable. We have assumed that 
$L_{bol}=\mu\dot{M}c^{2}$, where $\mu$ is the standard accretion
efficiency of 0.057 (see Paper-I), so PC2 is in effect
dominated by the mass accretion rate $\dot{M}$. The contribution
from the rest eigenvectors to the set of correlation
is small compared to PC1 and PC2, and are therefore not
important. It was also reported by Boroson (2002) that the PC1 from
the correlation matrix of optical emission line parameters is driven
predominantly by the Eddington ratio, while the PC2 from the same
matrix is dominated by the luminosity. Therefore, the PC1 and PC2
from our correlation matrix of the 11 parameters have a similar basis to
the two eigenvectors reported by Boroson (2002).

The inclusion of $\lambda_{Edd}$, M$_{BH}$ and L$_{bol}$ may bias the
outcome of PCA, so we performed an independent PCA by excluding 
$\lambda_{Edd}$, M$_{BH}$ and L$_{bol}$ from the correlation matrix
(Grupe 2004; Grupe 2011), and so the
remaining matrix only has a dimension of 8. The resultant eigenvectors were
again correlated with each of the 11 parameters. The results are listed in
Table~\ref{pca:table}. The two new principal eigenvectors
(PC1-excl and PC-excl in Table~\ref{pca:table}) are quite similar as PC1 and PC2
in terms of correlation outcomes with the 11 parameters. This confirms our previous
PCA finding that the three physical parameters $\lambda_{Edd}$, M$_{BH}$ and L$_{bol}$
drive the majority of the correlations.

\label{correlation:pca}
\begin{table}
 \centering
 \caption{The cross-correlation of the eigenvectors with the 11 key parameters. The coefficients given are from the Spearman's rank test. PC1 and PC2 are the principal components from the PCA on all the 11 parameters, while PC1-excl and PC2-excl are the principal components after excluding $\lambda_{Edd}$, M$_{BH}$ and L$_{bol}$ from the PCA. The row `Property' shows the percentage of correlations contained in the direction of that eigenvector.}
 \label{pca:table}
     \begin{tabular}{lcccc}
	\hline
	& PC1 & PC2 & PC1-excl &PC2-excl \\
	Property & 53\% & 25\% & 62\% & 14\%\\
	\hline
	$\Gamma_{2-10keV}$ & 0.71 & 0.03 & 0.75 & -0.14 \\
	$\kappa_{2-10keV}$ & 0.93 & 0.14 & 0.94 & 0.06\\
	$\kappa_{5100A}$ & 0.74 & 0.26 & 0.67 & 0.34\\
	$\lambda_{Edd}$ & 0.60 & 0.67 & 0.51 & 0.60\\
	H$\beta$ FWHM & -0.89& 0.17 & -0.90 & 0.21\\
	M$_{BH}$ & -0.69 & 0.48 & -0.65 & 0.34\\
	$\alpha_{ox}$ & 0.59 & 0.36 & 0.63 & 0.35 \\
	L$_{bol}$ & -0.06 & 0.99 & -0.10 & 0.81 \\
	L$_{2-10keV}$ & -0.60 & 0.76 & -0.65 & 0.67 \\
	$\alpha_{X}$ & 0.92 & -0.03 & 0.93 & -0.01 \\
	$\alpha_{UV}$ & -0.74 & -0.04 & -0.70 & -0.17 \\
	\hline
     \end{tabular}
\end{table}

\section{The Mean SEDs}
\label{best:sed:indicator}
\subsection{Diversity of the  Mean SEDs}
In Paper-I we derived the mean SED for the 12 NLS1s in our sample,
and compared this with the mean SEDs of another two groups of BLS1s.
We showed that the SED shape changes dramatically 
as the H$\beta$ FWHM increases. Since quasar SED 
are not uniform, modelling their spectra and evolution using a
single template such as that of Elvis et al. (1994), will mask out a large
dispersion in their intrinsic properties. In our study we aim to minimize this 
dispersion by grouping the SED based on each of the 11 key parameters
discussed in the previous section.
However, since $\alpha_{UV}$ and $\alpha_{X}$ are rough estimates of the
spectral shape in optical/UV and soft X-ray regions,
we excluded them from the mean SED calculation.

There are 51 AGNs in our sample. For each of
the 9 remaining parameters, we sorted the sources according to the parameter
value, and then classified the sample into three subsets so that each
subset includes 17 sources.
Then the BAL quasar PG 1004+130 was excluded from its subset.
The SEDs constructed using
Model-B in this paper were first corrected for redshift, and then
divided into 450 energy bins between 1 eV and 100 keV in the
logarithm. Within each subset we calculated the average luminosity in
every energy bin in logarithm, together with the 1$\sigma$
deviation.  Then we derived the mean SED for each group together with
the 1$\sigma$ dispersion.  The same calculation was repeated for each
of the 9 parameters, so that there are three mean SEDs for each 
parameter. No special note was made for the NLS1s because their
defining parameter, H$\beta$ FWHM, is one of the 9 parameters.

Figure~\ref{best-sed-indicator} shows our results. Each row displays
the three mean SEDs divided according to the parameter shown in the
panel title. The SEDs have all been renormalized to the
mean luminosity at 2500{\AA} of each subset. To highlight the
differences among these SEDs, we mark the locations of 2500{\AA}
and 2 keV by the vertical orange lines. The relative height
of these two lines directly reflects the value of $\alpha_{ox}$, and
the height of the bar at 2~keV shows the dispersion in $\alpha_{ox}$
since the co-added SED's are all normalized at 2500\AA. We also mark
the energy peak for each mean SED by the vertical thick purple
line. The mean values of other
parameters are given in each plot for comparison. 

We find that the mean SED changes in a similar way with all the parameters except
L$_{bol}$. The energy at which the disc emission peaks decreases
along with the ratio of luminosity in the disc compared to the
Comptonised components, and the 2-10~keV spectral slope.  Our
stringent sample selection means that these spectral differences are
intrinsic rather than due to absorption/extinction. 

If the SED changes are determined solely by one of the parameters
considered here, then binning based on that parameter should result in 
the minimum dispersion within each individual binned SED, and maximal
difference between the three SEDs spanning the range in that parameter.
However, the SED changes should ultimately be physically dependent on 
changes in M$_{BH}$ and $\lambda_{Edd}$. Since there are two dependent
variables no single parameter will completely determine the
behaviour. Hence the dispersion within each of the three binned SEDs
is minimized (and the difference between them is maximized)
for composite parameters which
depend on both $\lambda_{Edd}$ and M$_{BH}$ such as
$\kappa_{2-10keV}$, $\kappa_{5100}$ and H$\beta$ FWHM
rather than the fundamental physical parameter $\lambda_{Edd}$.
Future work with larger samples can improve on this study by selecting
a subsample of AGN with different $\lambda_{Edd}$ but similar
M$_{BH}$. Such mass selected samples would give the best comparison to
the SED changes in Black Hole Binaries (BHB), which all have the same
mass to within a factor of $\sim 2$.

\begin{figure*}
\centering
\includegraphics[scale=0.73,clip=]{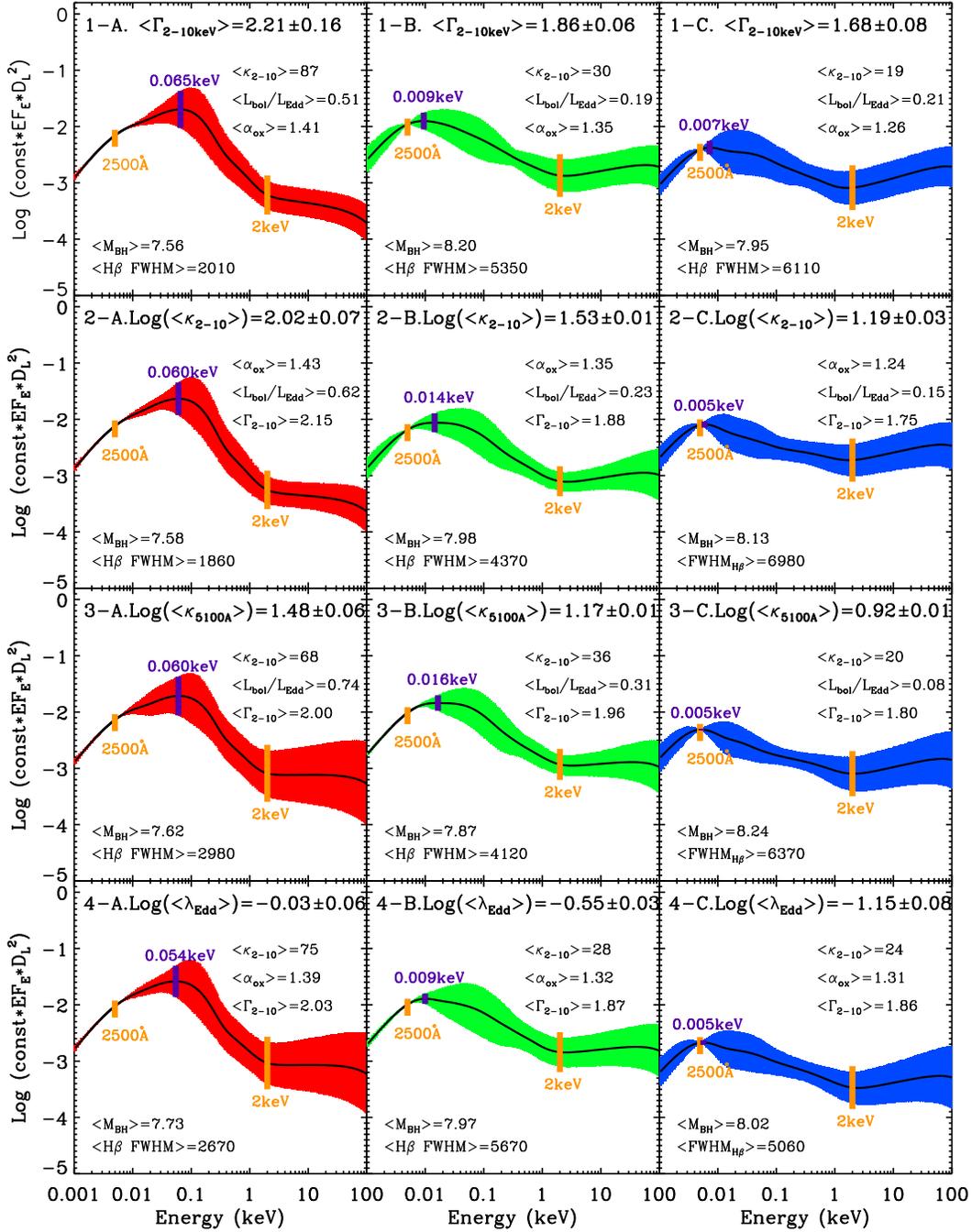}
\caption{The AGN mean SEDs based on different values of the 9 key parameters from Model-B fitting (i.e. including the effect of a colour temperature correction). For each parameter, the 51 sources are sorted according to the parameter value, and then are divided into three equal subsets so that each contains 17 sources. PG 1004+130 is excluded from its subset. Finally, a mean SED is constructed for each of the three subsets after renormalizing each individual SED to the mean luminosity at 2500{\AA} of that subset. The three panels (A, B, C) in each row show the mean SEDs for the subsets classified by the parameter shown in the panel title. In each panel the solid curve is the mean SED, while the shaded coloured region is the $\pm$1$\sigma$ deviation. The 2500 {\AA} and 2 keV positions are marked by the vertical solid orange lines, whose relative height indicates the value of $\alpha_{ox}$. The peak position of the SED is marked by the vertical solid purple line. The average values of some other parameters in that subset are also shown in the panel. Each mean SED has been rescaled by the same arbitrary constant on the Y-axis which is 1$\times$10$^{-46}$. Note that the energy ranges 1.4 eV $<$ E $<$ 6 eV and 0.3 keV $<$ E $<$ 10 keV, are covered by SDSS, OM and EPIC data respectively, while the SED in the rest energy bands is determined from an extrapolation of the best-fit model.}
\label{best-sed-indicator}
\end{figure*}
\addtocounter{figure}{-1}
\begin{figure*}
\centering
\includegraphics[scale=0.73,clip=]{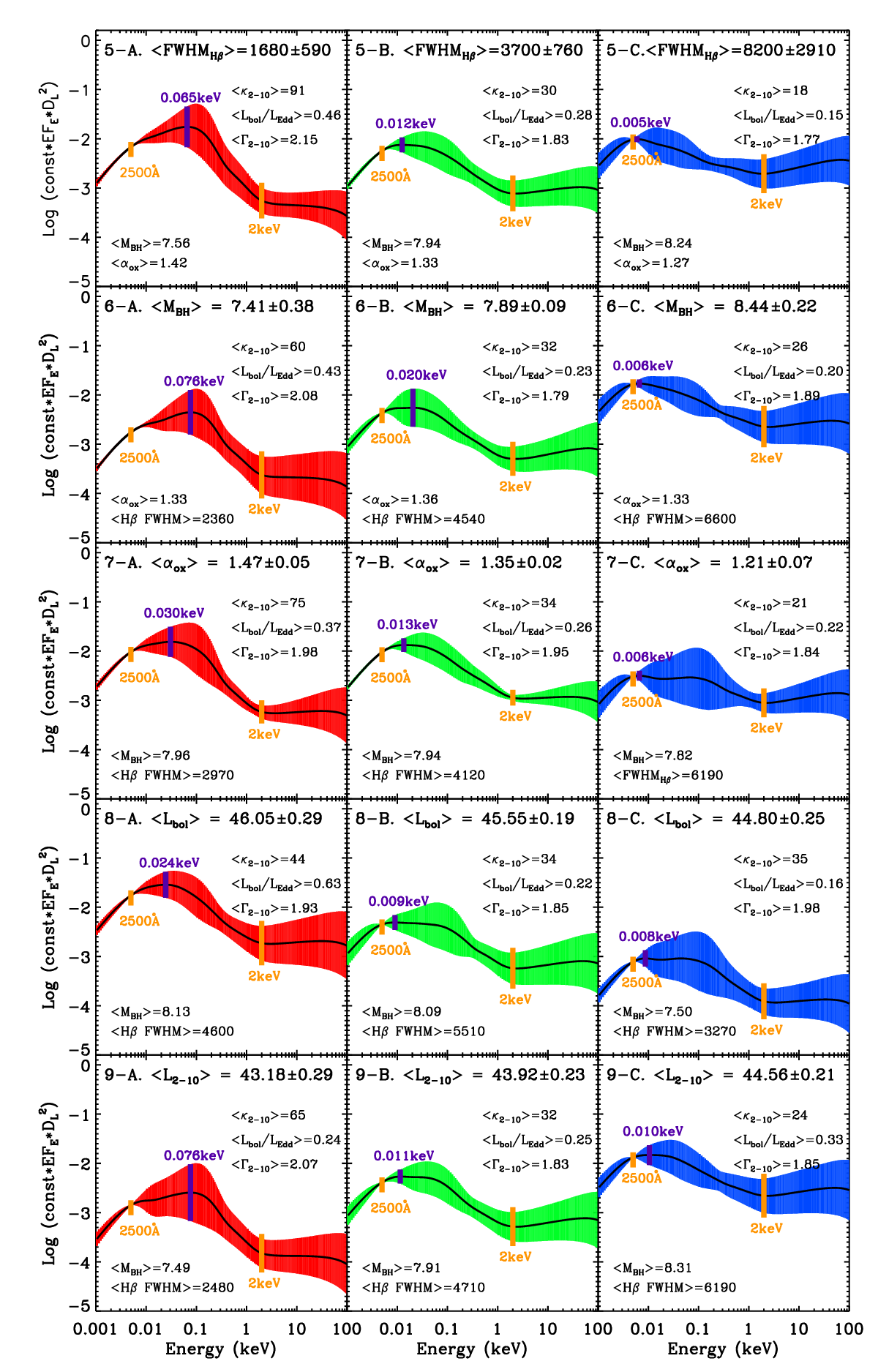}
\caption{{\it continued\/}}
\end{figure*}

\subsection{Discussion of the Mean SEDs}
Since our SED fitting is based on a physical model, we can
`correct' for the mass dependence of the SED shape to get an estimate
for the SED differences in AGN as a function solely of $\lambda_{Edd}$.
This is shown in Done et al. (2011) for a M$_{BH}=10^8 M_{\sun}$ and forms
the basis of a direct comparison with the BHB spectral states seen for
these M$_{BH}=10M_{\sun}$ systems as $\lambda_{Edd}$ changes.  This
has many superficial similarities to the dramatic state change seen in
BHB as their luminosity increases. The SED changes from a `low/hard
state' being dominated by Comptonisation, with a hard X-ray spectral
index $\Gamma_{2-10keV}<2$, and the disc component peaking at rather low
temperature, to a 'high/soft state' where the disc dominates the
luminosity and the X-ray spectral index is softer,
$\Gamma_{2-10keV}\sim 2-2.2$ (see e.g. the review by Done, Gierli\'{n}ski
\& Kubota 2007). However, this occurs at $\lambda_{Edd}\sim 0.02$ for
moderate changes in mass accretion rate (Maccarone 2003), an order of
magnitude lower than the spectral differences seen here in the AGN
(see also the discussion in Done et al. 2011). This
could indicate some subtle differences in the transition due to
the very different masses, but in BHB this transition is also
associated with the radio jet switching off (Fender, Belloni \& Gallo
2004). If the Compton dominated states in AGN correspond to the
low/hard state in BHB then we would expect them to be radio
loud. However, AGNs are radio-quiet by a factor of 10:1.  

This makes unlikely an identification of the two lower $\lambda_{Edd}$ AGN
templates as analogs to the low/hard (or intermediate state) seen in
BHB.  Instead, there is another state in BHB called the
`very high' or `steep power law state', where the disc also peaks at a
lower temperature than expected, and where the Comptonised component
contains a large fraction of the total luminosity (see e.g. the review by
Done, Gierli\'{n}ski \& Kubota 2007). However, this state has $\Gamma_{2-10keV}>2.5$ i.e. the
hard X-ray spectra are steep. Yet these AGN have $\Gamma_{2-10keV}<2$,
as well as a soft X-ray excess component. Therefore, to match the AGN
with the very high state would require that reflection and/or complex
absorption modify the spectrum, producing an apparent soft X-ray
excess and a hard power law from an intrinsically steep
spectrum. However, the time variability properties of individual objects
make it clear that these high mass, relatively low Eddington ratio
objects do indeed have two separate components. The intrinsically
hard power law which is more variable on short timescales, and a soft X-ray
excess which is relatively constant on these timescales but more
variable over longer times (e.g. Mkn509: Mehdipour et al. 2011; Noda et al. 2012). 

Therefore, the two lowest $\lambda_{Edd}$ spectra shown in
Figure~\ref{best-sed-indicator} do not look similar to {\it any\/} state
observed in BHB. Yet these sources span the typical QSO accretion rates
(e.g. Steinhardt \& Elvis 2010), and indeed our templates are very
similar to the mean SED in Elvis et al. (1994).
It seems that these most common QSO SED's
are not simply analogous to BHB accretion flows. Only the very rare
AGN SED's with the highest  $\lambda_{Edd}$ can be well matched to the BHB,
as they are similar to the high/soft state (see also Done et al. 2011).

\section{Discussion}

\subsection{Selection Effects}
\label{section:selection:effect}
Biases and systematics inherent in this sample have already been
discussed at length in Paper-I and Paper-II. The principal imposed
selection effect is that sources in our sample are bright nearby AGN
($z~<$ 0.4). The luminosity of our sources is higher than the average
among nearby sources, but only moderate with respect to samples
containing higher redshift sources (e.g. Green09; Lusso10).  Our
sample contains very few sources with L$_{2-10keV}~<$
5$\times$10$^{42}$ erg s$^{-1}$ or $\lambda_{Edd}~<$ 0.05.  As
discussed in Section~\ref{subsection:k210:eddr} and
Section~\ref{section:k210:l210}, those very low luminosity sources may
not follow the linear regression line in the $\kappa_{2-10keV}$
vs. $\lambda_{Edd}$ plot in Figure~\ref{k210:eddr}, and these sources
may populate the low L$_{2-10keV}$, low $\kappa_{2-10keV}$ region in
Figure~\ref{l210:k210:plot}.  It is also possible that these sources
may not follow other correlations reported in this paper, thereby
weakening the correlations between L$_{2-10keV}$ and M$_{BH}$ and
H$\beta$ FWHM. Further studies of large samples are required to test
such possibilities.

The weak anti-correlation found between $\lambda_{Edd}$ and M$_{BH}$
(Table~\ref{sum-cor-keypar:table}) also implies some selection effect.
Sources having both low black hole mass and low mass 
accretion rate are unlikely to
be included in our sample as they would be too faint. 
Hence low mass sources in our sample
will have relatively high $\lambda_{Edd}$. For higher mass sources,
their luminosity will peak when there is a considerable supply of gas around
them to be accreted. This occurs around redshift $\sim$2, and as the
available gas decreases, the mass accretion rate of high mass
AGN in the local universe is suppressed (so-called downsizing,
Fanidakis et al. 2010). Therefore, in the local universe high mass
sources should have low $\lambda_{Edd}$, resulting in the weak
anti-correlation found between M$_{BH}$ and $\lambda_{Edd}$
(Done et al. 2011). 

There are also redshift selection effects. Although comparison of
parameter correlations with previous work (based on larger samples)
results in a general consistency, the question remains whether there
could be a redshift dependence in the correlations we find. However, there are  
some evidences that redshift evolution in the spectral properties of AGN may 
not be strong (Fan 2006).

\subsection{Limitations of the Model and Uncertainties}
There is another underlying question, whether the correlations
found might arise artificially as a result of our model assumptions.
We will consider this point in two ways:\\
First, there are no direct constrains on the parameters
in our SED model. Compared with some previous work
(e.g. VF07, Lusso10, Grupe10), our spectral fitting employs the least
external constraints on the values of its parameters. The only parameter that 
is directly constrained is M$_{BH}$, whose value is restricted by the FWHM of
the intermediate and broad components of the H$\beta$ emission line.
However, this range often spans more than one order of magnitude, and the
best-fit M$_{BH}$ did not exceed the model limits for most sources
(see Table C1 of Paper-I). Therefore this constrain should not be 
strong enough to cause systematics. \\
Second, for previously known correlations such as $\lambda_{Edd}$
vs. $\kappa_{2-10keV}$, $\lambda_{Edd}$ vs. $\Gamma_{2-10keV}$
and H$\beta$ FWHM vs. $\Gamma_{2-10keV}$, our results are all
consistent with past studies based on a variety of AGN samples.
This suggests that the reported correlations should be intrinsic,
and that our results are not strongly contaminated by model assumptions.
As discussed in the previous sections, the differences between our results and
those previously reported are mainly due to two reasons.
One is our exclusion of highly obscured sources, which reduces the
non-intrinsic dispersion within these correlations.
The other reason is that our parameter
values are derived from a detailed spectral fitting, rather than
from simply applying scaling relations which will contain high 
uncertainties.

However, the range of values for each of the 11 key parameters discussed
previously could be dominated by model uncertainties, except
for $\Gamma_{2-10keV}$, whose measurement is relatively
model independent. For this reason we did not adopt the uncertainties
returned by {\tt Xspec}, because they must be negligible
compared with the model uncertainties. Such model uncertainties
are very difficult to estimate, and the values of the same parameter
derived from different SED models, can be quite different.
As an illustration, in the following paragraphs we will compare the 
parameter values before and after introducing a colour temperature
correction into our broadband SED model.

\begin{table*}
 \centering
   \caption{The average value of parameters from the best-fit SEDs using Model-A (without a colour temperature correction, Paper-I) and Model-B (with colour temperature correction). The values and their standard deviations are calculated separately for NLS1s, BLS1s and for the whole sample.}
   \label{tab:cc}
     \begin{tabular}{lcccccc}
\hline
&NLS1s&NLS1s&BLS1s&BLS1s&Whole Sample&Whole Sample\\
Model-&A&B&A&B&A&B\\
\hline
$<\kappa_{2-10keV}>$&127$^{+197}_{-77}$&86$^{+96}_{-45}$&29$^{+44}_{-17}$&30$^{+38}_{-17}$&41$^{+85}_{-27}$&38$^{+58}_{-23}$\\
\hline
$<\kappa_{5100A}>$&29$^{+37}_{-16}$&20$^{+13}_{-8}$&13$^{+17}_{-8}$&14$^{+14}_{-7}$&16$^{+23}_{-9}$&15$^{+14}_{-7}$\\
\hline
$<\lambda_{Edd}>$&0.95$^{+5.33}_{-0.80}$&0.35$^{+0.99}_{-0.26}$&0.27$^{+0.81}_{-0.20}$&0.25$^{+0.52}_{-0.17}$&0.36$^{+1.42}_{-0.29}$&0.27$^{+0.61}_{-0.19}$\\
\hline
$<$M$_{BH}>$&7.11$\pm$0.54&7.37$\pm$0.47&8.04$\pm$0.48&8.10$\pm$0.41&7.83$\pm$0.64&7.93$\pm$0.52\\
\hline
$<\alpha_{ox}>$&1.42$\pm$0.08&1.39$\pm$0.10&1.34$\pm$0.16&1.34$\pm$0.15&1.36$\pm$0.14&1.35$\pm$0.14\\
\hline
$<$L$_{bol}>$&45.19$\pm$0.54&45.02$\pm$0.49&45.59$\pm$0.52&45.61$\pm$0.53&45.49$\pm$0.55&45.47$\pm$0.57\\
\hline
   \end{tabular}
\end{table*}

\subsection{The Effect of the Colour Temperature Correction}
\label{section:color:correction}
\begin{figure*}
  \centering
  \includegraphics[scale=0.8,clip=]{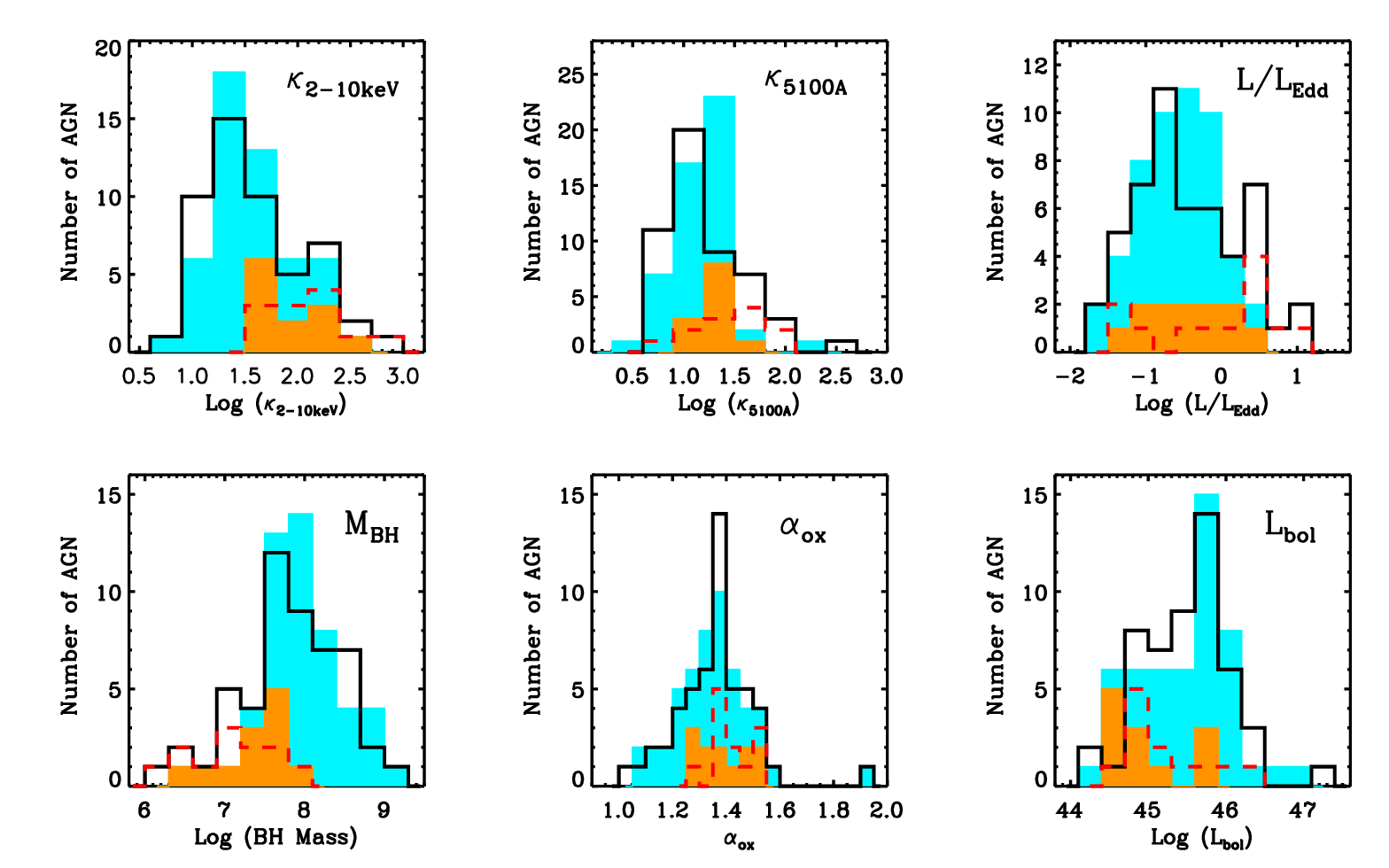}
  \caption{Comparison of parameter distributions between Model-A and Model-B SED fittings.
In each panel the shaded cyan histogram is based on our modified SED fitting using Model-B
(listed in Table~\ref{app:SED-key-parameters}), with the 12 NLS1s highlighted
by the shaded orange region. The solid black line shows the parameter histogram for 
Model-A fitting (listed in Table~3 of Paper-I), with the dashed red line indicating the 12 NLS1s.}
 \label{fig:cc}
\end{figure*}

The colour temperature correction is only important for sources having
both a low black hole mass and a high mass accretion rate
(see Done et al. 2011 and references therein). So it only
affects a small fraction of all the sources in our sample, mainly the
NLS1s. The main consequences of introducing a colour temperature
correction by using Model-B (i.e. {\it optxagnf \/} in {\tt Xspec}
v12) are that M$_{BH}$ increases, L$_{bol}$ decreases and so
$\lambda_{Edd}$ decreases. For example, the $\lambda_{Edd}$ of PG
2233+134 decreases significantly from 14 to 2.4 after using Model-B,
making it much less extreme.  $\kappa_{2-10keV}$ and $\kappa_{5100A}$ also
decrease due to the decrease of L$_{bol}$. $\alpha_{ox}$ changes
slightly but not significantly, because the
luminosity at 2500{\AA} is mainly constrained by the OM data (Paper-I).
Figure~\ref{fig:cc} compares the distribution of these parameters
between Model-A (Paper-I) and Model-B fitting. Table~\ref{tab:cc} lists
the average values of these parameters for NLS1s, BLS1s and the whole
sample, for both Model-A and Model-B fitting. This confirms that the
large differences in results from the Model-A and Model-B fittings are
mostly restricted to the NLS1s, whose colour temperature corrections
are significant.

In order to further investigate the differences between using Model-A
and Model-B, we redo all of the above cross-correlation analysis by
adopting Model-A parameter values from Paper-I, and then we compare the
statistical results in the appendices.
Appendix~\ref{app:section:plots} shows all of the correlation plots
that could in principle be modified by the differences between the
Model-A and Model-B fittings. In each plot the dashed gray lines are
for Model-B fitting (this paper), compared with the solid orange lines
for Model-A fitting. It is clear that
there are no significant changes in any of these correlation plots.
This is further confirmed by the correlation matrix for Model-A
fitting in Appendix~\ref{app:section:matrix}.  Performing a PCA on
this matrix, very similar eigenvectors and eigenvalues are obtained.
Appendix~\ref{app:section:mean:sed} shows the mean SEDs based on the
parameter values obtained from Model-A fitting, which does not include
the colour temperature correction.
Therefore, we conclude that use of the refined Model-B
compared with the original Model-A, does not alter the main results
reported in this paper, although for individual sources such as the
NLS1s, the refined model should be more realistic.

\subsection{The Correction for Radiation Pressure}
Marconi et al. (2008) suggested that the virial mass estimates should be corrected
for the effects of radiation pressure (RP), especially for those sources with high
Eddington ratios,  such as the NLS1s. In our study the M$_{BH}$ was not derived
directly from the virial mass, but was only constrained by the virial masses
calculated from the FWHM of the intermediate and broad components of the H$\beta$ line.
The final estimate of M$_{BH}$ is derived from the best-fit SED,
and so there is no need to correct for the
RP effect. However, we may still choose to derive the virial mass from using the FWHM of
H$\beta$ profile (narrow component subtracted),
and then correct these values for the RP effect.
Paper-I has listed and compared these masses (M$_{RP}$) with the best-fit masses
(M$_{BF}$) from the SED fitting using Model-A.
The distributions of these two estimates of masses
are very similar, except that the average M$_{RP}$ is 0.4 dex higher
compared with M$_{BF}$.

As discussed in previous section,
the SED fittings including the colour temperature correction
(Model-B) increases the average M$_{BF}$ by 0.27 dex for NLS1s,
and 0.05 dex for BLS1s (Table~\ref{tab:cc}).
So for the NLS1s using Model-B fitting, the average M$_{BF}$ is just 0.05 dex smaller
than  the average M$_{RP}$, while for the BLS1s the difference is still 0.34 dex.
Furthermore, no significant difference is found in Table~\ref{sum-cor-keypar:table}
if we substitute  M$_{RP}$ for M$_{BF}$,
and cross-correlate with the other 10 SED parameters.
The coefficients in Equation~\ref{gamma:bhmass:eqn:1}, \ref{gamma:bhmass:eqn:2}
only differ by less than 1 $\sigma$ when using M$_{RP}$ instead of M$_{BF}$.
This suggests that the difference between M$_{RP}$ and M$_{BF}$ is far less than 
the intrinsic dispersion in any of  the correlations,  
and so is not important in our correlation studies.

\subsection{The 4000 km s$^{-1}$ H$\beta$ FWHM Break}

The FWHM$_{H\beta}$ = 2000 km s$^{-1}$ is 
the conventional, but arbitrary value to 
distinguish between NLS1 and BLS1 (Goodrich 1989). 
Recently, the limit of  4000 km s$^{-1}$  
for the FWHM$_{H\beta}$ was claimed to be of special interest. For 
example, when AGNs are divided into two populations based on FWHM$_{H
\beta}$ = 4000 km s$^{-1}$ (population A: FWHM$_{H\beta}~\la$ 4000 
km s$^{-1}$ and B: FWHM$_{H\beta}~\ga$ 4000 km s$^{-1}$), it 
appeared that most radio loud sources are contained in population B
(Sulentic et al. 2008). Compared with the RL-RQ and the NLS1-BLS1 divisions,
this dividing line in FWHM also seems to be more 
effective in distinguishing the different SEDs (Sulentic et al. 2008).
Furthermore, Zhou10a 
reported that in the FWHM$_{H\beta}$ vs. $\Gamma_{2-10keV}$ 
correlation plot, there is a change in the slope at $\sim$4000 km s$^{-1}$, as 
consistent with our results. They also showed that the the broadest iron
K$\alpha$ lines, those with intrinsic width ${\sigma}>$ 0.5 keV)
have are all found in AGNs with 
FWHM$_{H\beta}<$ 4000 km s$^{-1}$.

In Section~\ref{group5}, we confirmed a slope change at FWHM$_{H\beta}$ $\simeq$ 4000
km s$^{-1}$ in FWHM$_{H\beta}$ vs. $\Gamma_{2-10keV}$ correlation plot.
All the three RL sources in our sample (purple symbols in
Figure~\ref{pi210-bhfwhm-bhmass}) have FWHM$_{H\beta}~>$ 4000 km s$^{-1}$.
To highlight the 27 population A sources in our sample, we use the
square symbol to identify these sources in all correlation plots,
and we show their distribution in the histograms as the green region.
The 12 NLS1s among them are shown as the red region.
The two-sided KS-Test was used to determine the significance of
the difference between the distributions of the sub-samples for
the NLS1-BLS1 division, and the population A-B division.
Among the 11 key parameters, the population A-B provides a
somewhat better division for the 7 parameters:
$\kappa_{2-10keV}$, $\kappa_{5100A}$, $\lambda_{Edd}$,
H$\beta$ FWHM, $\alpha_{ox}$, $\alpha_{X}$ and $\alpha_{UV}$;
while NLS1-BLS1 division is better for the other parameters (see
Table~\ref{population:A:B:KStwo}). Therefore, the population A-B
division for our RQ Type 1 AGN sample does not seem to be
significantly better than the NLS1-BLS1 division.

\section{Summary and Future Work}
\subsection{Summary of Principle Results}
In this third paper in the series,
we have studied the SED properties of our Type 1 AGN sample.
We employ a new broadband SED model ({\it optxagnf \/} in {\tt Xspec} v12),
which includes a colour temperature correction, to construct the
SED for each source in the sample. Various parameters were obtained from the
results of the SED fitting. A detailed statistical analysis was performed,
which can be divided into three major parts:

$\bullet$ In the first part we studied the diverse correlations found among 
the SED parameters.
We divided these parameters into five groups, according to previously reported
correlations. Within each group we conducted a detailed cross-correlation
analysis, and applied several regression methods. Our results are generally
in good agreement with previous work. However, as a consequence of 
the unobscured nature of the sample and the more
reliable parameter values that resulted from our refined spectral fitting,
we were able to reduce the non-intrinsic dispersion and so obtain
the intrinsic and better constrained correlations. 

\begin{table*}
 \centering
   \caption{The significance level of the difference in the NLS1-BLS1 pair (N1-B1) and Population A-B pair (Pop. A-B) for values of the 11 key parameters. The two-sided Kolmogorov-Smirnov test is applied. A smaller value suggests a greater difference within each sample division pair.}
   \label{population:A:B:KStwo}
     \begin{tabular}{@{}lccccccccccc@{}}
\hline
&$\Gamma_{2-10keV}$&$\kappa_{2-10}$&$\kappa_{5100A}$& $\lambda_{Edd}$ & FWHM &M$_{BH}$&$\alpha_{ox}$&L$_{bol}$&L$_{2-10keV}$&$\alpha_{X}$&$\alpha_{UV}$\\
\hline
N1-B1&2$\times$10$^{-5}$&5$\times$10$^{-5}$&1$\times$10$^{-1}$&5$\times$10$^{-1}$&1$\times$10$^{-7}$&1$\times$10$^{-5}$&2$\times$10$^{-1}$&4$\times$10$^{-3}$&7$\times$10$^{-7}$&1$\times$10$^{-4}$&1$\times$10$^{-2}$\\
Pop. A-B&1$\times$10$^{-4}$&7$\times$10$^{-6}$&5$\times$10$^{-3}$&8$\times$10$^{-3}$&4$\times$10$^{-11}$&1$\times$10$^{-3}$&1$\times$10$^{-2}$&5$\times$10$^{-1}$&1$\times$10$^{-2}$&3$\times$10$^{-5}$&3$\times$10$^{-3}$\\
\hline
   \end{tabular}
\end{table*}

For example, we confirmed the linear correlation in
Log(L$_{2500}$) vs. Log(L$_{2keV}$) with slope $\sim$1.
The correlation of Log($\kappa_{2-10keV}$) vs. $\alpha_{ox}$ can be
approximated using a second order polynomial. The correlation between
$\alpha_{ox}$ and Log($\lambda_{Edd}$) is weak and dominated by dispersion 
in intrinsic source properties. Some strong correlations have been
confirmed e.g. Log($\kappa_{2-10keV}$) vs. Log($\lambda_{Edd}$),
Log($\kappa_{5100}$) vs. Log($\lambda_{Edd}$), Log($\kappa_{5100}$) vs.
Log($\kappa_{2-10keV}$), $\Gamma_{2-10keV}$ vs. Log($\lambda_{Edd}$),
$\Gamma_{2-10keV}$ vs. Log($\kappa_{5100}$), Log(M$_{BH}$) vs.
Log(FWHM$_{H\beta}$) and Log($\kappa_{2-10keV}$) vs. Log(M$_{BH}$). 
The correlations in both $\Gamma_{2-10keV}$ vs. Log(FWHM$_{H\beta}$)
and $\Gamma_{2-10keV}$ vs. Log(M$_{BH}$) change slopes as
$\Gamma_{2-10keV}$ decreases to $\sim$1.8.
The break region is FWHM$_{H\beta}$ $\simeq$ 4000 km s$^{-1}$ and
Log(M$_{BH}$) $\simeq$ 8.0. $\Gamma_{2-10keV}$ is almost
independent of the FWHM$_{H\beta}$ after the break region, with a mean
value of 1.8.
By presenting the correlations between $\alpha_{X}$, $\alpha_{UV}$, $\alpha_{ox}$ and $L_{2-10keV}$,
we also confirmed that AGNs with bluer optical/UV spectra tend to have steeper X-ray spectra,
and their hard X-ray emission is also weaker.
Other strong (anti-)correlations were found for $\alpha_{X}$ and $\alpha_{UV}$, such as
$\alpha_{X}$ ($\alpha_{UV}$) vs. $\kappa_{2-10keV}$,
$\alpha_{X}$ ($\alpha_{UV}$) vs. $\kappa_{5100A}$,
$\alpha_{X}$ ($\alpha_{UV}$) vs. FWHM$_{H\beta}$
and $\alpha_{X}$ ($\alpha_{UV}$) vs. M$_{BH}$.

$\bullet$ In the second part of our work, we performed a systematic cross-correlation
study by producing the correlation matrix of the 11 key parameters:
$\Gamma_{2-10keV}$, $\kappa_{2-10keV}$, $\kappa_{5100A}$,
$\lambda_{Edd}$, FWHM$_{H\beta}$, M$_{BH}$, $\alpha_{ox}$,
L$_{bol}$, L$_{2-10keV}$, $\alpha_{X}$ and $\alpha_{UV}$. The PCA was performed on the correlation
matrix to discover the principal eigenvectors that drive the most correlations.
We found that the first two eigenvectors (PC1 and PC2) contain
$\sim$80\% of all correlations in the matrix. PC1 strongly
correlates with M$_{BH}$, while PC2 is dominated by L$_{bol}$.
In addition both PC1 and
PC2 well correlate with $\lambda_{Edd}$. Thus the two principle
eigenvectors are driven by M$_{BH}$, $\lambda_{Edd}$ and L$_{bol}$
(or equivalently $\dot{M}$). Our eigenvectors also have similar properties
to the two principal eigenvectors derived by Boroson (2002) based on 
their optical emission line study.

$\bullet$ In the third part we produced various mean SEDs classified by each of the
key parameters. The SED shapes are found to exhibit similar changes
with most parameters except L$_{bol}$. This explains the strong correlations found 
among these key parameters. A more detailed mean SED comparison suggests that
the dispersion within each of the three binned SEDs is minimized
(and the difference between them is maximized)
for composite parameters which
depend on both $\lambda_{Edd}$ and M$_{BH}$, such as
$\kappa_{2-10keV}$, $\kappa_{5100}$ and H$\beta$ FWHM.
This is because the SED change is not determined solely by any one
of these key parameters. It should ultimately depend on both 
$\lambda_{Edd}$ and M$_{BH}$.

$\bullet$ To test the robustness of the main results from our
correlation study, we looked at the 
black hole masses corrected for the effect of radiation pressure.
We found no significant differences
from using our best-fit black hole masses.
We also compared the correlation results
between Model-A (without a colour temperature correction)
and Model-B (including a colour temperature correction) fitting,
and found that they were all very similar.

$\bullet$ The population A-B division for AGNs was compared with
the NLS1-BLS1 division, but it did not prove to be a better
AGN classification method.

\subsection{Future Work}
Our sample is limited to relatively high $\lambda_{Edd}$, with few objects 
below $\lambda_{Edd}=0.05$. These (predominantly LINER) objects
are the ones expected to be the counterparts of the
low/hard state in BHB. Another important extension would be 
to increase the sample size and include rare higher mass objects
with high Eddington ratios. This would allow
the SEDs to be co-added for different $\lambda_{Edd}$ at a given
(fixed) black hole mass, thus providing a direct comparison with the BHB states.

The major result of this study is that the SEDs of AGN exhibit a very wide range,
most plausibly as a function of mass accretion rate for a given mass black hole. 
This clearly shows that so-called unified schemes,
where AGN have intrinsically identical spectra which are  
modified by orientation dependent obscuration, are an over simplification 
of the actual situation. In fact, unobscured AGN can have quite different SED shapes
depending on $\lambda_{Edd}$, and $M_{BH}$.

Although not widely appreciated, this is broadly expected by analogy of
AGN with BHB. The stellar mass black holes
clearly show a dramatic change in spectral shape with $\lambda_{Edd}$,
but unlike AGN, these changes can be tracked in a single object because of  
the much shorter timescale for variability. However, while the highest
$\lambda_{Edd}$ spectra appear similar to the disc dominated
`high/soft state' seen in BHB, the more typical AGN (with an SED similar to the 
standard quasar SED template in Elvis et al. (1994)), do not appear to 
have SED properties which match with any spectral state known in BBB.

This might indicate a physical break in the properties of the accretion flow between
stellar mass and supermassive black holes. The most obvious change in 
physical conditions between these two mass regimes is that AGN discs
are strong in the UV, and so are capable of powering substantial mass loss via 
a UV line driven wind. A consequence of mass loss in the wind is that the
accretion rate is no longer constant as a function of radius, 
causing an intrinsic change in the structure of the accretion flow
(e.g. Proga, Stone \& Kallman 2000). 
Emission/absorption/scattering processes in the wind can also change the 
observed properties of the spectrum (Sim et al. 2010). Further work on 
theoretical disc models including these effects will show whether
standard AGN accretion flows are indeed sculpted by a wind.

\section*{Acknowledgements}
We are very grateful to Dirk Grupe for his useful comments and suggestions.
C. Jin acknowledges financial support through Durham Doctoral Fellowship.
This work is partially based on data from SDSS,
whose funding is provided by the Alfred P. Sloan Foundation,
the Participating Institutions, the National Science Foundation,
the U.S. Department of Energy, the National Aeronautics and Space Administration,
the Japanese Monbukagakusho, the Max Planck Society,
and the Higher Education Funding Council for England.
This work is also partially based on observations obtained with XMM-Newton,
an ESA science mission with instruments and contributions directly
funded by ESA Member States and the USA (NASA).

\appendix
\onecolumn
\centering

\section{Summary of References for SED Parameter Correlations}
\begin{table*}
 \centering
  \begin{minipage}{175mm}
   \caption{A summary of references about the correlations among the most important AGN SED parameters. The upper right triangle region shows some most recent works about each correlations pair, while the lower left triangle region shows which cross-correlations are studied in this paper. `$\surd$' means this cross-correlation pair is studied in this paper, `$\times$' means not studied. The `FWHM' is the FWHM of narrow component subtracted H$\beta$ profile.}
   \label{former:study:summary:table}
   \begin{tabular}{@{}lccccccccccc@{}}
   \hline
   & $\Gamma_{2-10}$&$\kappa_{2-10}$&$\kappa_{5100}$& $\lambda_{Edd}$ &FWHM &M$_{BH}$&$\alpha_{ox}$&L$_{bol}$&L$_{2-10}$&L$_{2keV}$&L$_{2500}$\\
   \hline
   &&&&&Zhou10a&&&&&&\\
   $\Gamma_{2-10}$&---&Zhou10b&---&S08; S06&G10; S08&S06&Green09&---&---&Green09&---\\
   &&&&&S06; G99&&&&&&\\
   \hline
   $\kappa_{2-10}$&$\surd$&---&---&L10; V09&---&---&L10&M04; H07&V07; H07&---&---\\
   &&&&V07&&&&&M04&&\\
   \hline
   $\kappa_{5100}$&$\surd$&$\surd$&---&V07; R06&---&---&---&---&---&---&---\\
   \hline
   $\lambda_{Edd}$&$\surd$&$\surd$&$\surd$&---&Paper-I&F10&L10; V09&---&---&---&---\\
   &&&&&G10&&S08; V07&&&&\\
   \hline
   FWHM&$\surd$&$\surd$&$\surd$&$\surd$&---&(RM)$^{*}$&G10&Paper-I&Paper-I&---&---\\
   \hline
   M$_{BH}$&$\surd$&$\surd$&$\surd$&$\surd$&$\surd$&---&---&Woo02&---&---&---\\
   \hline
   $\alpha_{ox}$&$\surd$&$\surd$&$\surd$&$\surd$&$\surd$&$\surd$&---&---&---&L10&L10; G10\\
   &&&&&&&&&&Green09&S08; V07\\
   \hline
   L$_{bol}$&$\surd$&$\surd$&$\surd$&$\surd$&$\surd$&$\surd$&$\surd$&---&H07; M04&---&---\\
   \hline
   L$_{2-10}$&$\surd$&$\surd$&$\surd$&$\surd$&$\surd$&$\surd$&$\surd$&$\surd$&---&---&---\\
   \hline
   &&&&&&&&&&&G10; L10\\
   L$_{2keV}$&$\times$&$\times$&$\times$&$\times$&$\times$&$\times$&$\surd$&$\times$&$\times$&---&Green09\\
   &&&&&&&&&&&S08; V07\\
   \hline
   L$_{2500}$&$\times$&$\times$&$\times$&$\times$&$\times$&$\times$&$\surd$&$\times$&$\times$&$\surd$&---\\
   \hline
   \end{tabular}
   (RM)$^{*}$: the reverberation mapping studies, e.g. Kaspi et al. (2000), Peterson et al. (2004);
    Green09: Green et al. (2009); G99: Grupe et al. (1999); G10: Grupe et al. (2010);
    H07: Hopkins, Richards \& Hernquist (2007); F10: Fanidakis et al. (2010); Paper-I: Jin et al. 2011;
    L10: Lusso et al. (2010); M04: Marconi et al. (2004); S06: Shemmer et al. (2006); S08:
    Shemmer et al. (2008); V07: Vasudevan \& Fabian (2007); V09: Vasudevan \& Fabian (2009);
    Woo02: Woo \& Urry (2002); Zhou10a: Zhou \& Zhang (2010); Zhou10b: Zhou \& Zhao (2010).
 \end{minipage}
\end{table*}

\clearpage
\section{Parameter Values from Model-B Fitting ({\it optxagnf\/}: with Colour Correction) }
\begin{table*}
 \centering
  \begin{minipage}{165mm}
   \caption{Broadband SED fitting parameters using Model-B, and the fitting outputs (L$_{bol}$, {\it f}$_{d}$, {\it f}$_{c}$, {\it f}$_{p}$).
ID: object number, the same as Table~1 in Paper-I; N$_{H,gal}$ and N$_{H,int}$: the fixed galactic and free intrinsic neutral hydrogen column densities in 10$^{20}$ $cm^{-2}$; $\Gamma_{pow}$: the powerlaw component's slope in the SED fitting, (*) denotes the objects whose powerlaw slopes hit the uplimit of 2.2 and were fixed there; Fpl: the fraction of powerlaw component in the total reprocessed disc emission; R$_{cor}$: corona (truncation) radius in unit of Gravitational radii (r$_{g}$) within which all disc emission is reprocessed into the Comptonisation and powerlaw components; T$_{e}$: temperature of the Compton up-scattering electron population; Tau: optical depth of the Comptonisation component; log(M$_{BH}$): the best-fit black hole mass; log($\dot{M}$): total mass accretion rate; L$_{bol}$: bolometric luminosity integrated from 0.001 keV to 100 keV; {\it f}$_{d}$, {\it f}$_{c}$, {\it f}$_{p}$: luminosity fractions of disc emission, soft Comptonisation and hard X-ray Compotonisation components in the bolometric luminosity; $\chi^{2}$: the reduced $\chi^{2}$ of the broadband SED fitting.}
   \label{app:SED-fitting-parameters}
     \begin{tabular}{@{}ccccccccccccccc@{}}
\hline
ID & N$_{H,gal}$ & N$_{H,int}$ & $\Gamma_{pow}$ & Fpl & R$_{cor}$ & T$_{e}$ & Tau & {\it log}(M$_{BH}$) & {\it log}($\dot{M}$) & L$_{bol}$ &{\it f\/}$_{d}$ & {\it f\/}$_{c}$ & {\it f\/}$_{p}$ & $\chi^{2}$\\

 & $\times{\it 10}^{\it 20}$ & $\times{\it 10}^{\it 20}$ & & & $r_{g}$ & {\it keV} & & {\it M}$_{\sun}$ & $g$ $s^{-1}$ & ${\it 10}^{\it 44}$ & & & & {\it reduced} \\

\hline
1 & 1.79 & 0.00 & 1.74 & 0.69 & 100. & 0.246 & 17.4 & 8.61 & 26.08 & 62.2 & 0.19 & 0.25 & 0.56 & 1.00 \\
2 & 2.43 & 1.28 & 1.78 & 0.39 & 100. & 0.212 & 16.3 & 7.84 & 25.25 & 9.11 & 0.19 & 0.49 & 0.32 & 0.97 \\
3 & 6.31 & 9.44 & 1.85 & 0.25 & 10.2 & 0.214 & 12.2 & 7.61 & 25.92 & 42.8 & 0.87 & 0.10 & 0.03 & 1.14 \\
4 & 3.49 & 2.80 & 1.66 & 0.50 & 100. & 0.317 & 15.2 & 8.78 & 25.45 & 14.4 & 0.19 & 0.41 & 0.40 & 1.16 \\
5 & 3.53 & 5.08 & 2.20 & 0.36 & 69.0 & 0.202 & 14.4 & 7.94 & 26.37 & 119 & 0.32 & 0.44 & 0.24 & 1.04 \\
6 & 4.24 & 0.00 & 1.90 & 0.46 & 20.6 & 0.348 & 12.2 & 8.50 & 26.41 & 130 & 0.59 & 0.22 & 0.19 & 1.09 \\
7 & 1.33 & 0.00 & 2.20 & 0.29 & 10.5 & 0.149 & 33.4 & 7.58 & 26.01 & 52.9 & 0.26 & 0.53 & 0.21 & 1.16 \\
8 & 3.12 & 3.94 & 1.79 & 0.15 & 23.7 & 0.598 & 6.91 & 7.54 & 25.44 & 14.3 & 0.58 & 0.35 & 0.06 & 1.42 \\
9 & 1.30 & 0.59 & 1.72 & 0.71 & 16.5 & 0.267 & 15.0 & 7.40 & 25.70 & 25.9 & 0.84 & 0.05 & 0.11 & 20.7 \\
10 & 1.74 & 0.46 & 1.90 & 0.32 & 100. & 0.295 & 13.7 & 8.34 & 26.24 & 89.9 & 0.19 & 0.55 & 0.26 & 1.40 \\
11 & 1.72 & 0.57 & 1.49 & 0.49 & 22.5 & 0.775 & 8.40 & 7.92 & 26.02 & 54.2 & 0.65 & 0.18 & 0.17 & 0.96 \\
12 & 1.20 & 1.13 & 1.65 & 0.48 & 19.5 & 0.385 & 11.0 & 7.82 & 25.32 & 10.7 & 0.65 & 0.18 & 0.17 & 1.13 \\
13 & 3.56 & 0.00 & 1.38 & 0.87 & 11.4 & 0.142 & 17.9 & 9.20 & 26.50 & 161 & 0.90 & 0.01 & 0.09 & 1.33 \\
14 & 1.76 & 0.00 & 1.72 & 0.71 & 100. & 0.294 & 16.0 & 8.24 & 25.85 & 36.0 & 0.19 & 0.23 & 0.58 & 1.07 \\
15 & 1.31 & 3.71 & 2.20 & 0.09 & 100. & 0.764 & 4.33 & 6.30 & 25.28 & 9.72 & 0.80 & 0.18 & 0.02 & 3.67 \\
16 & 1.70 & 2.06 & 1.80 & 0.31 & 100. & 0.242 & 15.1 & 8.30 & 25.93 & 43.9 & 0.19 & 0.56 & 0.25 & 1.40 \\
17 & 0.65 & 2.51 & 1.68 & 0.14 & 88.9 & 0.474 & 8.27 & 7.79 & 26.12 & 68.4 & 0.35 & 0.56 & 0.09 & 1.06 \\
18 & 1.45 & 0.00 & 2.20 & 0.24 & 11.9 & 0.260 & 13.6 & 8.12 & 26.33 & 110 & 0.51 & 0.37 & 0.12 & 1.48 \\
19 & 3.70 & 1.63 & 1.98 & 0.19 & 31.4 & 0.144 & 20.6 & 7.71 & 24.92 & 4.28 & 0.37 & 0.52 & 0.12 & 1.03 \\
20 & 1.91 & 3.14 & 2.20 & 0.36 & 12.1 & 0.186 & 22.7 & 6.80 & 24.87 & 3.81 & 0.94 & 0.04 & 0.02 & 1.55 \\
21 & 1.77 & 0.00 & 1.78 & 0.75 & 23.1 & 0.205 & 19.5 & 7.98 & 26.09 & 63.0 & 0.61 & 0.10 & 0.29 & 3.45 \\
22 & 2.75 & 7.95 & 1.85 & 0.21 & 47.7 & 0.116 & 22.1 & 7.84 & 25.50 & 16.2 & 0.34 & 0.52 & 0.14 & 1.09 \\
23 & 1.59 & 0.00 & 1.39 & 0.45 & 95.8 & 0.628 & 9.70 & 8.00 & 25.06 & 5.92 & 0.22 & 0.43 & 0.35 & 0.99 \\
24 & 1.63 & 0.66 & 1.86 & 0.94 & 30.8 & 0.144 & 54.9 & 8.26 & 25.96 & 46.7 & 0.48 & 0.03 & 0.49 & 1.89 \\
25 & 2.34 & 0.51 & 1.79 & 0.40 & 16.8 & 0.354 & 12.4 & 8.43 & 26.41 & 131 & 0.56 & 0.27 & 0.17 & 1.88 \\
26 & 2.31 & 6.32 & 2.09 & 0.03 & 13.5 & 0.291 & 10.6 & 7.70 & 26.08 & 61.0 & 0.46 & 0.52 & 0.02 & 1.14 \\
27 & 2.75 & 0.00 & 2.03 & 0.22 & 36.8 & 0.194 & 17.0 & 7.86 & 25.28 & 9.75 & 0.43 & 0.45 & 0.12 & 1.13 \\
28 & 1.45 & 5.52 & 1.73 & 0.60 & 72.8 & 0.315 & 11.5 & 7.96 & 26.25 & 90.4 & 0.26 & 0.30 & 0.45 & 1.16 \\
29 & 1.18 & 4.24 & 2.11 & 0.12 & 17.8 & 0.363 & 7.33 & 7.87 & 26.13 & 68.8 & 0.49 & 0.45 & 0.06 & 1.19 \\
30 & 1.87 & 1.94 & 2.20 & 0.36 & 12.1 & 0.228 & 17.9 & 7.27 & 25.39 & 12.5 & 0.94 & 0.04 & 0.02 & 1.03 \\
31 & 0.84 & 0.00 & 1.66 & 0.54 & 100. & 0.400 & 13.0 & 8.70 & 25.89 & 39.5 & 0.19 & 0.37 & 0.43 & 0.99 \\
32 & 0.90 & 0.02 & 1.82 & 0.44 & 100. & 0.361 & 12.9 & 7.62 & 25.14 & 7.00 & 0.19 & 0.46 & 0.35 & 1.42 \\
33 & 1.07 & 0.00 & 2.17 & 0.57 & 12.9 & 0.244 & 16.1 & 7.92 & 25.93 & 43.4 & 0.78 & 0.10 & 0.13 & 1.20 \\
34 & 1.83 & 0.85 & 1.90 & 0.33 & 100. & 0.252 & 14.8 & 8.71 & 26.06 & 59.4 & 0.19 & 0.54 & 0.26 & 1.11 \\
35 & 1.76 & 2.30 & 1.83 & 0.83 & 70.8 & 0.178 & 17.7 & 7.67 & 26.27 & 95.9 & 0.19 & 0.14 & 0.67 & 1.02 \\
36 & 1.18 & 3.77 & 2.20 & 0.22 & 30.1 & 0.624 & 5.58 & 7.00 & 24.90 & 4.05 & 0.75 & 0.20 & 0.05 & 1.62 \\
37 & 1.82 & 0.00 & 2.04 & 0.38 & 100. & 0.219 & 17.2 & 8.23 & 25.91 & 42.1 & 0.19 & 0.50 & 0.31 & 1.33 \\
38 & 1.42 & 0.26 & 1.61 & 0.97 & 100. & 0.229 & 31.1 & 7.79 & 24.55 & 1.82 & 0.19 & 0.02 & 0.79 & 1.22 \\
39 & 1.36 & 3.45 & 2.08 & 0.11 & 38.1 & 0.259 & 13.2 & 7.34 & 24.94 & 4.50 & 0.40 & 0.53 & 0.07 & 0.98 \\
40 & 0.77 & 2.01 & 1.92 & 0.06 & 22.4 & 1.150 & 4.75 & 7.88 & 25.81 & 32.8 & 0.59 & 0.39 & 0.02 & 1.51 \\
41 & 1.81 & 0.46 & 1.88 & 0.39 & 14.0 & 0.354 & 11.9 & 8.14 & 26.33 & 109 & 0.52 & 0.30 & 0.19 & 1.31 \\
42 & 2.86 & 3.26 & 1.84 & 0.41 & 100. & 0.083 & 31.2 & 7.74 & 24.71 & 2.64 & 0.19 & 0.47 & 0.33 & 1.01 \\
43 & 2.69 & 0.93 & 1.71 & 0.58 & 100. & 0.469 & 10.7 & 8.07 & 26.17 & 75.9 & 0.31 & 0.29 & 0.40 & 1.26 \\
44 & 2.78 & 7.77 & 2.26 & 0.04 & 9.96 & 0.218 & 13.5 & 7.56 & 26.15 & 73.2 & 0.53 & 0.45 & 0.02 & 1.56 \\
45 & 1.46 & 2.23 & 1.93 & 0.49 & 44.5 & 0.198 & 17.2 & 8.78 & 26.86 & 369 & 0.40 & 0.30 & 0.30 & 2.45 \\
46 & 4.02 & 0.54 & 1.81 & 0.81 & 100. & 0.207 & 20.2 & 8.56 & 25.61 & 20.8 & 0.19 & 0.15 & 0.66 & 1.12 \\
47 & 3.78 & 16.5 & 1.85 & 0.25 & 85.8 & 0.115 & 29.0 & 7.96 & 25.68 & 24.6 & 0.19 & 0.61 & 0.20 & 0.97 \\
48 & 2.11 & 0.72 & 1.84 & 0.19 & 31.2 & 0.475 & 9.28 & 7.47 & 24.99 & 5.01 & 0.70 & 0.24 & 0.06 & 1.18 \\
49 & 4.90 & 0.35 & 2.20 & 0.33 & 71.0 & 0.211 & 19.6 & 7.73 & 25.18 & 7.85 & 0.25 & 0.50 & 0.25 & 1.15 \\
50 & 4.51 & 0.00 & 2.20 & 0.80 & 9.10 & 0.590 & 7.57 & 8.38 & 27.17 & 750 & 0.98 & 0.00 & 0.01 & 2.21 \\
51 & 2.91 & 1.45 & 1.77 & 0.95 & 100. & 0.136 & 31.8 & 7.60 & 25.37 & 11.9 & 0.19 & 0.04 & 0.77 & 1.41 \\
\hline
   \end{tabular}
  \\
 \end{minipage}
\end{table*}

\begin{table*}
 \centering
  \begin{minipage}{175mm}
   \caption{Key Parameters from Model-B fitting.
ID: object number, the same as Table~1 in Paper-I; $\Gamma_{2-10keV}$: the slope of the single powerlaw fitted to 2-10 keV spectrum. L$_{2-10keV}$: 2-10 keV luminosity (in 10$^{44}$ erg s$^{-1}$); $\kappa_{2-10keV}$: the 2-10keV bolometric correction coefficient; $\lambda$L$_{2500\AA}$: the monochromatic luminosity at 2500{\AA} (in 10$^{43}$ erg s$^{-1}$); $\nu$L$_{2keV}$: the monochromatic luminosity at 2keV (in 10$^{43}$ erg s$^{-1}$); $\alpha_{ox}$: the optical X-ray spectral index; $\lambda$L$_{5100}$: the monochromatic luminosity at 5100{\AA} (in 10$^{44}$ erg s$^{-1}$); $\kappa_{5100}$: the 5100{\AA} bolometric correction coefficient; FWHM$_{H\beta}$: the narrow component subtracted H$\beta$ FWHM; L$_{bol}$/L$_{Edd}$: the Eddington Ratio; $\alpha_{X}$: the soft X-ray slope between 0.2-2 keV (corrected for Galactic and intrinsic absorption), assuming $F_{\nu}~\propto~\nu^{-\alpha}$; $\alpha_{UV}$: the optical/UV slope between 1700-6500{\AA} (corrected for Galactic and intrinsic reddening and absorption).}
   \label{app:SED-key-parameters}
     \begin{tabular}{@{}ccccccccccccc@{}}
\hline
ID & $\Gamma_{2-10keV}$ & L$_{2-10keV}$ & $\kappa_{2-10keV}$ & $\lambda$L$_{2500\AA}$ & $\nu$L$_{2keV}$ & $\alpha_{ox}$ & $\lambda$L$_{5100}$ & $\kappa_{5100}$ & FWHM$_{H\beta}$ & L$_{bol}$/L$_{Edd}$ & $\alpha_{X}$ & $\alpha_{UV}$\\

 & & $\times{\it 10}^{\it 44}$ & & $\times{\it 10}^{\it 43}$ & $\times{\it 10}^{\it 43}$ & & $\times{\it 10}^{\it 44}$ & & {\it km s}$^{\it -1}$ & & & \\

\hline
1 & 1.69$\pm$0.06 & 4.941 & 12.6 & 84.3 & 25.8 & 1.20 & 8.15 & 7.65 & 13000 & 0.12 & 1.09 & 1.34 \\
2 & 1.67$\pm$0.10 & 0.469 & 19.5 & 18.9 & 2.48 & 1.34 & 0.791 & 11.5 & 6220 & 0.10 & 1.34 & 0.14 \\
3 & 1.77$\pm$0.07 & 0.289 & 148 & 38.9 & 1.77 & 1.51 & 1.35 & 31.7 & 2310 & 0.80 & 2.37 & 0.14 \\
4 & 1.80$\pm$0.11 & 0.567 & 25.5 & 13.3 & 3.15 & 1.24 & 1.91 & 7.55 & 10800 & 0.019 & 1.23 & 1.68 \\
5 & 2.10$\pm$0.22 & 2.284 & 52.2 & 132 & 13.1 & 1.38 & 5.48 & 21.8 & 2720 & 1.1 & 1.88 & 0.066 \\
6 & 1.93$\pm$0.18 & 4.855 & 26.9 & 281 & 27.4 & 1.39 & 14.8 & 8.87 & 5430 & 0.32 & 1.34 & 0.35 \\
7 & 2.39$\pm$0.22 & 0.267 & 199 & 43.0 & 2.38 & 1.48 & 1.95 & 27.1 & 1980 & 1.1 & 2.49 & 0.15 \\
8 & 1.84$\pm$0.04 & 0.418 & 34.3 & 16.4 & 2.82 & 1.29 & 0.539 & 26.6 & 2840 & 0.31 & 1.59 & 0.12 \\
9 & 1.76$\pm$0.01 & 0.839 & 30.9 & 20.2 & 5.29 & 1.22 & 0.113 & 230 & 3030 & 0.80 & 1.46 & 0.10 \\
10 & 1.92$\pm$0.05 & 3.532 & 25.5 & 199 & 23.0 & 1.36 & 7.59 & 11.9 & 4810 & 0.32 & 1.40 & 0.21 \\
11 & 1.71$\pm$0.11 & 1.811 & 30.0 & 70.5 & 9.06 & 1.34 & 3.75 & 14.5 & 5640 & 0.50 & 1.12 & 0.28 \\
12 & 1.68$\pm$0.23 & 0.502 & 21.3 & 19.4 & 1.58 & 1.42 & 1.04 & 10.3 & 4390 & 0.13 & 1.38 & 0.24 \\
13 & 1.37$\pm$0.12 & 0.751 & 215 & 697 & 3.03 & 1.91 & 42.6 & 3.79 & 10800 & 0.078 & 0.73 & 0.65 \\
14 & 1.69$\pm$0.04 & 3.189 & 11.3 & 50.1 & 17.0 & 1.18 & 3.91 & 9.22 & 7060 & 0.16 & 1.06 & 0.85 \\
15 & 2.35$\pm$0.12 & 0.042 & 234 & 2.09 & 0.397 & 1.28 & 0.204 & 47.7 & 988 & 3.7 & 3.24 & 0.085 \\
16 & 1.78$\pm$0.07 & 1.502 & 29.3 & 107 & 8.27 & 1.43 & 4.26 & 10.3 & 3560 & 0.17 & 1.48 & 0.30 \\
17 & 1.80$\pm$0.20 & 0.779 & 88.0 & 82.7 & 3.70 & 1.52 & 3.31 & 20.7 & 2250 & 0.86 & 1.87 & -0.17 \\
18 & 2.23$\pm$0.08 & 1.254 & 88.0 & 153 & 9.68 & 1.46 & 6.11 & 18.1 & 2310 & 0.64 & 1.74 & 0.18 \\
19 & 1.98$\pm$0.18 & 0.084 & 51.0 & 8.66 & 0.496 & 1.48 & 0.443 & 9.69 & 2000 & 0.064 & 1.98 & 0.21 \\
20 & 2.34$\pm$0.12 & 0.053 & 71.9 & 2.29 & 0.468 & 1.26 & 0.215 & 17.7 & 774 & 0.47 & 2.27 & 0.11 \\
21 & 1.70$\pm$0.04 & 3.856 & 16.4 & 84.7 & 20.5 & 1.24 & 2.22 & 28.5 & 6090 & 0.51 & 1.19 & 0.19 \\
22 & 1.70$\pm$0.09 & 0.396 & 41.0 & 27.3 & 2.15 & 1.42 & 0.983 & 16.5 & 7050 & 0.18 & 1.84 & 0.15 \\
23 & 1.80$\pm$0.19 & 0.145 & 41.1 & 11.0 & 0.912 & 1.42 & 0.708 & 8.39 & 1980 & 0.046 & 1.37 & 0.65 \\
24 & 1.83$\pm$0.18 & 4.735 & 9.88 & 94.9 & 24.4 & 1.23 & 6.64 & 7.05 & 13900 & 0.20 & 1.01 & 0.39 \\
25 & 1.88$\pm$0.03 & 3.054 & 43.1 & 261 & 19.6 & 1.43 & 8.44 & 15.6 & 4980 & 0.37 & 1.44 & 0.26 \\
26 & 2.09$\pm$0.25 & 0.362 & 169 & 56.0 & 2.57 & 1.51 & 2.04 & 30.0 & 1720 & 0.94 & 2.47 & 0.22 \\
27 & 1.94$\pm$0.04 & 0.277 & 35.3 & 19.1 & 2.52 & 1.34 & 0.988 & 9.89 & 4310 & 0.10 & 1.50 & 0.19 \\
28 & 1.71$\pm$0.14 & 2.951 & 30.7 & 115 & 13.4 & 1.36 & 4.80 & 18.9 & 4240 & 0.77 & 1.54 & 0.035 \\
29 & 2.00$\pm$0.12 & 0.726 & 95.0 & 78.1 & 4.93 & 1.46 & 3.25 & 21.3 & 3560 & 0.71 & 2.03 & 0.19 \\
30 & 2.46$\pm$0.09 & 0.146 & 85.5 & 10.2 & 1.26 & 1.35 & 0.452 & 27.7 & 954 & 0.52 & 2.10 & 0.11 \\
31 & 1.69$\pm$0.14 & 2.420 & 16.4 & 53.2 & 11.9 & 1.25 & 6.49 & 6.10 & 6810 & 0.061 & 1.12 & 1.49 \\
32 & 1.88$\pm$0.03 & 0.464 & 15.1 & 11.4 & 2.96 & 1.23 & 0.512 & 13.7 & 3100 & 0.13 & 1.22 & 0.14 \\
33 & 2.14$\pm$0.21 & 1.157 & 37.6 & 60.5 & 7.50 & 1.35 & 4.03 & 10.8 & 5690 & 0.40 & 1.42 & 0.26 \\
34 & 1.90$\pm$0.14 & 2.489 & 23.9 & 141 & 13.5 & 1.39 & 10.8 & 5.53 & 3310 & 0.089 & 1.28 & 1.05 \\
35 & 1.76$\pm$0.07 & 3.918 & 24.5 & 72.9 & 52.7 & 1.05 & 3.59 & 26.8 & 2790 & 1.6 & 1.21 & 0.19 \\
36 & 2.20$\pm$0.08 & 0.091 & 44.7 & 3.18 & 0.651 & 1.26 & 0.244 & 16.6 & 1890 & 0.31 & 2.06 & 0.11 \\
37 & 1.95$\pm$0.08 & 1.768 & 23.9 & 88.7 & 12.3 & 1.33 & 5.39 & 7.82 & 3960 & 0.19 & 1.49 & 0.31 \\
38 & 1.55$\pm$0.09 & 0.175 & 10.4 & 1.57 & 0.770 & 1.12 & 0.197 & 9.26 & 6630 & 0.023 & 0.66 & 1.58 \\
39 & 2.17$\pm$0.20 & 0.079 & 57.4 & 5.55 & 0.726 & 1.34 & 0.233 & 19.4 & 991 & 0.16 & 1.84 & 0.15 \\
40 & 2.02$\pm$0.06 & 0.468 & 70.3 & 46.6 & 3.46 & 1.43 & 2.05 & 16.0 & 2790 & 0.34 & 1.80 & 0.17 \\
41 & 1.94$\pm$0.05 & 2.444 & 44.7 & 156 & 15.9 & 1.38 & 6.26 & 17.4 & 2610 & 0.60 & 1.46 & 0.21 \\
42 & 1.76$\pm$0.11 & 0.158 & 16.7 & 4.96 & 0.803 & 1.30 & 0.265 & 9.99 & 4920 & 0.037 & 1.49 & 0.47 \\
43 & 1.74$\pm$0.07 & 4.524 & 16.8 & 121 & 26.0 & 1.26 & 4.36 & 17.4 & 4550 & 0.50 & 1.21 & 0.10 \\
44 & 2.25$\pm$0.05 & 0.236 & 311 & 52.5 & 2.08 & 1.54 & 2.36 & 31.1 & 1070 & 1.5 & 2.94 & 0.11 \\
45 & 1.82$\pm$0.06 & 17.502 & 21.1 & 840 & 100. & 1.35 & 30.4 & 12.2 & 10900 & 0.47 & 1.37 & 0.40 \\
46 & 1.81$\pm$0.12 & 2.175 & 9.60 & 19.2 & 10.4 & 1.10 & 2.97 & 7.04 & 9930 & 0.044 & 1.01 & 1.77 \\
47 & 1.45$\pm$0.25 & 0.868 & 28.4 & 40.8 & 4.39 & 1.37 & 0.931 & 26.5 & 4100 & 0.21 & 2.18 & 0.21 \\
48 & 2.03$\pm$0.11 & 0.101 & 49.5 & 7.14 & 0.728 & 1.38 & 0.278 & 18.1 & 1190 & 0.13 & 1.76 & 0.13 \\
49 & 2.40$\pm$0.22 & 0.200 & 39.4 & 14.4 & 1.69 & 1.36 & 0.719 & 10.9 & 1340 & 0.11 & 1.79 & 0.13 \\
50 & 2.41$\pm$0.18 & 3.299 & 228 & 834 & 27.8 & 1.57 & 29.5 & 25.5 & 2200 & 2.4 & 1.82 & 0.24 \\
51 & 1.67$\pm$0.03 & 1.659 & 7.20 & 13.2 & 8.22 & 1.08 & 0.624 & 19.1 & 11100 & 0.23 & 0.99 & 0.37 \\
\hline
   \end{tabular}
  \\
 \end{minipage}
\end{table*}

\clearpage
\section{Parameter Cross-correlations Using Values from Model-A Fitting ({\it optxagn\/}: without Color Correction)}
Similar cross-correlation plots as reported in previous sections but use parameter values from Model-A fitting in Paper-I. In each panel, the various point symbols show Model-A data. The solid orange line is the regression result for these Model-A data using the same regression methods as in previous sections. The dashed gray line is our result from previous sections based on Model-B data. All dotted lines are from the other literatures as indicated in previous sections.
\label{app:section:plots}
\begin{table*}
\centering
  \begin{tabular}{cc}
  \includegraphics[bb=54 144 594 650,scale=0.37,clip=1]{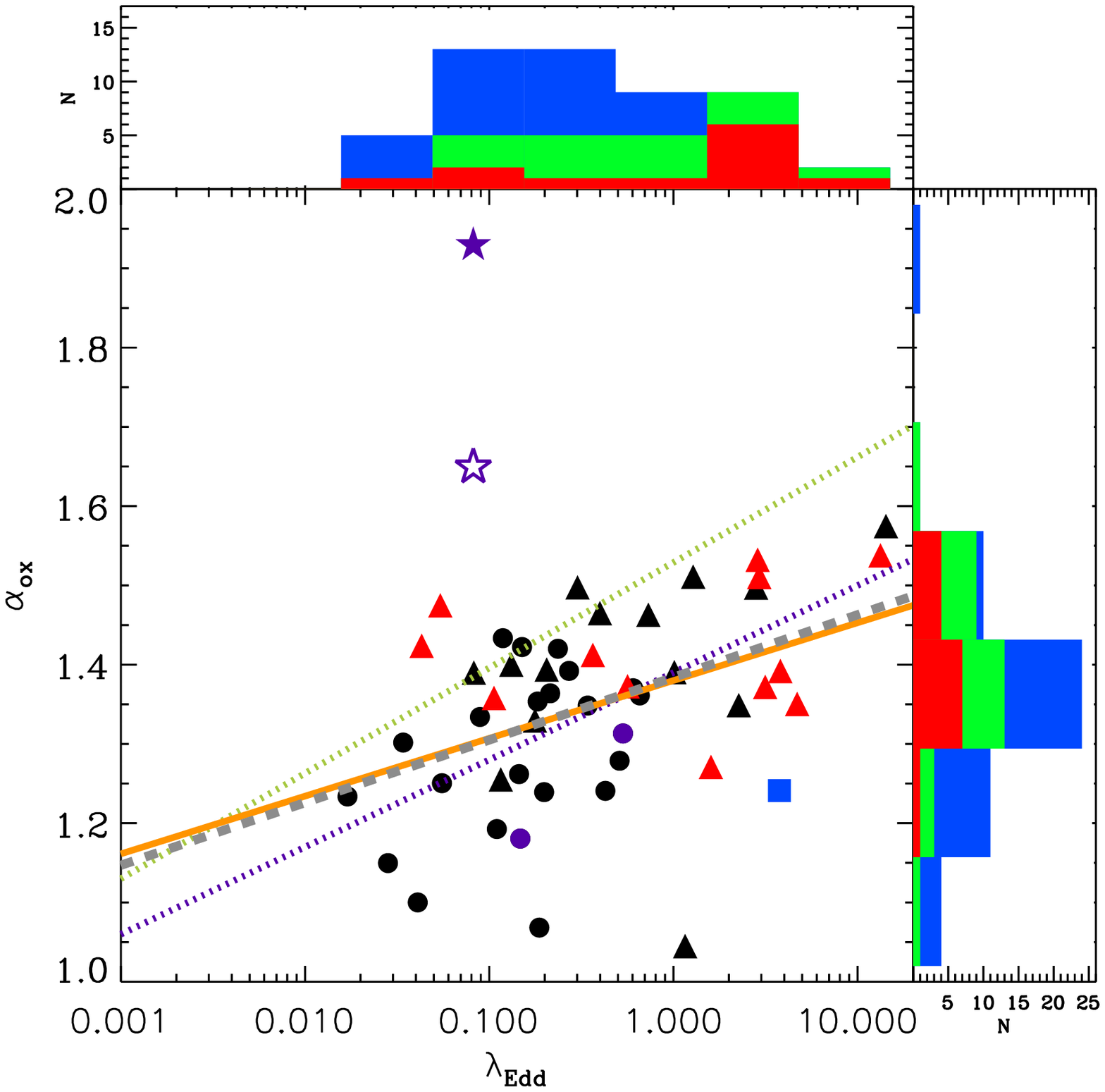}&\includegraphics[bb=54 144 594 650,scale=0.37,clip=1]{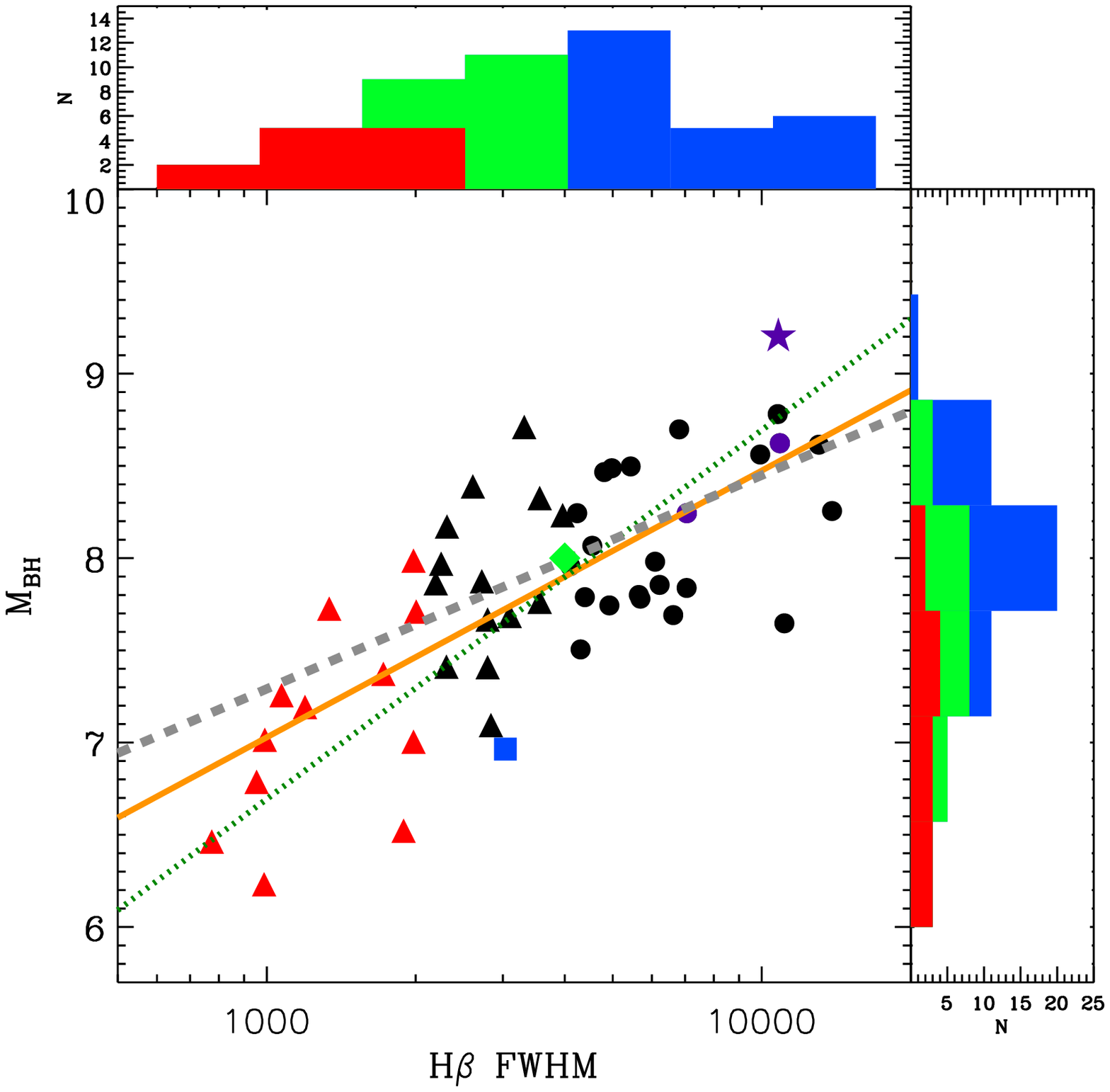}\\
  \ref{app:section:plots}1&\ref{app:section:plots}2\\
  \includegraphics[bb=54 144 594 650,scale=0.37,clip=1]{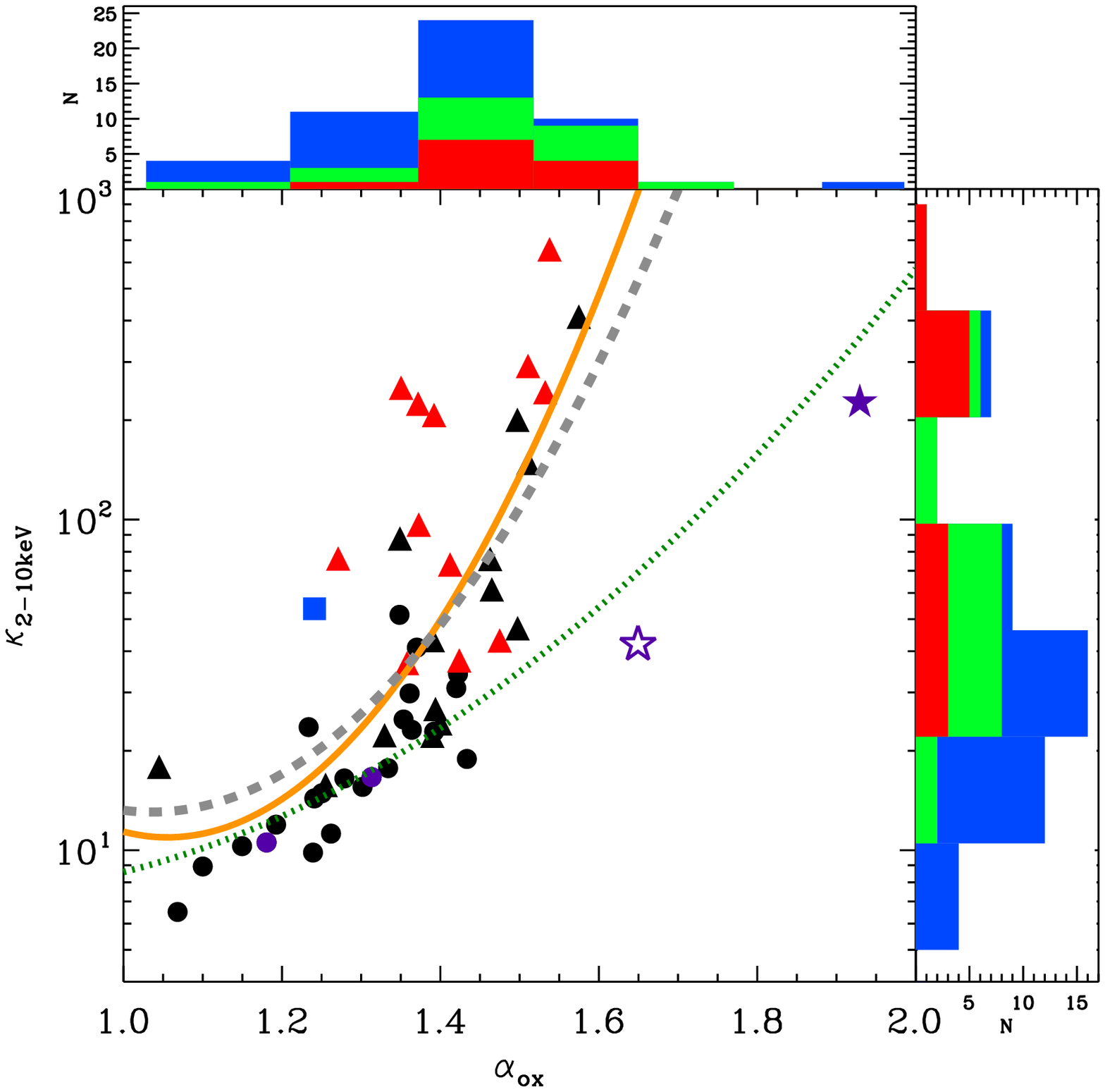}&\includegraphics[bb=54 144 594 650,scale=0.37,clip=1]{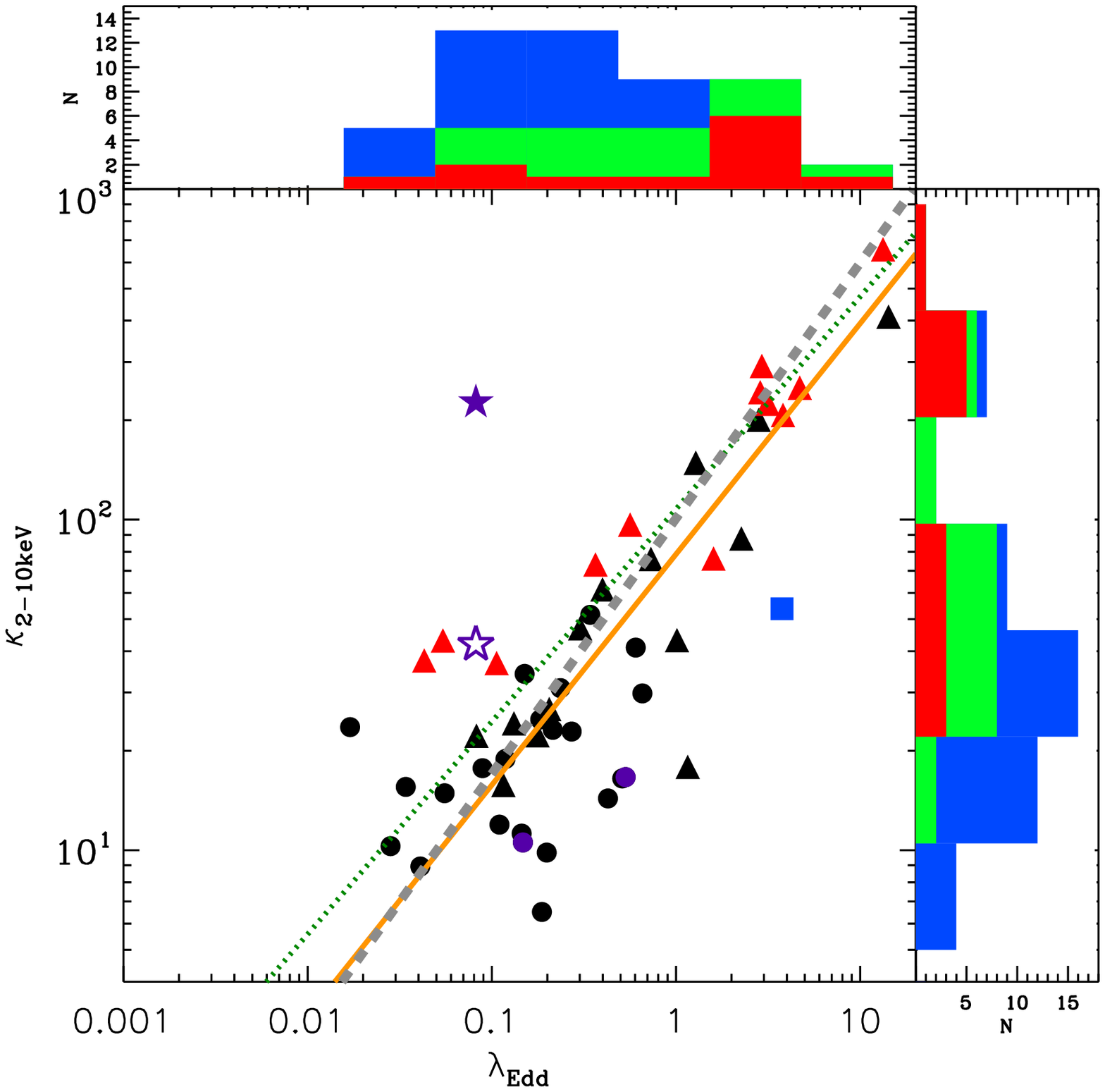}\\
  \ref{app:section:plots}3&\ref{app:section:plots}4\\
  \includegraphics[bb=54 144 594 650,scale=0.37,clip=1]{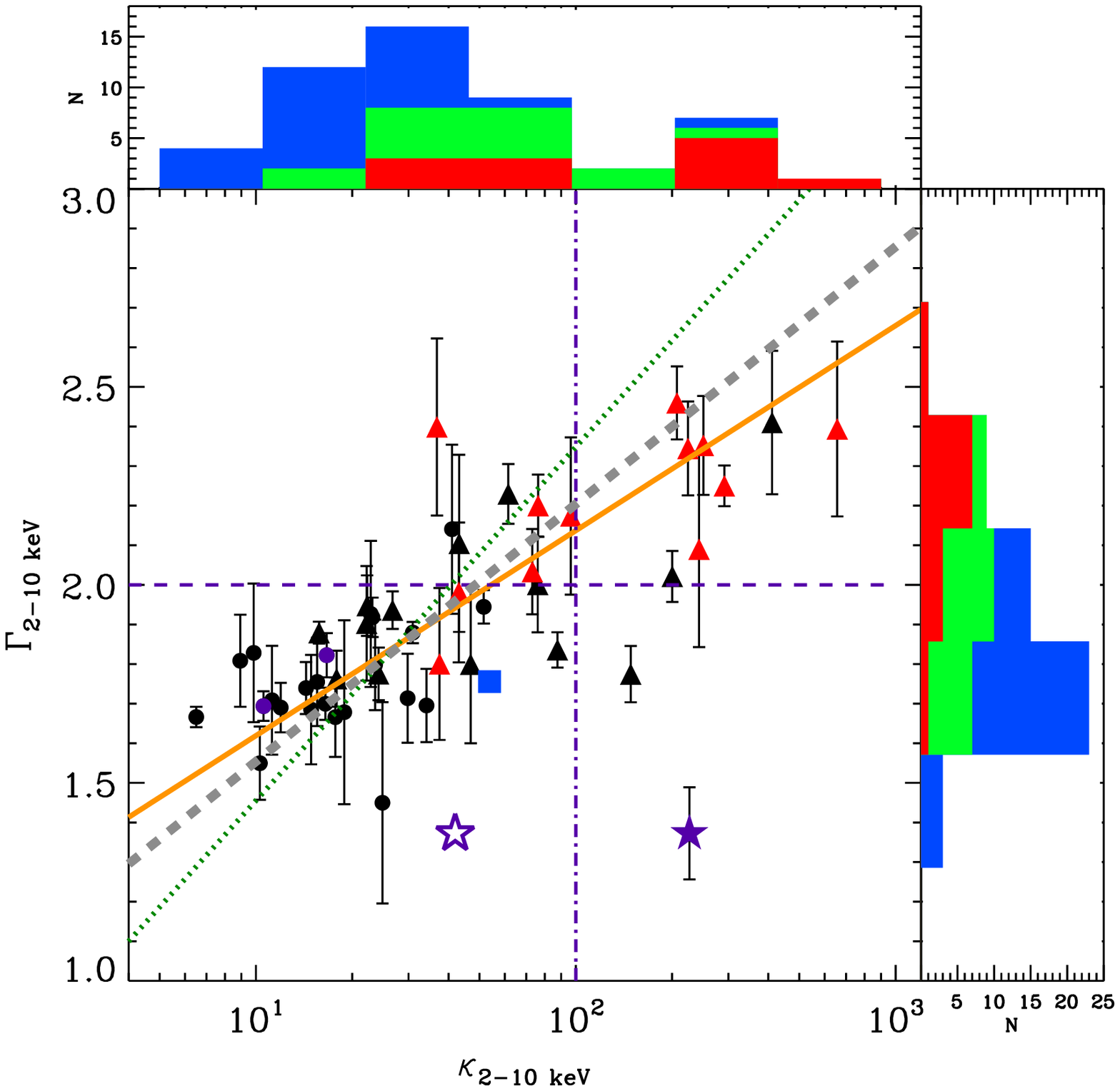}&\includegraphics[bb=54 144 594 650,scale=0.37,clip=1]{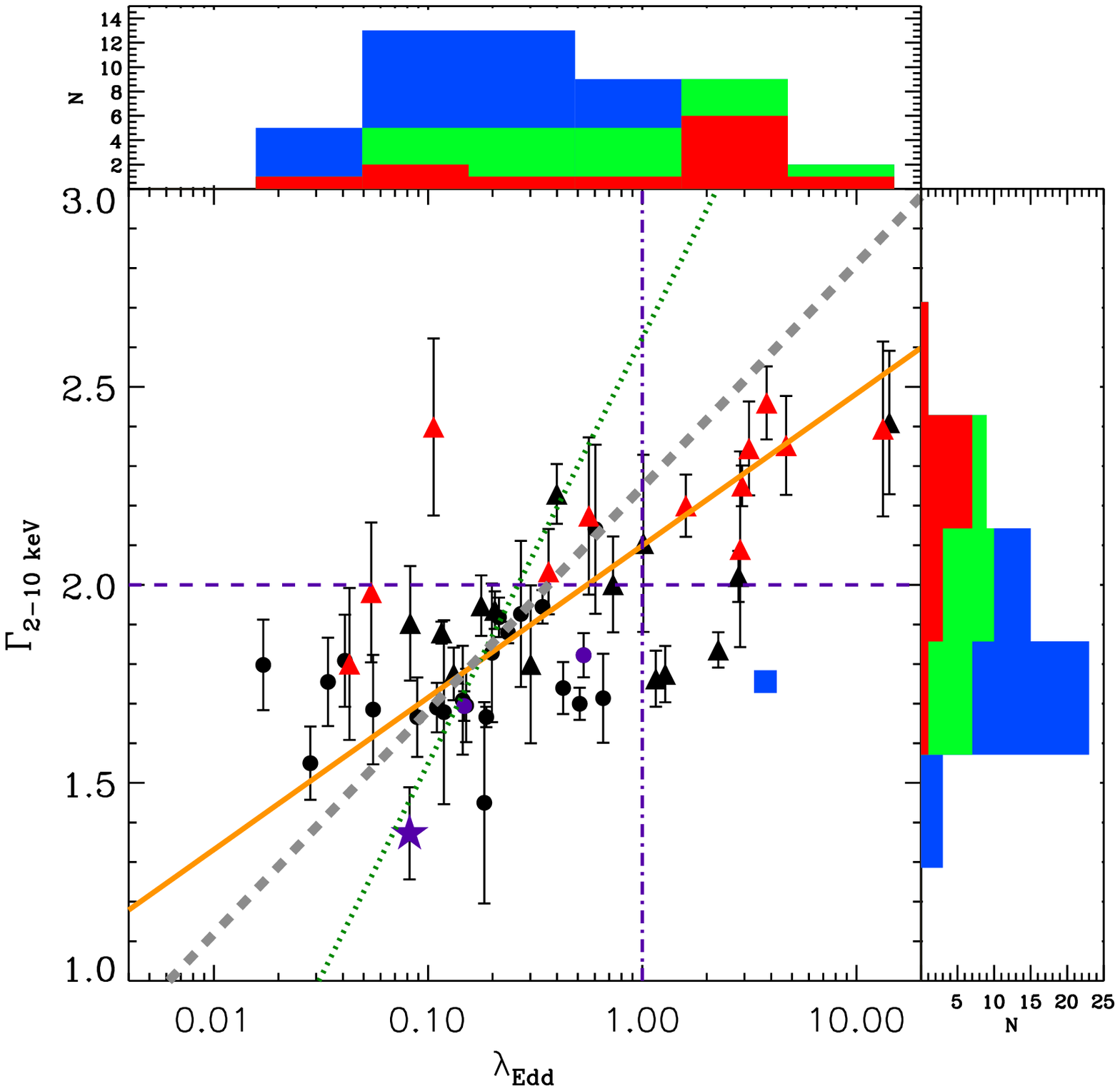}\\
 \ref{app:section:plots}5&\ref{app:section:plots}6\\
  \end{tabular}
\end{table*}

\begin{table*}
\centering
  \begin{tabular}{cc}
  \includegraphics[bb=54 144 594 650,scale=0.37,clip=1]{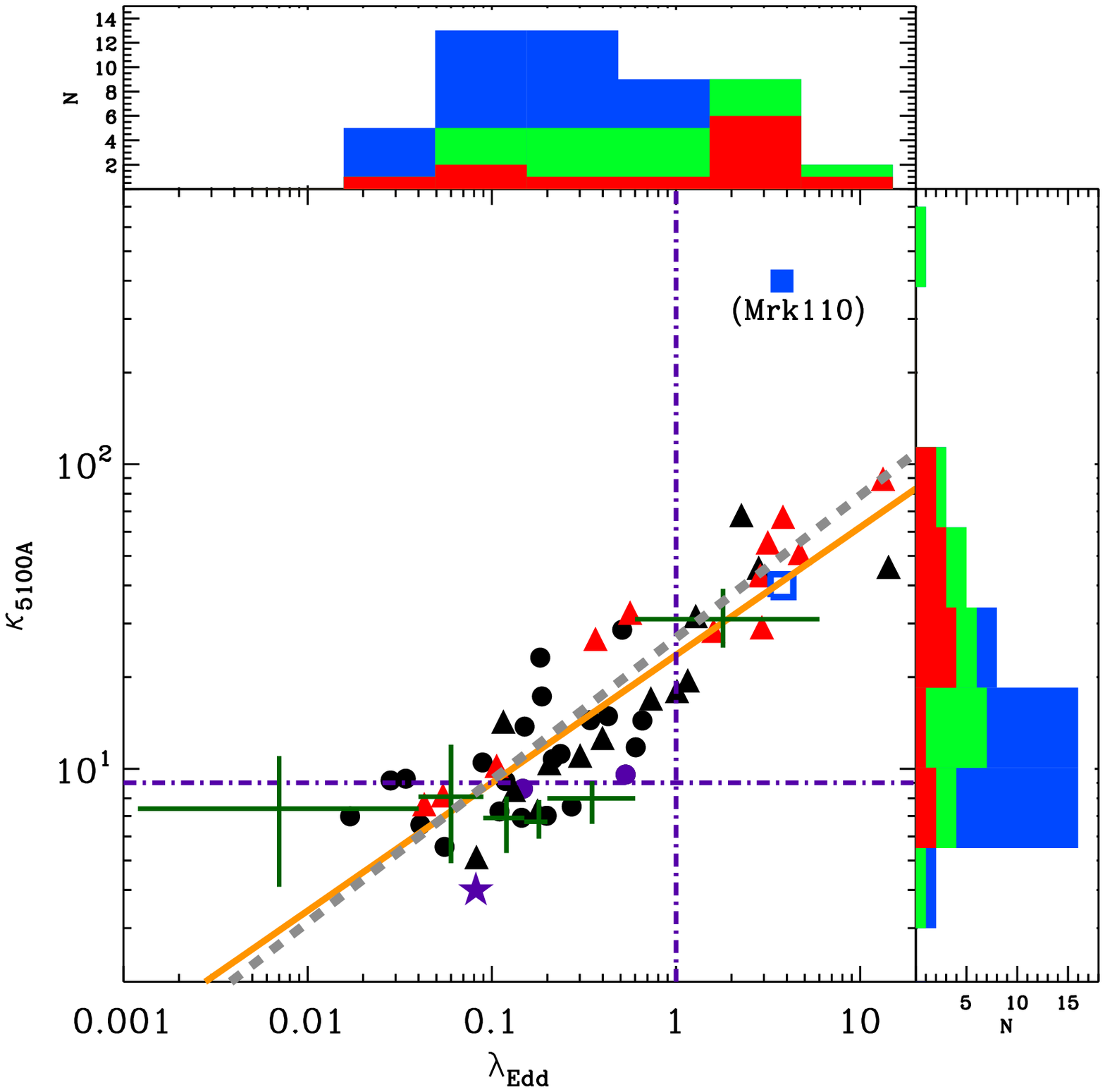}&\includegraphics[bb=54 144 594 650,scale=0.37,clip=1]{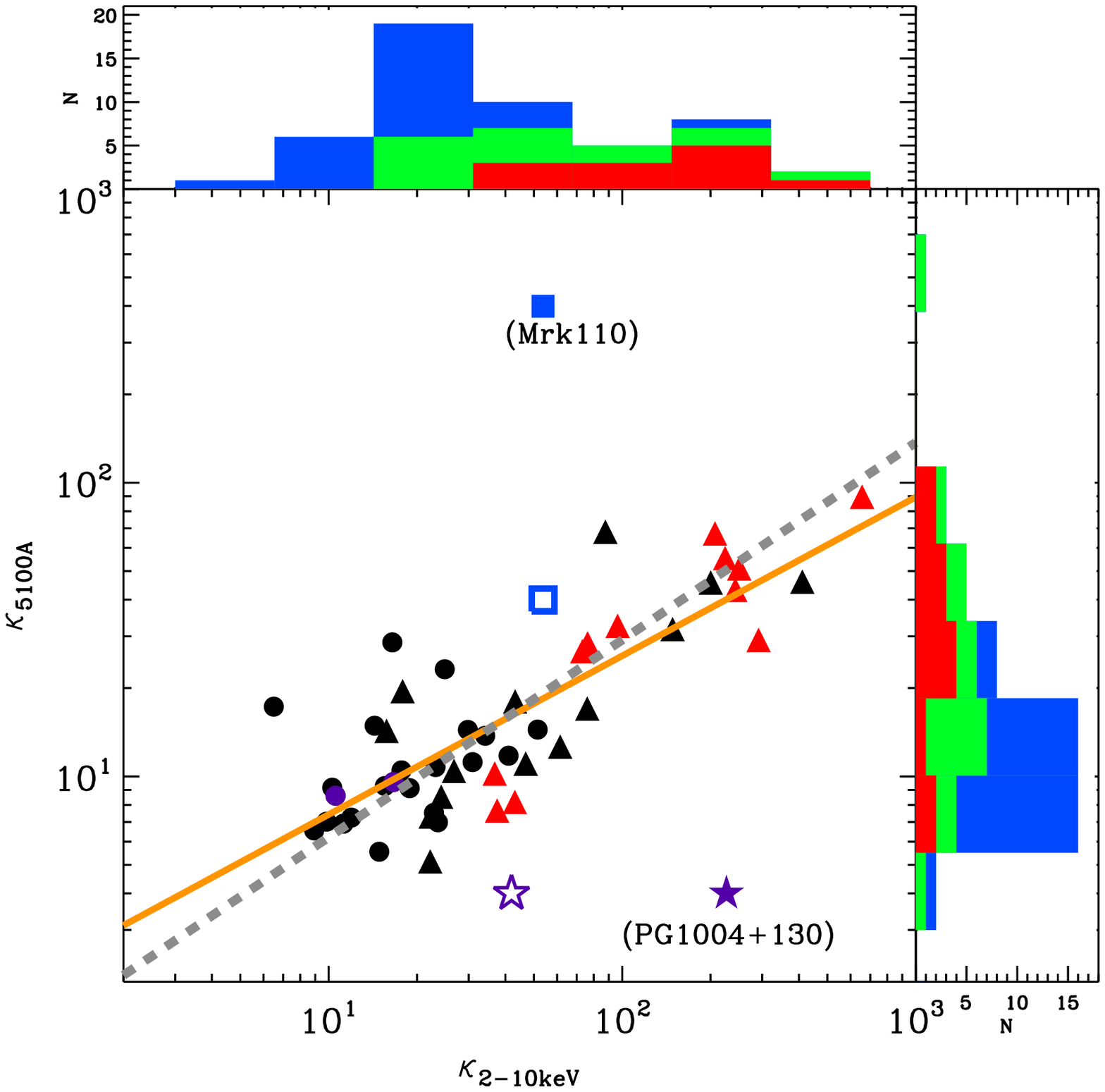}\\
 \ref{app:section:plots}7&\ref{app:section:plots}8\\
  \end{tabular}
\end{table*}

\begin{table*}
\centering
\begin{tabular}{c}
\includegraphics[scale=0.47,clip=]{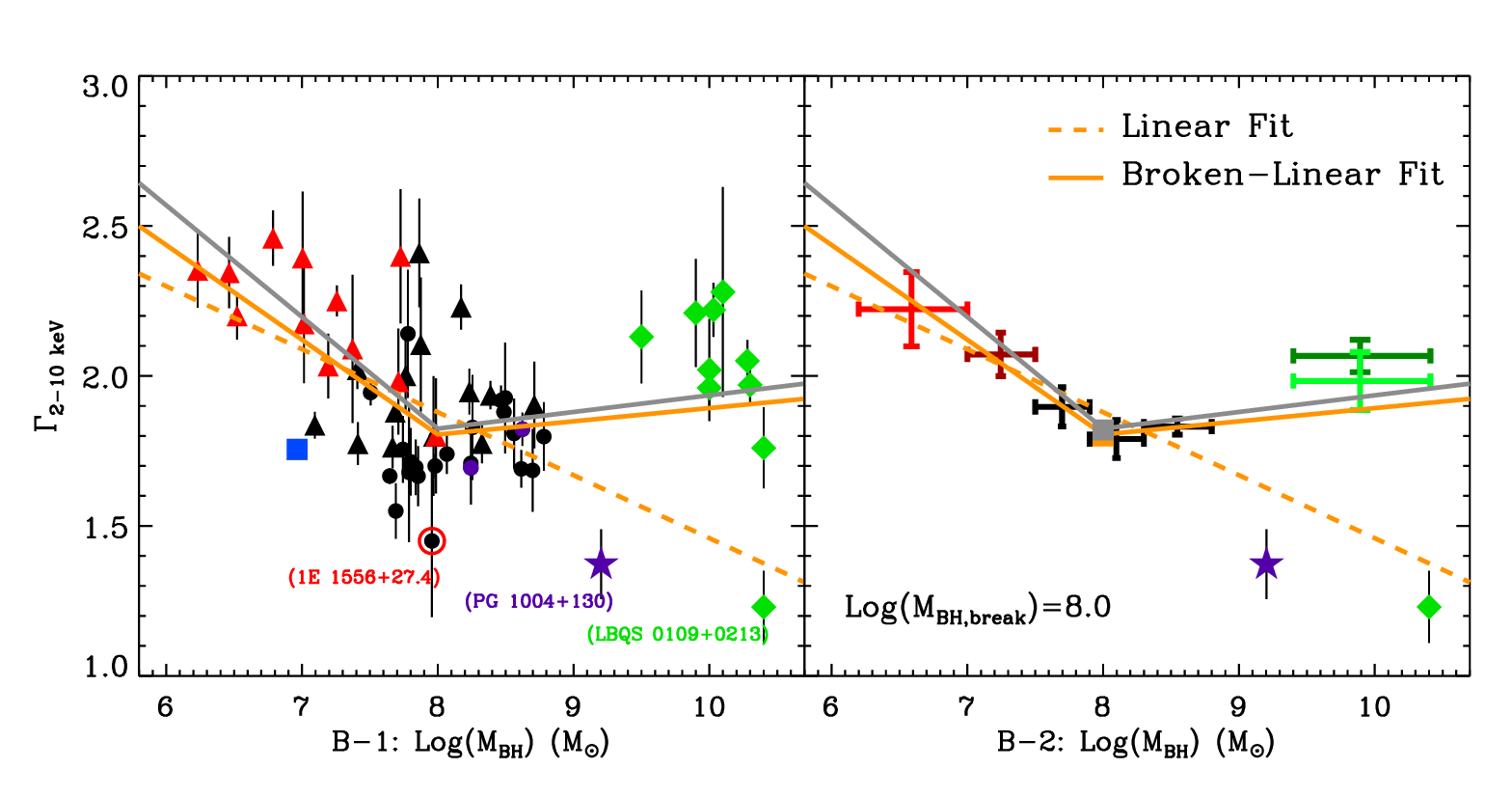}\\
 \ref{app:section:plots}9\\
\end{tabular}
\end{table*}

\begin{table*}
\centering
\begin{tabular}{c}
\includegraphics[scale=0.47,clip=1,angle=90]{app_keypar_cor5.ps}\\
 \ref{app:section:plots}10\\
\end{tabular}
\end{table*}

\clearpage
\section{Parameter Correlations Matrix Using Values from Model-A Fitting ({\it optxagn\/}: without Color Correction)}
\label{app:section:matrix}
\begin{table*}
 \centering
  \begin{minipage}{175mm}
   \caption{The cross-correlation matrix of the 9 key parameters based on Model-A data from Jin et al. (2011). ID 1$\sim$9 are given to each parameter. $\rho_{s}$ is the Spearman's rank coefficient. $d_{s}^{1}$ is probability of random distribution in logarithm. $\alpha$ and $\beta$ are the bisector regression line coefficients assuming Y=$\beta$X+$\alpha$. The coefficients in the upper right triangle region assumes the vertical parameters to be X. The coefficients in the lower left triangle region assumes the horizontal parameters to be X.}
   \label{app:sum-cor-keypar:table}
     \begin{tabular}{@{}lcccccccccc@{}}
\hline

& &$\Gamma_{2-10keV}$&$\kappa_{2-10keV}$&$\kappa_{5100A}$& $\lambda_{Edd}$ &H$\beta$ FWHM &M$_{BH}$&$\alpha_{ox}$&L$_{bol}$&L$_{2-10keV}$\\
&&&&&&{\it km s$^{-1}$\/}&{\it M$_{\sun}$\/}&&{\it 10\/}$^{+44}$&{\it 10\/}$^{+44}$\\
&&&{\it log\/}&{\it log\/}&{\it log\/}&{\it log\/}&{\it log\/}&&{\it log\/}&{\it log\/}\\
\hline
ID&&(1)&(2)&(3)&(4)&(5)&(6)&(7)&(8)&(9)\\

\hline

(1) & $\rho_{s}$ & 1 & 0.76 & 0.46 & 0.60 & -0.72 & -0.43 & 0.50 & 0.22 & -0.38 \\
 & $d_{s}^{1}$ & -$\infty$ & -10. & -3. & -5. & -8. & -3. & -4. & -1. & -2. \\
 & $\beta$ & 1 & 1.93$\pm$0.17 & 1.26$\pm$0.11 & 2.61$\pm$0.27 & -1.27$\pm$0.12 & -2.19$\pm$0.30 & 0.56$\pm$0.06 & 1.42$\pm$0.37 & -2.02$\pm$0.33 \\
 & $\alpha$ & 0 & -2.13$\pm$0.32 & -1.24$\pm$0.21 & -5.47$\pm$0.52 & 5.99$\pm$0.24 & 12.02$\pm$0.58 & 0.28$\pm$0.12 & -1.26$\pm$0.70 & 3.77$\pm$0.65 \\
\hline
(2) & $\rho_{s}$ & 0.76 & 1 & 0.72 & 0.72 & -0.80 & -0.62 & 0.76 & 0.17 & -0.57 \\
 & $d_{s}^{1}$ & -10. & -$\infty$ & -8. & -8. & -11. & -6. & -10. & -1. & -5. \\
 & $\beta$ & 0.52$\pm$0.05 & 1 & 0.67$\pm$0.05 & 1.40$\pm$0.09 & -0.66$\pm$0.06 & -1.23$\pm$0.14 & 0.26$\pm$0.03 & 1.06$\pm$0.07 & -1.24$\pm$0.13 \\
 & $\alpha$ & 1.10$\pm$0.07 & 0 & 0.12$\pm$0.08 & -2.69$\pm$0.18 & 4.60$\pm$0.09 & 9.76$\pm$0.21 & 0.94$\pm$0.04 & -0.21$\pm$0.15 & 1.86$\pm$0.19 \\
\hline
(3) & $\rho_{s}$ & 0.46 & 0.72 & 1 & 0.85 & -0.59 & -0.80 & 0.37 & 0.17 & -0.41 \\
 & $d_{s}^{1}$ & -3. & -8. & -$\infty$ & -14. & -5. & -11. & -2. & -1. & -3. \\
 & $\beta$ & 0.79$\pm$0.07 & 1.49$\pm$0.10 & 1 & 2.11$\pm$0.15 & -0.98$\pm$0.09 & -1.84$\pm$0.17 & 0.49$\pm$0.09 & 1.23$\pm$0.19 & -1.68$\pm$0.18 \\
 & $\alpha$ & 0.98$\pm$0.09 & -0.18$\pm$0.13 & 0 & -2.96$\pm$0.19 & 4.71$\pm$0.11 & 10.01$\pm$0.19 & 0.76$\pm$0.11 & 0.02$\pm$0.24 & 1.88$\pm$0.23 \\
\hline
(4) & $\rho_{s}$ & 0.60 & 0.72 & 0.85 & 1 & -0.53 & -0.58 & 0.42 & 0.51 & -0.12 \\
 & $d_{s}^{1}$ & -5. & -8. & -14. & -$\infty$ & -4. & -5. & -3. & -4. & -0. \\
 & $\beta$ & 0.38$\pm$0.04 & 0.71$\pm$0.05 & 0.47$\pm$0.03 & 1 & -0.51$\pm$0.06 & -0.89$\pm$0.09 & 0.23$\pm$0.05 & 0.83$\pm$0.06 & -0.97$\pm$0.04 \\
 & $\alpha$ & 2.10$\pm$0.04 & 1.91$\pm$0.03 & 1.40$\pm$0.03 & 0 & 3.31$\pm$0.06 & 7.41$\pm$0.09 & 1.45$\pm$0.03 & 1.85$\pm$0.08 & -0.56$\pm$0.12 \\
\hline
(5) & $\rho_{s}$ & -0.72 & -0.80 & -0.59 & -0.53 & 1 & 0.65 & -0.59 & 0.08 & 0.63 \\
 & $d_{s}^{1}$ & -8. & -11. & -5. & -4. & -$\infty$ & -6. & -5. & -0. & -6. \\
 & $\beta$ & -0.79$\pm$0.07 & -1.51$\pm$0.15 & -1.02$\pm$0.09 & -1.96$\pm$0.23 & 1 & 1.85$\pm$0.17 & -0.43$\pm$0.06 & 1.11$\pm$0.15 & 1.89$\pm$0.14 \\
 & $\alpha$ & 4.71$\pm$0.26 & 6.94$\pm$0.54 & 4.79$\pm$0.34 & 6.48$\pm$0.82 & 0 & 1.27$\pm$0.59 & 2.87$\pm$0.20 & -2.46$\pm$0.55 & -6.82$\pm$0.47 \\
\hline
(6) & $\rho_{s}$ & -0.43 & -0.62 & -0.80 & -0.58 & 0.65 & 1 & -0.25 & 0.31 & 0.76 \\
 & $d_{s}^{1}$ & -3. & -6. & -11. & -5. & -6. & -$\infty$ & -1. & -2. & -10. \\
 & $\beta$ & -0.46$\pm$0.06 & -0.82$\pm$0.09 & -0.54$\pm$0.05 & -1.12$\pm$0.11 & 0.54$\pm$0.05 & 1 & -0.39$\pm$0.12 & 0.95$\pm$0.09 & 1.04$\pm$0.08 \\
 & $\alpha$ & 5.50$\pm$0.50 & 7.97$\pm$0.73 & 5.43$\pm$0.41 & 8.33$\pm$0.91 & -0.69$\pm$0.38 & 0 & 4.38$\pm$0.91 & -5.93$\pm$0.70 & -8.23$\pm$0.64 \\
\hline
(7) & $\rho_{s}$ & 0.50 & 0.76 & 0.37 & 0.42 & -0.59 & -0.25 & 1 & 0.30 & -0.32 \\
 & $d_{s}^{1}$ & -4. & -10. & -2. & -3. & -5. & -1. & -$\infty$ & -1. & -2. \\
 & $\beta$ & 1.79$\pm$0.20 & 3.87$\pm$0.40 & 2.03$\pm$0.37 & 4.42$\pm$1.00 & -2.33$\pm$0.31 & -2.58$\pm$0.78 & 1 & 2.81$\pm$0.96 & -3.09$\pm$0.77 \\
 & $\alpha$ & -0.49$\pm$0.27 & -3.64$\pm$0.54 & -1.55$\pm$0.51 & -6.43$\pm$1.36 & 6.70$\pm$0.42 & 11.30$\pm$1.07 & 0 & -2.32$\pm$1.30 & 4.06$\pm$1.05 \\
\hline
(8) & $\rho_{s}$ & 0.22 & 0.17 & 0.17 & 0.51 & 0.08 & 0.31 & 0.30 & 1 & 0.65 \\
 & $d_{s}^{1}$ & -1. & -1. & -1. & -4. & -0. & -2. & -1. & -$\infty$ & -6. \\
 & $\beta$ & 0.70$\pm$0.18 & 0.94$\pm$0.07 & 0.82$\pm$0.13 & 1.21$\pm$0.09 & 0.90$\pm$0.13 & 1.05$\pm$0.10 & 0.36$\pm$0.12 & 1 & 1.14$\pm$0.13 \\
 & $\alpha$ & 0.89$\pm$0.26 & 0.20$\pm$0.13 & -0.01$\pm$0.20 & -2.24$\pm$0.15 & 2.21$\pm$0.20 & 6.26$\pm$0.18 & 0.83$\pm$0.18 & 0 & -1.79$\pm$0.19 \\
\hline
(9) & $\rho_{s}$ & -0.38 & -0.57 & -0.41 & -0.12 & 0.63 & 0.76 & -0.32 & 0.65 & 1 \\
 & $d_{s}^{1}$ & -2. & -5. & -3. & -0. & -6. & -10. & -2. & -6. & -$\infty$ \\
 & $\beta$ & -0.50$\pm$0.08 & -0.81$\pm$0.08 & -0.60$\pm$0.07 & -1.03$\pm$0.04 & 0.53$\pm$0.04 & 0.96$\pm$0.08 & -0.32$\pm$0.08 & 0.88$\pm$0.10 & 1 \\
 & $\alpha$ & 1.87$\pm$0.04 & 1.50$\pm$0.06 & 1.12$\pm$0.05 & -0.57$\pm$0.12 & 3.60$\pm$0.04 & 7.93$\pm$0.06 & 1.31$\pm$0.03 & 1.58$\pm$0.07 & 0 \\
\hline
   \end{tabular}
 \end{minipage}
\end{table*}

\clearpage
\section{Mean AGN SEDs Using Model-A Fitting ({\it optxagn\/}: without Color Correction)}
\label{app:section:mean:sed}
\begin{figure*}
\centering
\includegraphics[scale=0.73,clip=]{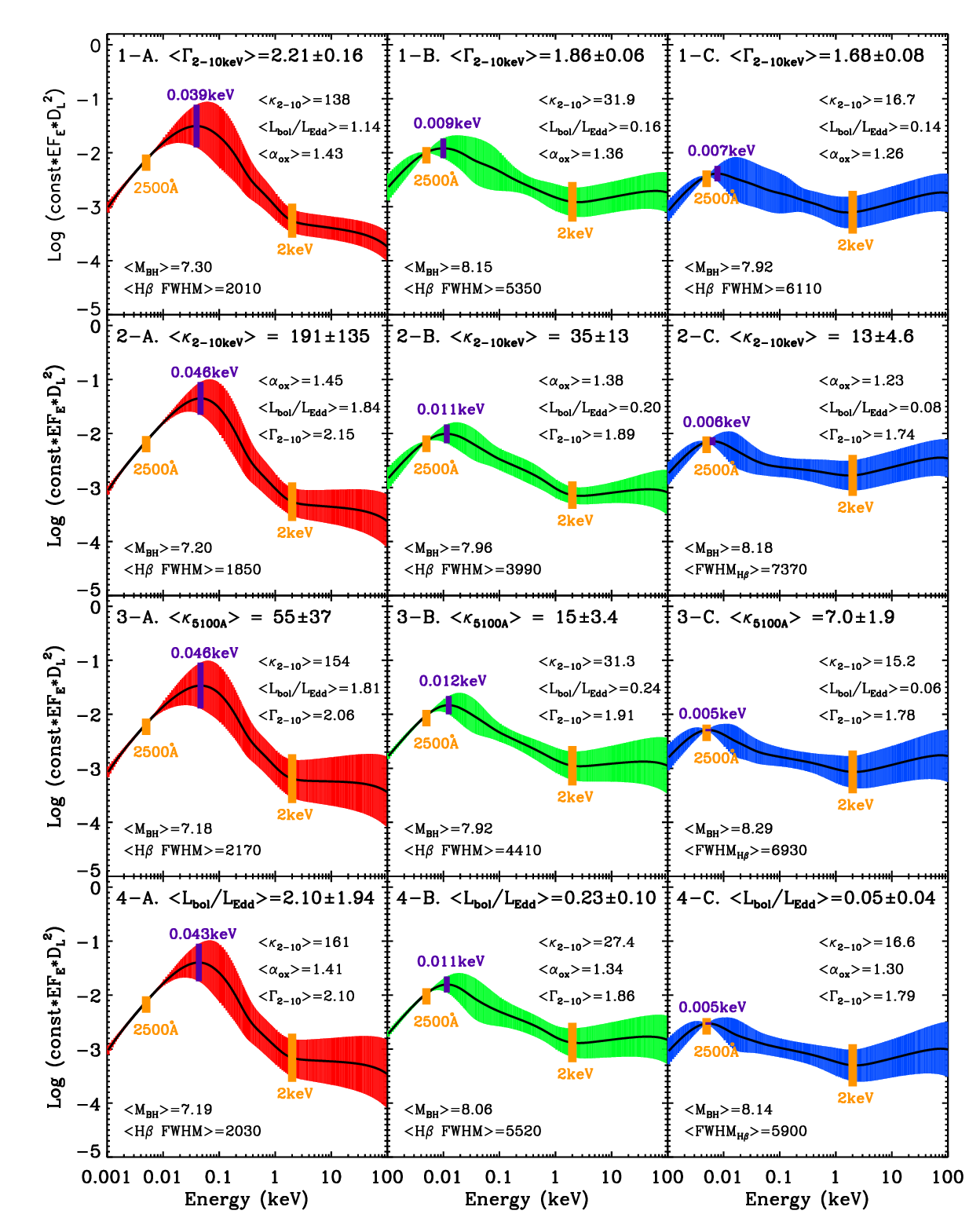}
\caption{The AGN mean SEDs based on different values of the 9 key parameters from Model-A fitting in Paper-I (i.e. without effect of color temperature correction). For each parameter, the 51 sources are sorted according to the parameter value, and are then divided into three subsets evenly so that each subset contains 17 sources. PG 1004+130 is excluded from its subset. Finally, a mean SED is constructed for each of the three subsets after renormalizing each individual SED to the mean 2500{\AA} luminosity of that subset. The three panels (A, B, C) in each row show the mean SEDs of the subsets classified by the parameter shown in the panel title. In each panel, the solid curve is the mean SED, while the color shaded region is the $\pm$1$\sigma$ deviation. The 2500 {\AA} and 2 keV positions are marked by the vertical solid orange lines, whose related height shows the value of $\alpha_{ox}$, The SED peaking position is also marked by the vertical solid purple line. The average values of some other parameters in that subset are also shown in the panel. All the mean SEDs have been rescaled by the same arbitrary constant in the Y-axis which is 1.3$\times$10$^{-46}$. Note that the energy ranges 1.4 eV $<$ E $<$ 6 eV and 0.3 keV $<$ E $<$ 10 keV are covered by SDSS, OM and EPIC data, while the SED in the rest energy bands is determined by the extrapolating of the best-fit model.}
\label{app:best-sed-indicator:old}
\end{figure*}
\addtocounter{figure}{-1}
\begin{figure*}
\centering
\includegraphics[scale=0.73,clip=]{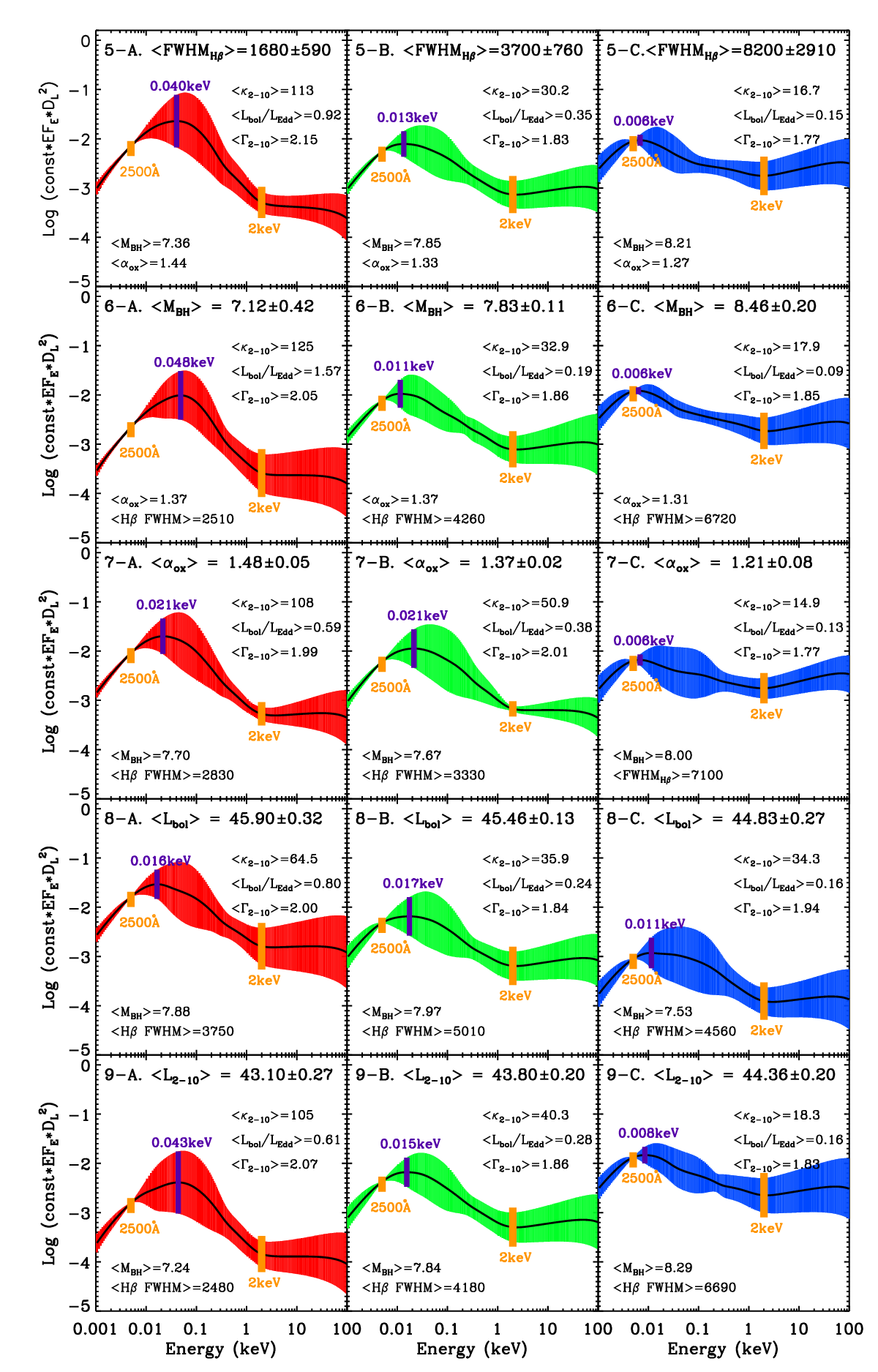}
\caption{{\it continued}}
\end{figure*}


\end{document}